\documentclass[12pt]{article}

\usepackage{lscape,multirow}
\usepackage{graphics,amsmath, amssymb,multirow, mathrsfs,graphicx,bm,colordvi,verbatim}
\usepackage{setspace, natbib}
\usepackage{caption, tabularx}
\usepackage{enumitem,colortbl}
\usepackage{url}
\usepackage{tablefootnote, ragged2e}
\usepackage{mathptmx}
\usepackage{pdflscape}
\usepackage{amsthm}
\usepackage{changepage}
\usepackage{tikz, tikz-cd, tikz-dependency}
\usepackage{float}
\usepackage{paralist}
\usepackage{booktabs}
\usepackage{tabularx}
\usepackage{graphicx}
\usepackage{subcaption}
\usepackage{setspace}

\newcommand{\sym}[1]{\rlap{#1}}
\newcommand{\citeay}[1]{\citeauthor{#1}, \citeyear{#1}}

\usepackage{hyperref}
\hypersetup{
    colorlinks=true,       
    linkcolor=blue,          
    citecolor=blue,        
    filecolor=magenta,      
    urlcolor=blue           
}

\usepackage{booktabs}
\usepackage{longtable}

\usepackage{dcolumn}
\newcolumntype{d}[1]{D..{#1}} 

\newcolumntype{Y}{>{\centering\arraybackslash}X}

\usepackage[left=1in,right=1in,top=1in, bottom=1in]{geometry}

\newcommand{\st}{\begin{eqnarray}}
    \newcommand{\nd}{\end{eqnarray}}
\newcommand{\stt}{\begin{eqnarray*}}
    \newcommand{\ndd}{\end{eqnarray*}}

\usepackage{setspace}
\usepackage{siunitx}

\graphicspath{{./figures/}}

\begin{document}
\begin{titlepage}
\title{\bf The Value of Information from Sell-side Analysts\thanks{Linying Lv is at Washington University in St. Louis (\url{llyu@wustl.edu}). I am grateful to John Barrios, Romain Boulland (discussant), William Cassidy, Olivier Dessaint, Philip Dybvig, Richard Frankel, Zhiyu Fu, Xiao Han, Songrun He, Jean Helwege, Jiantao Huang, Jared Jennings, Zachary Kaplan, Hong Liu, Semyon Malamud, Asaf Manela, Matilde Faralli, Xiumin Martin, Stefan Nagel, Andreas Neuhierl, Michaela Pagel, Christine Parlour, Michael Wittry, Liyan Yang, Zilong Zhang, Guofu Zhou, and Qifei Zhu for their valuable comments. I also thank the participants at the AFA 2025 conference, AFFECT 2025 workshop, 7th Future of Financial Information Conference, SoFiE 2024 workshop, Wolfe Research 8th Global Quant and Macro Conference, the 1st Workshop on Large Language Models and Generative AI for Finance, and seminar attendees at Washington University in St. Louis for their helpful comments and discussions.}}
\author{Linying Lv}
\date{First draft October 2024. This version June 2025.}
\maketitle
\thispagestyle{empty}
\begin{center}
    {\bf Abstract}
\end{center}
\begin{spacing}{1.5}
I examine the value of information from sell-side analysts by analyzing a large corpus of their written reports. Using embeddings from state-of-the-art large language models, I show that qualitative information in analyst reports explains above 10\% of contemporaneous stock returns out-of-sample, a value that is more economically significant than quantitative forecasts. I then perform a Shapley value decomposition to assess how much each topic within the reports contributes to explaining stock returns. The results show that analysts' income statement analyses account for more than half of the reports' explanatory power. Expressing these findings in economic terms, I estimate that early acquisition of analyst reports can yield significant profits. Analyst information value peaks in the first week following earnings announcements, highlighting their vital role in interpreting new financial data.
\end{spacing} 
\end{titlepage}

\newpage

\begin{quote}
\emph{``No one ever made a decision because of a number. They need a story.''}  
\begin{flushright}
― Daniel Kahneman
\end{flushright}
\end{quote}
\section{Introduction}

Sell-side analysts play a crucial role in financial markets by producing and processing information for investors. Their research reports, which include both quantitative forecasts and qualitative analyses, are widely used by investors in decision-making (\citeay{barber2001can}; \citeay{kong2021whose}). This raises a fundamental question: Do analysts generate value for their clients, or are they merely peddling expensive noise?

Quantifying the value of analyst information, especially that contained in written reports, has been a persistent challenge in finance literature due to the unstructured nature of text. Prior studies have shed light on the incremental information value of report text through sentiment analysis (\citeay{asquith2005information}; \citeay{huang2014evidence}; \citeay{huang2018analyst}). However, analyst reports are meant to provide contextual information beyond the simple sentiment dimension. The context has been perceived to be at least as important as the quantitative outputs. According to the annual survey of \textit{Institutional Investor} magazine, investors consistently rank ``Written Reports'' as more valuable than earnings forecasts and stock recommendations. Yet, empirical evidence supporting this survey feedback is limited, and it remains unclear what specific discussion holds the most value for investors.

The advent of large language models (LLMs) has significantly enhanced our ability to quantify textual meaning beyond sentiment and topics, allowing us to capture both contextual information and the reasoning logic within the text (see, for example: \citeay{chen2022expected}; \citeay{li2024dissecting}). Consider the following two statements: ``Reported EPS of \$0.94 misses consensus of \$1.40 and our estimate of \$1.17.'' and ``Reported 1Q12 MS franchise sales are below our estimate, as sales were negatively impacted by unfavorable distribution channel dynamics.'' While both sentences analyze earnings in a negative tone, the latter provides additional contextual information and reasoning. The question now becomes: When presented with both, do investors prioritize the explicit figures or the explanatory narrative?

I investigate the research question through three general steps: analyst output representation, econometric modeling, and value decomposition. For quantitative information, I consider three numerical forecasts, including stock recommendations, earnings forecasts, and target prices. For qualitative information, I extract contextualized representations by mapping each report into a structured LLM embedding space.

With structured representation, I evaluate how well analyst information explains stock returns as evidence of information content. Due to the high dimensionality of LLM embeddings, projecting stock returns on them bears overfitting risk. To address this issue, I employ the Ridge regression. The simple linear model penalizes large coefficients while still preserving the model's ability to capture essential signals that move the market.

I find that the textual information generates an out-of-sample $R^2$ of 10.19\% for three-day cumulative abnormal returns ($CAR_{[-1,+1]}$), surpassing the 9.01\% explained by numerical information. When both textual and numerical measures are combined, the explainable variation in $CAR$ increases to 12.28\%, which is significantly higher than using either information type alone. These findings remain robust across various LLM representations, machine learning algorithms, and $CAR$ windows, highlighting the distinct and economically meaningful information content embedded in analyst report narratives. 

The informativeness of analyst reports varies substantially depending on whether they are accompanied by forecast revisions. Reports that revise earnings forecasts or recommendations convey significantly more value to investors than those that simply reiterate prior views. For reports containing recommendation revisions, the combination of numerical and textual information generates an out-of-sample $R^2$ of 22.63\%. In contrast, reports merely reiterating previous earnings forecasts yield statistically insignificant out-of-sample $R^2$.

Previous research emphasizes analysts' role in processing information around corporate earnings announcements (e.g., \citeay{livnat2012information}, \citeay{keskek2014analyst}, \citeay{kim2015management}, \citeay{lobo2017effect}, and \citeay{barron2017earnings}). Consistent with this literature, my analysis shows that the information content of analyst reports peaks in the first week following earnings releases. Comparing models trained separately on post-earnings and non-earnings periods, I find that the out-of-sample $R^2$ approximately doubles during the one-week earnings announcement window. The substantial increase highlights the incremental value analysts add by interpreting and contextualizing earnings news for investors.

To address the concern that the explanatory power may simply reflect ``piggybacking'' on earnings conference call information or post-earnings-announcement drift (PEAD), I control for both earnings surprises and the latest conference call transcripts. The results confirm that analyst reports provide distinct and valuable insights.

An unresolved question in the literature, highlighted by \citet{bradshaw2017financial}, is which qualitative aspects of analyst reports investors find most informative. This study addresses this gap by developing a systematic approach to quantify the relative importance of various types of report content. Specifically, I leverage the additive structure of text embeddings and implement Shapley value decomposition to attribute the explanatory power of analyst reports across 17 prevalent topics. The results reveal that Income Statement Analyses, especially interpretations of realized earnings, account for over half of this explanatory power. These results reinforce analysts' critical role in interpreting financial data and constructing the narratives essential for investment decisions.

To translate the statistical informativeness of analyst reports into economic significance, I adopt the \citet{Kadan_Manela_Infoval} framework to quantify the dollar value of analyst information for strategic investors. Their multi-agent model provides a realistic basis for assessing the economic impact of analyst report dissemination. The information value represents the total expected profits accruing to strategic investors who either: (1) gain early access to analyst views, or (2) independently derive insights comparable to those from analysts. Intuitively, the measure reflects both explainable return variance (the uncertainty reduction from analyst information) and the price impact (the cost of trading on such information).

The analysis indicates substantial economic value in analyst information. For an average S\&P 100 stock, the estimated profit over a three-day window for strategic investors with early access (two days prior to public release) is approximately \$0.34 million from numerical information, \$0.38 million from textual information, and \$0.47 million when both sources are combined. Assuming an average of 15 report days per year, this implies an annualized information value of \$6.89 million, a conservative lower bound according to \citet{Kadan_Manela_Infoval}. The value is notably higher for large-cap stocks, reports from bold analysts, and those released promptly following earnings announcements.

My paper relates to three strands of literature. First, this paper offers several extensions to prior research that highlight the incremental information in report narratives and the interpretative role of analysts (e.g., \citeay{asquith2005information}; \citeay{huang2014evidence}; \citeay{huang2018analyst}). Specifically, I show that more contextualized textual representations not only provide significant incremental information but can also surpass the explanatory power of analysts' traditional quantitative forecasts. Additionally, I demonstrate that the informational value derived from the report text extends substantially beyond simple sentiment measures. To illustrate, transitioning from basic textual tone measures to comprehensive LLM embeddings increases the out-of-sample $R^2$ from approximately 3\% to over 10\%. 

Furthermore, to my knowledge, this is the first study to systematically identify and quantify the specific qualitative attributes in analyst reports that investors find most salient. The research complements the existing literature by providing a more granular understanding of which textual elements drive the value of analyst reports and when that value peaks, such as after earnings announcements or when accompanying forecast revisions.

Second, this paper makes a distinct methodological contribution to the expanding literature on AI and machine learning in finance. While the application of LLMs in financial paradigms is promising (e.g., \citeay{chen2022expected}; \citeay{li2024dissecting}; \citeay{lopez2023can}; \citeay{jha2024chatgpt}), their inherent `black-box' nature presents significant hurdles for model validation and reliability. To address this critical interpretability gap, this paper implements a Shapley value decomposition framework specifically designed for LLM embeddings plus topic modeling. By leveraging the additive structure of these embeddings, this approach systematically attributes an LLM's performance to constituent textual features, thereby enhancing model transparency and offering a robust pathway toward explainable AI in financial applications.

Third, this paper offers new insights into analyst tipping. It provides a rigorous quantification of the economic dollar value accruing from privileged early access to research, a practice of brokerage firms providing early research access to high-commission clients (\citeay{irvine2007tipping}; \citeay{christophe2010informed}). While prior work (e.g., \citeay{green2006value}; \citeay{kadan2018trading}) documents significant abnormal returns for clients with early access, this study further translates the information advantages into tangible monetary terms. The dollar value offers a more concrete measure for understanding the incentive of selective disclosure and the scale of wealth transfer involved.

The rest of the paper is organized as follows. Section \ref{method} introduces the methodology, measures of analyst information value, and the interpretation framework. Section \ref{empirical} describes the data and empirical results. Section \ref{conclu} concludes.

\section{Methodology}\label{method}

This section describes the key methodologies: using large language models for text representation and topic modeling, and the decomposition techniques to evaluate the information value of individual topics.

\subsection{Text Representation}

A core step in analyzing the information value of analyst reports is constructing structured, numerical representations of textual content. LLMs offer a significant advancement by providing richer, more contextualized linguistic representations than traditional methods. 

As illustrated in Figure \ref{fig:transformer} in the online Appendix, the process begins with text tokenization, where text segments are mapped to unique identifiers based on a predefined vocabulary. These tokens are then processed through a transformer architecture. The transformer's key output consists of dense, continuous vectors that capture both semantic and syntactic information. To generate report-level representations, I compute embeddings as the average across all tokens and transformer layers:
\begin{equation}
y^{emb} = \frac{1}{N_kN_l}\sum_{l=1}^{N_l}\sum_{i=1}^{N_k}{e}_{i, l},
\end{equation}
where $e_{i,l}$ denotes the contextual embedding of token $i$ generated by layer $l$, $N_k$ is the total number of tokens in the report, $N_l$ is the number of layers from which embeddings are aggregated.

The embedding representation provides two salient advantages over traditional sentiment measures. First, LLMs position each token within a high-dimensional semantic space. This maintains complex relationships both within and across sentences, producing tone measurements that are substantially more nuanced and reliable than those generated by dictionary-based approaches. Second, these embeddings encode context-dependent meaning and narrative structure, thereby conveying information that extends well beyond simple positive-negative classifications and conveys far richer linguistic signals.

For this study, I use MetaAI’s pre-trained LLaMA-2-13B model and its associated tokenizer. With an input length limit of 4,096 tokens, the model accommodates almost all analyst reports, which have a median length of 1,393 tokens and a mean of 2,055 tokens. The resulting contextual embeddings serve as the foundation for downstream tasks such as topic attribution and explaining stock returns.

\subsection{Topic Modeling} \label{topic}

This section introduces the methodology used to identify prevalent themes in analyst reports and to classify each sentence into a single, primary topic. The initial challenge is selecting a topic modeling approach that effectively avoids ad-hoc classifications and ensures economically meaningful interpretations. I exclude manual topic construction to minimize researcher subjectivity.

Given the availability of sentence-level embeddings, clustering these embeddings based on semantic proximity is a straightforward procedure to consider. Accordingly, I expriment a standard K-Means algorithm for initial assessment.\footnote{Topic classification results from the K-Means experimentation are available upon request.} The unsupervised clustering approach is not adopted as the final methodology for two primary reasons. First, due to the industry-sensitive nature of analyst reports, K-Means tended to generate topics heavily aligned with particular sectors (e.g., pharmaceuticals, banking), whereas this study seeks to identify common thematic categories that span across industries. Second, as also noted by \citet{aleti2025news}, topics derived from purely data-driven clustering tend to be messy and difficult to interpret from an economically meaningful perspective.

To provide a set of distinct and interpretable topics, this study uses targeted prompting of the ChatGPT-4o model. The objective of this approach is to identify a comprehensive set of primary topics consistently discussed in analyst reports and to assign each sentence to one of these predefined categories.

In the first step, I use a random sample of 100 analyst reports and request ChatGPT-4o model to analyze the content and identify high-level categories with the following prompt:

\vspace{0.1in}
\noindent
\begin{tabular}{|p{16cm}|}
\hline
\textbf{Prompt 1:} Please read the provided text file of sell-side analyst reports carefully. What are high-level, mutually exclusive topics covered in these reports? Make sure that each sentence from the text file can be assigned to one of the topics. Here is the report content: \{text\}.\\
\hline
\end{tabular}
\vspace{0.1in}

It ends up with 16 categories defined in Table \ref{tab: topic}. A category labeled ``None of the Above'' is included to accommodate sentences that do not fit into any of the 16 topics. Table \ref{tab: topic_example} offers illustrative sentences of each category.

In the second step, I build a sentence classifier to assign each sentence to one topic. To begin with, I employ the GPT-4-turbo model to categorize each sentence in a random sample of 100 analyst reports into a single, most relevant topic.

\vspace{0.1in}
\noindent
\begin{tabular}{|p{16cm}|}
\hline
\textbf{Prompt 2:} Please read the following sentence from a sell-side analyst report of the company \{firm\} (\{ticker\}) carefully. Determine which category of information it belongs to in the following 17 categories: \{categories\}. Pay attention to the sentence in the context of the report. \\
Output your response in JSON format. \\
Here is the sentence from the analyst report: \{sentence\}.\\
\hline
\end{tabular}
\vspace{0.1in}

\begin{figure}[H]
    \begin{center}
    \includegraphics[width = \textwidth]{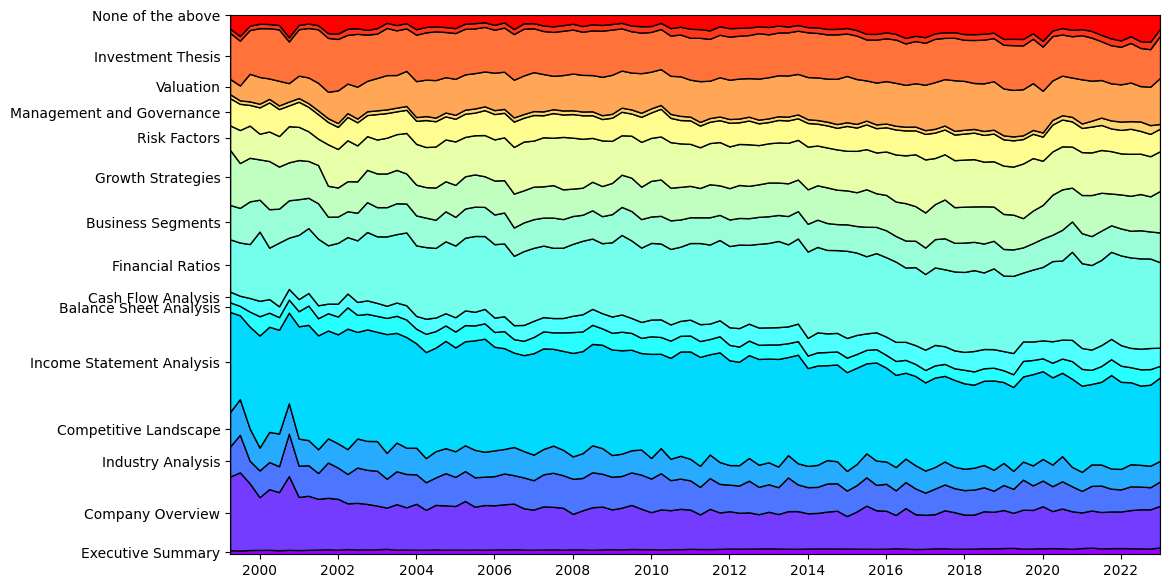}
    \end{center}
    \caption{Distribution of Topics over Time}
    \label{fig:stack}
    {\footnotesize This figure shows the distribution of report sentences across 17 topics from 2000Q1 to 2023Q4. The stacked plot illustrates the proportional composition of sentences over time. The topic categories are Executive Summary, Company Overview, Industry Analysis, Competitive Landscape, Income Statement Analysis, Balance Sheet Analysis, Cash Flow Analysis, Financial Ratios, Business Segments, Growth Strategies, Risk Factors, Management and Governance, ESG Factors, Valuation, Investment Thesis, Appendices and Disclosures, and None of the Above.}
\end{figure}

I then fine-tune a BERT-based classifier using the labeled sample and classify all 6,975,114 sentences in the sample. The distribution of topics is relatively stable across years, as illustrated in Figure \ref{fig:stack}. Income Statement Analyses and Financial Ratios are the most frequently discussed topics, accounting for 17.23\% and 15.65\% of all sentences, respectively. These are followed by Risk Factors (8.32\%), Valuation (8.10\%), and Investment Thesis (7.87\%), which collectively form the next tier of frequently discussed topics. While ESG discussions are minimal overall (0.23\%), they increase notably after 2020, reflecting growing market interest in sustainability issues.

I validate the topic modeling procedure based on three key evidence. First, the fine-tuned BERT classifier demonstrates high performance, achieving 89\% accuracy on the test sample. Given the challenge of distinguishing among 17 topic categories, it demonstrates a notably strong performance. Second, the 16 GPT-generated topics provide comprehensive coverage of the report content. Less than 5\% of sentences fall into the fallback category ``None of the Above'', indicating that the topic set captures nearly all relevant content. Third, Figure \ref{fig: a1} presents word clouds showing the most frequent terms within each topic. The word frequencies align closely with expected topic content. For example, ``Income Statement Analysis'' is strongly associated with terms like revenue, EPS, and sales, confirming the semantic coherence of the classification.

\subsection{Shapley Value Decomposition} \label{shap}

Next, I propose a methodology to decompose the value of different topics within analyst reports. As a starting point, a simple approach is aggregating the embeddings of tokens corresponding to each topic and measure their corresponding value independently. However, this method has two significant drawbacks. First, the self-attention architecture of transformer models creates cross-topic contamination. It means, each token's embedding incorporates contextual information from all other tokens in the report, including those assigned to different topics. Second, topics are not independent. Their interactions can be a primary driver of value. For example, the combination of a valuation discussion and an earnings analysis may be more informative than either alone.

To mitigate the first concern of cross-topic contamination, I construct sentence-segmented embeddings. Specifically, each sentence is processed by the LLM individually. The procedure ensures that the resulting token embeddings $\left(e_i\right)$ are contextualized only by the words within their own sentence. A report-level embedding $\left(y^{emb}\right)$ is then constructed as the token-weighted average of these sentence embeddings:
\begin{equation}
y^{emb} = \sum_{i=1}^{n} \frac{Token_i}{\sum_{i=1}^{n}Token_i} S_i^{emb},
\end{equation}
where $S_i^{emb}$ is the average token embeddings of sentence $i$, $n$ is the number of sentences in the report, and $Token_i$ is the number of tokens in sentence $i$.

To account for topic interdependency, I implement the Shapley Additive exPlanations (SHAP) framework. Originating in cooperative game theory \citep{shapley1953value}, the Shapley value fairly allocates a collective payoff among contributing members. This concept is widely used in machine learning to ensure fair attribution of model predictions to individual features \citep{lundberg2017unified, chen2022expected}.

The SHAP framework is ideal for topic value attribution for two reasons. First, it satisfies the additivity (or efficiency) property, ensuring that the total information content of a report is fully distributed among all constituent topics without omissions or double-counting. Second, it inherently accounts for topic interdependencies by evaluating the marginal contribution of each topic across all possible topic combinations.

The Shapley value $\left(\varphi_p\right)$ for a specific topic $p$, which quantifies its contribution to the overall out-of-sample explanatory power ($R_{\text {oos }}^2$), is defined as:
\begin{equation}
\label{eq:shapley_value}
\varphi_p(R_{oos}^2) = \sum_{S \subseteq P \setminus{p}} \frac{|S|!(|P|-|S|-1)!}{|P|!} \left[ R_{oos}^2(S \cup {p}) - R_{oos}^2(S) \right],
\end{equation}
where $P$ is the set of all topics in a report. The summation is over all subsets $S$ of topics excluding $p$. The term $R_{\text {oos }}^2(S)$ denotes the out-of-sample $R^2$ of a model whose input is the report embedding constructed using only sentences belonging to topics in the subset $S$.\footnote{I define missing topic embeddings as zero vectors, effectively removing them from the model input. For a given subset of topics $S$, the corresponding embedding is constructed using the token-weighted average formula above, but summed only over sentences classified into topics within $S$.}

\section{Data and Empirical Results}\label{empirical}

This section describes the data and presents empirical results of the analyst information value. I begin with an examination of analyst information content and its decomposition across topics. I then translate these findings into economic terms by quantifying the strategic dollar value of analyst information.

\subsection{Data}\label{data}

The initial sample comprises 223,091 sell-side analyst reports of S\&P 100 constituent stocks from 2000 to 2023 downloaded from the Mergent Investext database. After converting the reports from their original PDF format to plain text using Adobe Acrobat, the extracted text is segmented into individual sentences. To focus on stock-relevant discussions, I fine-tune a BERT classifier to identify and remove boilerplate content.\footnote{Boilerplate segments typically contain disclosures about the brokerage firm and analyst research team and do not reflect analyst opinions. Following \citet{li2024dissecting}, I create a training set by sampling one report from each of the top 20 brokerage firms and manually labeling sentences. The fine-tuned BERT model achieves 97.31\% accuracy on a held-out test set.}

The Investext database provides detailed metadata for analyst reports, including analyst identities and release dates. Reports are linked to I/B/E/S forecasts by matching analyst names, firm tickers, and the matching window. I first match lead analyst names to unique analyst identifiers (AMASKCD) following the procedure in \citet{li2024dissecting}. I then associate each report with corresponding earnings forecasts, target prices, and stock recommendations in I/B/E/S. Since multiple reports may be issued between a forecast announcement and revision, I adopt the windowing strategy of \citet{huang2014evidence}, linking reports to forecasts issued between two days prior to the I/B/E/S announcement date (ANNDATS) and two days after the revision date (REVDATS).\footnote{I/B/E/S records both the forecast announcement date (ANNDATS) and the revision date (REVDATS), which marks when the forecast becomes outdated.} 

To isolate the distinct information content of analyst reports, I exclude reports released on earnings announcement days. The final matched sample consists of 122,252 analyst reports on S\&P 100 stocks, authored by 1,305 analysts across 140 brokerage firms. Table \ref{tab:sum1} provides summary statistics of analyst reports across years and the Fama-French 12 industries, including the number of reports, unique brokerage firms, and analysts. The table also reports average report length, measured by page and token counts. The temporal distribution of reports follows the pattern documented by \citet{bonini2023value}, with a peak in 2013 followed by a steady decline, likely driven by regulatory changes such as the Dodd-Frank Act and MiFID II \citep{bradshaw2017financial}.

To quantify market reactions, I estimate stock-level price impacts using intraday data from the NYSE TAQ database for the out-of-sample period of January 2, 2015, to December 28, 2023. All monetary values are adjusted for inflation using the Consumer Price Index (CPI) and are normalized to December 2020.

Additionally, I match each analyst report with the latest earnings conference call transcript to control for concurrent events. Transcripts for S\&P 100 firms are sourced from Seeking Alpha and matched to firm-quarter earnings announcement dates (RDQ) from Compustat, resulting in a sample of 7,909 firm-quarter earnings conference calls from 2005 to 2023.\footnote{Transcripts are available at \href{https://seekingalpha.com/earnings/earnings-call-transcripts}{https://seekingalpha.com/earnings/earnings-call-transcripts}.} A comparative analysis is presented in Section \ref{ea}.

I include a set of firm-specific numerical features in the machine learning models, categorized as follows. Firm characteristics are sourced from CRSP and Compustat.\footnote{Detailed variable definitions are provided in Table \ref{tab:numdef}, and their summary statistics are in Table \ref{tab:numss}.}

\textbf{Report-Level Measures}: $REC_{REV}$ denotes recommendation revision, calculated as the current report's recommendation minus the last recommendation in I/B/E/S issued by the same analyst for the same firm. $EF_{REV}$ represents earnings forecast revision, calculated as the current report's EPS forecast minus the prior EPS forecast in I/B/E/S issued by the same analyst for the same firm, scaled by the stock price 50 trading days prior to the report’s release. $TP_{REV}$ measures target price revision, calculated as the difference between current and previous target prices in I/B/E/S for the same analyst-firm pair, scaled by the stock price 50 trading days prior to release. $Boldness$ is a binary indicator set to 1 when revisions are either above both the consensus and previous forecasts or below both. $SR$ denotes stock recommendations in the I/B/E/S rating system (1 = Strong Buy, 5 = Strong Sell). $ERet$ represents the 12-month return forecast, calculated as the 12-month price target scaled by the stock price 50 trading days prior to release. $PriorCAR$ measures cumulative abnormal return over [-10, -2] window preceding the report release.

\textbf{Firm-Level Measures}: $Size$ is quantified by market equity. Book-to-Market Ratio ($BtoM$) represents the ratio of book value to market value. Earnings surprise ($SUE$) is calculated as the actual earnings per share (EPS) minus the last consensus EPS forecast prior to the earnings announcement date (EAD), with $AbsSUE$ denoting its absolute value. $Miss$ is a binary indicator equal to 1 if $SUE < 0$. $TradingVolume$ represents the stock's trading volume on recent earnings announcement days, standardized by shares outstanding. Distance to Default ($DD$) is sourced from the NUS Credit Research Initiative (CRI). $Fluidity$ measures firms' product market competition following \citet{hoberg2014product}.

\textbf{Industry-Level Measures}: $IndustryRecession$ is a binary indicator set to one if the stock's Fama-French 48 industry experiences negative monthly returns in the bottom quintile.

\textbf{Macroeconomic Measures}: $TimeTrend$ is a temporal variable measuring the number of years since the start of the sample period.

\subsection{Information Content of Analyst Reports}

I start by investigating market reactions to analyst quantitative and qualitative outputs. Following \citet{asquith2005information}, I consider $R^2$ as a proxy for the information content of analyst reports, which quantifies how well analyst information explains contemporaneous stock returns. I estimate Ridge regression models annually using an expanding window. The specification is given by:
\vspace{-0.05in}
\begin{equation}\label{ridge_1}
CAR_{[-1,+1], i t}=\beta_0+\beta^{\prime} y_{ijt}^{\mathrm{AI}}+\epsilon_{i j t},
\end{equation}
\begin{equation}\label{ridge_2}
\hat{\beta}=\underset{\beta}{\operatorname{argmin}}\left\{\left\|C A R_{[-1,+1], i t}-\beta_0-y_{i j t}^{\mathrm{AI}} \beta\right\|_2^2+\theta\|\beta\|_2^2\right\},
\end{equation}
where $CAR_{[-1,+1], it}$ is the three-day window cumulative abnormal return of stock $i$ excess to market return surrounding reports' release day $t$, and $y_{ijt}^{\mathrm{AI}}$ denotes the structured information representation extracted from analyst report $j$ of stock $i$ at day $t$. Abnormal returns are calculated as the raw return minus the buy-and-hold return on the NYSE/AMEX/NASDAQ value-weighted market return. For reports issued on non-trading days, $t=0$ is adjusted to the next available trading day on CRSP.

In-sample $R^2$ can be inflated due to overfitting, particularly when employing machine learning techniques and high-dimensional embeddings. Following \cite{gu2020empirical}, I calculate the out-of-sample $R^2$ using:
\begin{equation}
R_{\mathrm{oos}}^2 = 1 - \frac{\sum_{(i, t) \in \mathcal{T}}\left(CAR_{[-1,+1], i t} - \widehat{CAR}_{[-1,+1], i t}\right)^2}{\sum_{(i, t) \in \mathcal{T}} CAR_{[-1,+1], i t}^2},
\end{equation}
where $\mathcal{T}$ is the test set of $(i,t)$ which has not been used in training and validation. $R_{\mathrm{oos}}^2$ aggregates estimation across analyst reports and captures the proportion of return variance attributable to analyst information. The initial training sample comprises 60\% of the dataset, with out-of-sample testing spanning 2015 to 2023.

I implement a modified version of the \citet{diebold2002comparing} (DM) test to compare the statistical significance of the $R_{\mathrm{oos}}^2$. Given the strong cross-sectional dependence inherent in the stock-level analysis, I follow \citet{gu2020empirical} and compare the cross-sectional means of squared prediction errors to evaluate the improvement over a benchmark.

\subsubsection{Quantitative and Qualitative Information}

Table \ref{tab:base} evaluates the information content of numerical versus textual measures. Panel A presents the performance of four input configurations: (1) forecast revisions, encompassing analyst revisions on recommendations, earnings forecasts, and target prices; (2) 17 numerical measures, forecast revisions plus the report-level, firm-level, industry-level, and macroeconomic indicators detailed in Section \ref{data}; (3) text embeddings, 5,120-dimensional full-context representations; and (4) a hybrid input combining forecast revisions with report embeddings. 

Column (1) reveals an $R^2_{\mathrm{OOS}}$ of 9.01\% for the forecast revisions, consistent with existing literature \citep{brav2003empirical, asquith2005information, bradshaw2013sell} that documents the price impact of analyst revisions. The incorporation of 14 additional numerical features in column (3) yields a modest improvement to an $R^2_{\mathrm{OOS}}$ of 9.08\%. Both revision and numerical input models significantly outperform the zero benchmark, as evidenced by the DM t-statistics reported in columns (2) and (4).

Columns (5) and (6) show that text embeddings alone achieve an $R^2_{\mathrm{OOS}}$ of 10.19\% with a t-statistic of 8.20, highlighting the substantial information content of report text. The explanatory power of textual inputs remains relatively stable across the 2015–2023 period, with the exception of a dip during the 2020 pandemic recession. A DM test comparing revision-only and text-only inputs yields a t-statistic of 1.66, indicating that report text offers comparable or marginally larger information content relative to quantitative forecasts.

One potential concern is that the explanatory power might stem from the numbers mentioned in the text. To address this concern, I conduct robustness checks using text embeddings generated by reports with all numbers removed. Table \ref{tab:rmnum} presents the performance of models using number-free text embeddings. The resulting $R^2_{\mathrm{OOS}}$ of 10.95\% exceeds that of quantitative revision measures, with a t-statistic of 2.40, suggesting that the superior performance of text-based models is not driven by embedded numerical information.

To ensure the results aren't look-ahead biased by the models having seen future information (\citeay{sarkar2024lookahead}), I evaluate model performance in 2023. Since all the language models in this study were trained only on pre-2023 text, this setting ensures a truly out-of-sample test. The $R^2_{\mathrm{OOS}}$ of 9.30\% reinforces the robustness of the main findings and confirms that the explanatory power is not driven by information leakage.

Columns (7) and (8) report the model performance when combining forecast revisions with text embeddings. The combination achieves an $R^2_{\mathrm{OOS}}$ of 12.28\%, representing improvements of 3.27\% and 2.09\% over revision-only and text-only models, respectively. The DM t-statistics of 3.95 and 3.77 demonstrate that both the report text and forecast revisions contain distinct, complementary information valued by investors.

To validate that the findings are not contingent on a specific model choice, I test their robustness across both alternative language models used to generate embeddings and the machine learning algorithms used for estimation.

First, I assess the sensitivity to the choice of LLMs by replicating the analysis with embeddings from several alternatives: BERT, OpenAI's text-embedding-3-small, and LLaMA-3-8B model. As shown in Panel B of Table \ref{tab:base}, while smaller models like BERT and OpenAI underperform the LLaMA-2 benchmark, the similar-size LLaMA-3 model achieves a comparable $R^2_{\mathrm{OOS}}$ of 9.66\%.

Second, I test the LLaMA-2 embeddings across three conceptually distinct machine learning algorithms: Partial Least Squares (PLS), XGBoost, and Neural Networks (NNs). The results, summarized in Table \ref{tab:ml}, show that the significant explanatory power is successfully identified by all models, confirming that the main conclusions are not an artifact of using Ridge regression. While NNs achieve the highest performance, the strong results from linear models like PLS and Ridge indicate that a substantial portion of the information captured in the embeddings is linearly related to stock returns.

Collectively, the tests demonstrate that the significant, market-moving information contained in analyst report narratives is robustly captured across a range of language models and machine learning algorithms.

To ensure the findings are not contingent on the specific three-day event window used in the main analysis, I test the robustness of results to alternative return horizons. I re-estimate Ridge models using several different dependent variables for the CAR, including a one-day window $\left(AR_{[0]}\right)$, a two-day window $\left(CAR_{[0,+1]}\right)$, and a five-day window $\left(CAR_{[-2,+2]}\right)$. As presented in Table \ref{tab:car_alt}, the explanatory power of the textual embeddings remains stable and economically significant across all tested windows (out-of-sample $R^2_{\mathrm{OOS}}$ ranges from 6.92\% to 7.57\%), confirming that the central finding is robust to the choice of the contemporaneous return measurement window.

I next explore whether the information content of analyst reports varies systematically across industries. Table \ref{tab: ind} presents the performance of text-based models across the Fama-French 12 industry classifications. A pronounced heterogeneity emerges. The model demonstrates substantial explanatory power (defined as $R^2_{\mathrm{OOS}} > 10\%$) in five industries: Shops, Manufacturing, Chemicals, Durables, and a diversified ``Other'' category. Conversely, the information content is significantly lower in sectors such as Non-Durables, Telecommunications, Energy, and Utilities.

This cross-sectional pattern, also visualized in Figure \ref{fig:iv_ind}, supports the economic intuition that the value of detailed analyst reports is greatest for firms with higher operational complexity or informational uncertainty. In contrast, for industries operating in more stable, regulated, or commoditized environments, the incremental value of qualitative analyst insights appears to be more limited.

To assess the economic significance and directly compare the information content of different sources, I conduct a series of ``horse race'' regressions. The dependent variable is the actual three-day cumulative abnormal return, $CAR_{[-1,+1], it}$, and the primary independent variables are the out-of-sample predictions generated by Ridge regressions. The OLS specification is:
\begin{equation}\label{equation:ols}
CAR_{[-1,+1],it} = \alpha + \beta_{1}\widehat{CAR}_{txt,it} + \beta_{2}\widehat{CAR}_{rev, it} +\epsilon_{it}.
\end{equation}

This model incorporates out-of-sample $CAR_{[-1,+1]}$ estimations using either report text embeddings or forecast revisions as input. The estimates for 2015, for example, are derived from Ridge regression models trained on 2000-2014 data. The $\widehat{CAR}_{txt}$ condenses the 5120-dimension text information into a single number, while $\widehat{CAR}_{rev}$ aggregates the information from the three forecast revision measures into a single variable.

Table \ref{tab:ols}, Panel B, presents the regression results.\footnote{The sample for this analysis is restricted to observations with valid data for all current and prior forecast revision variables.} As a baseline, Column (1) confirms that the individual revision measures for recommendations, earnings forecasts, and target prices are all significant predictors of abnormal returns, consistent with prior literature \citep{brav2003empirical, asquith2005information}.

The subsequent columns conduct the main ``horse race''. A model with only the estimated CAR from text $\left(\widehat{CAR}_{txt}\right)$ as a regressor (Column 2) achieves an $R^2$ of $10.5\%$. A model with only the estimated CAR from revisions $\left(\widehat{CAR}_{rev}\right)$ yields an $R^2$ of $8.9 \%$ (Column 4). When both predictors are included in the same regression (Column 5), the adjusted $R^2$ rises to $14.7\%$. The coefficients on both $\widehat{CAR}_{txt}$ and $\widehat{CAR}_{rev}$ remain highly statistically significant, underscoring that the textual content of analyst reports provides substantial explanatory power over and above that contained in forecast revisions.

The economic magnitudes of these effects are also significant. Based on the coefficients from standardized variables, a one-standard-deviation increase in the predicted CAR from report text $\left(\widehat{C A R}_{t x t}\right)$ is associated with a 120 basis point increase in the actual 3-day abnormal return. The equivalent impacts for recommendation, earnings, and target price revisions are 10, 40, and 90 basis points, respectively.

Finally, Columns (6) and (7) test the robustness of these findings by including controls of the latest earnings surprises ($SUE$) and recent stock performance ($PriorCAR$). The stability of the coefficients on both $\widehat{CAR}_{txt}$ and $\widehat{CAR}_{rev}$ indicates that the information captured by analyst reports is distinct from and incremental to publicly available information about recent earnings and past returns.

\subsubsection{The Timing of Analyst Reports around Earnings Announcements} \label{ea}

Building on prior research examining the interplay between analyst reports and corporate disclosures (\citeay{frankel2006determinants}; \citeay{chen2010relationship}), this section investigates the information content of analyst reports around earnings announcements.

Following \citet{chen2010relationship}, I group analyst reports into 13 weekly bins relative to the earnings announcement date. Panel A in Table \ref{tab:ea} presents the $R^2_{\mathrm{OOS}}$ and DM t-statistics for each bin. Report issuance peaks in the week immediately following earnings announcements. This is also the period when reports are most informative. The $R_{\mathrm{OOS}}^2$ reaches nearly $10\%$ (DM t-statistic $=10.17$ ), indicating significant market reaction to analysts' timely interpretations of earnings news.

To account for the possibility that analyst information differs depending on its timing relative to an EA, I estimate separate Ridge models for reports issued within different event windows. This approach allows the models to learn distinct relationships between text and returns for reports published in information-rich versus information-sparse periods. Panel B of Table \ref{tab:ea} compares the performance of models trained on reports from varying windows ($1,2,3$, and $7$ days post-EA). The models trained on reports issued most promptly after earnings consistently outperform those trained on other periods. Notably, the $R_{\mathrm{OOS}}^2$ declines from $11.84\%$ to $7.29\%$ as the training sample window expands from one to seven days, suggesting that the market places the highest value on immediate, timely analysis following corporate disclosures.

A critical question is whether analysts provide unique insights or merely act as conduits, reiterating information already disclosed in earnings conference calls. To test this, I conduct a direct comparison of the information content of analyst reports and the latest earnings conference call transcripts.\footnote{Given the significant length of conference call transcripts (median 13,029 tokens), I employ the LLaMA-3-8B model for this analysis, as its 8,192-token context window is larger than the 4,096-token limit of the LLaMA-2-13B model used elsewhere. Texts are truncated at these limits.} I deliberately focus on reports released one day after an EA, creating a ``hard test'' for the analyst's contribution, as this is when both analyst reports and contemporaneous returns are more likely to be influenced by earnings conference call information. I estimate Ridge regressions that combine both information sources:

$$
C A R_{[-1,+1], i t} =\beta_0+\beta^{\prime} y_{ijt}^{{\text {AI+EA}}}+\epsilon_{ijt},
$$
$$
\hat{\beta} = \underset{\beta}{\operatorname{argmin}} \left\{ \left\| C A R_{[-1,+1], i t} - \beta_0 - y_{ijt}^{{\text{AI+EA}}} \beta \right\|_2^2 + \theta \|\beta\|_2^2 \right\},
$$
where $y_{i j t}^{\mathrm{AI}+\mathrm{EA}}$ is created by concatenating the embeddings from analyst report $j$ for stock $i$ on day $t$ with the embeddings from that stock's most recent earnings conference call.

Panel C of Table \ref{tab:ea} presents the results. A model using only the text embeddings of earnings conference call transcripts achieves an $R^2_{\mathrm{OOS}}$ of $4.20\%$ (Column 1). The transcript data starts in 2005. For a fair comparison, I retrain analyst report models over the same 2005-2023 sample period. This model achieves an $R_{\mathrm{OOS}}^2$ of $9.72\%$ (Column 2), substantially outperforming the explanatory power of the earnings conference call transcripts alone.

Finally, a model that combines embeddings from both analyst reports and earnings conference calls yields an $R_{\mathrm{OOS}}^2$ of $11.96\%$ (Column 3). The highly significant improvement over the transcript-only model (DM t-statistic = 5.12) demonstrates the substantial, incremental information content provided by analyst reports.

These findings collectively refute the notion that analysts merely restate earnings information. Instead, they emphasize analysts' role in distilling and contextualizing corporate disclosures in ways that are uniquely informative to the market.

\subsubsection{What Content in Analyst Reports is Valued?}

Having established that analyst report text contains significant value-relevant information, this section investigates which specific content categories drive this informativeness. Using the Shapley value methodology, I decompose the text-based $R^2_{\mathrm{OOS}}$ across the 17 report topics pre-defined in Table \ref{tab: topic}. To minimize the confounding effects of cross-topic attention, the analysis uses sentence-segmented embeddings rather than full-context embeddings. This adjustment results in a 2.25 percentage point drop in $R^2_{\mathrm{OOS}}$, a figure that quantifies the value of inter-sentence contextuality.

\begin{figure}[H]
    \begin{center}
    \includegraphics[width = 0.9\textwidth]{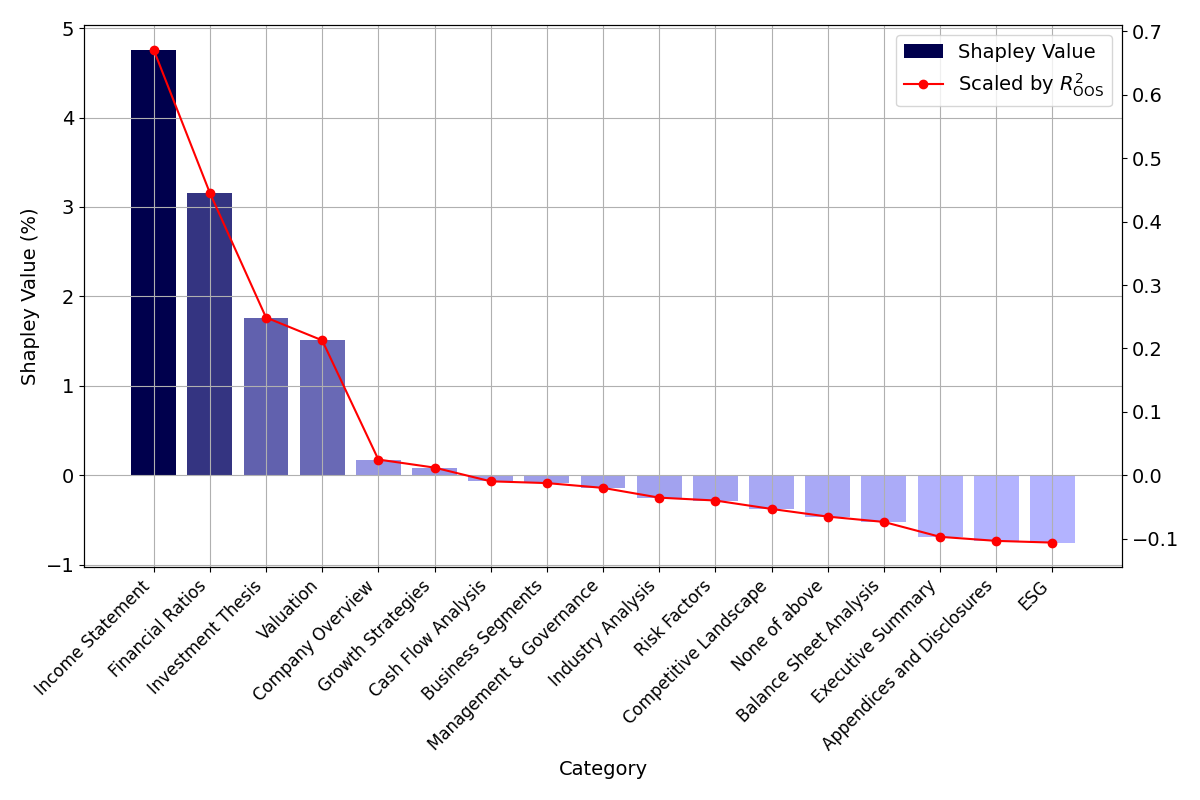}
    \end{center}
    \caption{Topic Importance via Shapley Value Decomposition}
    \label{fig:shap}
    {\footnotesize This figure shows the relative importance of 17 topics discussed in analyst reports, calculated using the Shapley value decomposition method. The total explanatory power of analyst reports ($R^2_{\mathrm{OOS}}$) is decomposed into the contributions from each topic, represented by their Shapley values. The red line plots each topic's contribution as a proportion of the total $R^2_{\mathrm{OOS}}$.}
\end{figure}

Figure \ref{fig:shap} illustrates the Shapley value decomposition, revealing a clear hierarchy in topic importance. Income Statement Analysis emerges as the single most valuable component. The Shapley value indicates it accounts for a relative contribution of 67\% to the total $R^2_{\mathrm{OOS}}$, followed by Financial Ratio Analysis as the second most crucial topic. A second tier of importance includes Investment Thesis and Valuation, with Shapley values of 1.76\% and 1.51\%. The remaining topics exhibit minimal or even negative Shapley values, suggesting little contribution to reports' market-moving power.

The robustness of this topic hierarchy is confirmed through several additional tests. Normalizing the Shapley values by sentence count and token length (Figure \ref{fig:shap_scale}) confirms that the dominance of Income Statement Analysis is not merely an artifact of size effect. Furthermore, this ranking remains remarkably consistent across all Fama-French 12 industries (Figure \ref{fig:shap_ind}). The Online Appendix (Figure \ref{fig:shap_yearly}) shows that while the absolute predictive power of text fluctuated over time, dipping during the COVID-19 pandemic, the relative contribution of Income Statement Analysis remained stable, underscoring its enduring information value.

To disentangle whether the topic importance stems from analyst expertise in certain areas (a supply-side effect) or from consistent investor demand for specific information (a demand-side effect), I examine how topic importance varies across different analyst and report characteristics. Figure \ref{fig: shap_ana} presents the Shapley value decompositions for six subsamples.

The topic hierarchy proves to be remarkably consistent across all comparison groups: experienced versus inexperienced analysts; analysts at large versus small brokerage firms; and reports that primarily use DCF-based versus multiple-based valuation models. This pervasive consistency suggests that the dominance of certain topics, such as Income Statement Analysis, does not merely reflect analysts' differing abilities, professional experience, or chosen valuation methodologies. Rather, the evidence indicates that the hierarchy is driven by the market's broad and persistent demand for these specific types of fundamental analysis.

Given its paramount importance, I conduct a deeper analysis of the ``Income Statement Analysis'' topic. I examine analysts' dual roles: a discovery role, where they generate or report otherwise unavailable information, and an interpretation role, where they process and contextualize available information into more digestible signals \citep{huang2018analyst, blankespoor2020disclosure}. I sub-classify each sentence in this topic along two dimensions: (1) Information Acquisition vs. Interpretation and (2) Realized vs. Expected Income, using the classification framework detailed in Prompt 3.

\vspace{0.1in}
\noindent
\begin{longtable}{|p{16cm}|}
\hline
\textbf{Prompt 3:} Please read the following sentence from a sell-side analyst report for the company \{firm\} (\{ticker\}) and classify it based on two criteria: \\
\endfirsthead

\hline
\textbf{Prompt 3:} Please read the following sentence from a sell-side analyst report for the company \{firm\} (\{ticker\}) and classify it based on two criteria: \\
\endhead

\hline
\endfoot

\hline
\endlastfoot
\textbf{Information Type}:    \\
    Information Acquisition (0): Sentences that directly report quantitative financial data, such as earnings, revenue, expenses, or other metrics. Example: `Reported EPS of \$1.40 beat consensus of \$0.94 and our estimate of \$1.17.' \\
    Information Interpretation (1): Sentences that provide analysis, context, or interpretation of the financial data, such as discussing trends, comparing performance to forecasts, assessing market impacts, or considering strategic implications. Example: `Biogen reported 1Q12 MS franchise sales below our estimate and the Street, as Avonex sales were negatively impacted by unfavorable distribution channel dynamics.'    \\
    \textbf{Time Reference}:    \\
    Realized Income (0): Sentences referring to concrete, historical results from completed periods. Example: `The company's revenue for the fiscal year 2022 was \$500 million.' \\
    Expected Income (1): Sentences containing subjective predictions or expectations for future periods. Example: `The company expects to achieve revenue of \$550 million in fiscal year 2023 based on current market conditions.' \\      
    Output your classification as two comma-separated numbers: the first for Information Type (0 for Acquisition, 1 for Integration) and the second for Time Reference (0 for Actual, 1 for Forecast). \\
    Sentence to classify: \{report\_sent\}. \\
\hline
\end{longtable}
\vspace{0.1in}

\begin{figure}[H]
    \begin{center}
        \includegraphics[width=0.8\textwidth]{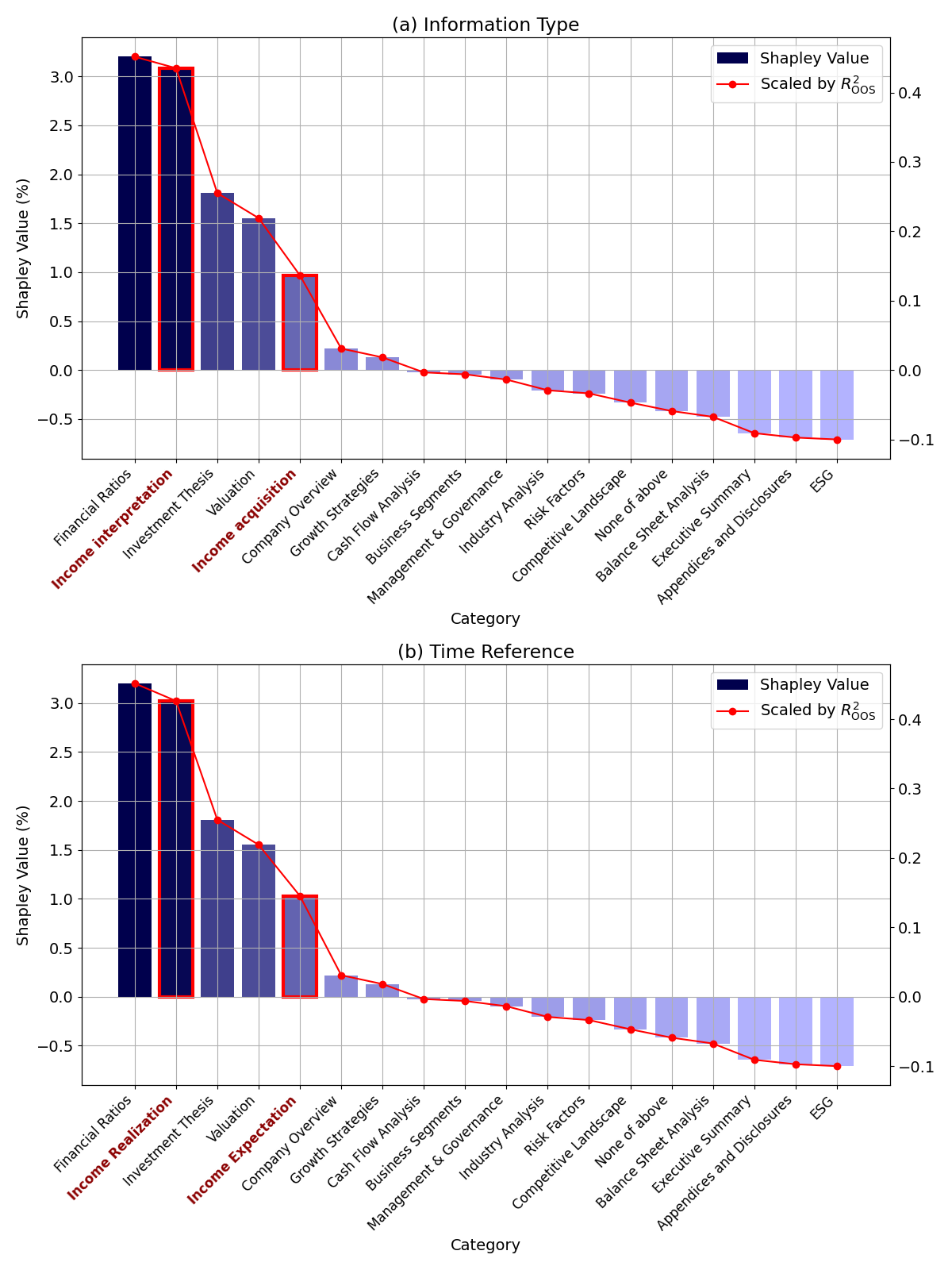}
    \end{center}
    \caption{Decomposition of Income Statement Analysis by Sub-Topic}\label{fig:subtopic}
    {\footnotesize 
    This figure shows the relative importance of sub-topics within income statement analyses, calculated using the Shapley value decomposition method. Panel (a) categorizes sub-topics by information type (income acquisition vs. interpretation), while Panel (b) categorizes them by time reference (realized vs. expected income). The total explanatory power of analyst reports ($R^2_{\mathrm{OOS}}$) is decomposed into the contributions from each topic, represented by their Shapley values. The red line plots each topic's contribution as a proportion of the total $R^2_{\mathrm{OOS}}$.}
\end{figure}

Figure \ref{fig:subtopic} presents the results of this granular decomposition. Analysts' interpretation of income statement data accounts for nearly three times the explanatory power of their acquisition and reporting of raw data. This suggests that investors value analytical context and insight far more than mere financial numbers. Similarly, discussions of realized (historical) income contribute three times as much to the $R^2_{\mathrm{OOS}}$ as discussions of expected (future) income, indicating the market's strong preference for analysis grounds in actual financial outcomes over forward-looking speculation.

These results reinforce the findings from Section \ref{ea} that the market reacts most strongly to reports issued immediately following earnings announcements. This section reveals that the most valued content within those reports is the interpretation of realized performance. Taken together, the evidence suggests that the primary value-add of sell-side analysts lies in their ability to rapidly digest complex, newly-released corporate disclosures and provide timely, insightful interpretations of realized income.

\subsubsection{Comparison of Revision and Reiteration Reports}

This section investigates the differential informativeness of analyst reports conditional on whether they contain forecast revisions or simple reiterations. While the market impact of forecast revisions is well-documented (e.g., \citeay{gleason2003analyst}; \citeay{asquith2005information}; \citeay{bradshaw2021soft}), I test whether the overall information content of analyst reports is significantly greater in revision versus reiteration events.

Following the classification methodology of \citet{huang2014evidence}, I identify a report as a revision report if it is released within two days that a new forecast is recorded in the I/B/E/S database. This classification is applied independently to recommendations, target prices, and earnings forecasts, acknowledging that a single report may contain a revision for one forecasting target (e.g., target price) while reiterating another (e.g., recommendation).

Table \ref{tab:reiteration} presents the out-of-sample performance, estimated separately for revision and reiteration report subsamples. Descriptive statistics reveal that revisions are relatively common for earnings forecasts (56.6\%) and target prices (31.1\%) but are rare for stock recommendations (2.3\%). The subsequent analysis demonstrates that the market response is overwhelmingly concentrated in revision reports.

Among the three forecasting targets, recommendation revisions are the most potent information events. For reports containing upgrades or downgrades, the combined numerical and textual information achieves an exceptionally high $R_{\mathrm{OOS}}^2$ of $22.63\%$. This finding aligns with prior literature documenting the significant market impact of recommendation changes \citep{loh2011analyst, bradley2014analysts}.

Reports accompanied by target price revisions also show strong explanatory power. While reiteration reports produce a modest combined $R^2_{\mathrm{OOS}}$ of 3.58\%, revision reports attain 20.88\% (t-stat = 11.18). This stark contrast underscores the incremental information value of specific target price changes beyond the coarser recommendations \citep{bradshaw2002use, brav2003empirical}.

Earnings forecast revisions follow a similar pattern. Reiteration reports offer little explanatory power, whereas revision reports achieve a combined $R^2_{\mathrm{OOS}}$ of 16.37\% (t-stat = 11.42), confirming that the market responds specifically to forecast revision events.

In summary, the evidence in Table \ref{tab:reiteration} consistently demonstrates that the informational value of analyst reports is heavily concentrated in those that revise explicit quantitative forecasts. Among these, the rarest recommendation changes, appear to convey the strongest signal to the market.

\subsubsection{Comparison of Analyst Report Sentiment Measures}

This section conducts a sentiment analysis to serve two primary purposes: first, to benchmark the performance of an LLM-based sentiment classifier (BERT) against a Naive Bayes algorithm used in \citet{huang2014evidence}; and second, to assess the incremental information content of high-dimensional text embeddings relative to a single sentiment dimension.

Specifically, I examine tone derived from a Naive Bayes classifier (${Tone}_{{NB}}$) and a BERT-based sentiment classifier (${Tone}_{{BERT}}$). Both models are trained on a manually labeled dataset of 10,000 sentences categorized as positive, neutral, or negative. I calculate report-level sentiment scores for each report as the number of positive sentences minus the number of negative sentences, scaled by the total number of sentences.

Panel A of Table \ref{tab:tone} presents OLS regressions of the CAR on these tone measures. In Column (1), I replicate the original analysis of \citet{huang2014evidence}. The coefficient on ${Tone}_{{NB}}$ is 0.022, closely matching their estimate of 0.021. A one-standard-deviation increase in this tone measure is associated with a 54 basis point increase in CAR, comparable to the 41 basis point impact they document over a two-day window. Column (2) adds granularity by separating tone into Income Statement and non-Income Statement content, both yielding statistically significant effects. Column (3) confirms the robustness of these findings with firm and year fixed effects. 

Columns (4)-(6) show that the BERT-based tone measure has substantially stronger explanatory power. A one-standard-deviation increase in ${Tone}_{BERT}$ corresponds to a 95 basis point increase in CAR. In a direct ``horse race'' regression including both measures (Column 7), the coefficient on ${Tone}_{{BERT}}$ remains large and highly significant, while the coefficient on ${Tone}_{NB}$ becomes insignificant. This indicates that the more advanced, context-aware BERT model captures more value-relavant text sentiment than the Naive Bayes classifier.

Panel B of Table \ref{tab:tone} compares the $R^2_{\mathrm{OOS}}$ across distinct inputs. Tone measures alone offer modest explanatory power, with $R^2_{\mathrm{OOS}}$ ranging from 0.05\% to 3.68\%. Combining tone with forecast revision measures increases performance to 9.49\%. Further integrating text embeddings lifts it to 12.27\%. Notably, adding tone measures to a model already incorporating text embeddings yields no significant incremental improvement.

Together, the findings highlight two key advantages of LLMs. First, for the specific task of classifying sentiment, LLMs like BERT more accurately capture nuanced tone and generate stronger signals than traditional NLP methods. Second, LLM embeddings are informationally superior to single sentiment scores. The fact that tone adds no value once embeddings are included suggests that the high-dimensional embeddings already capture the sentiment dimension, in addition to the complex, context-rich information that simple sentiment indicators miss.

\subsection{Information Value of Analyst Reports}

To translate the statistical measure of $R^2_{\mathrm{OOS}}$ into economic terms, I follow the framework of \citet{Kadan_Manela_Infoval}. The framework develops an empirical measure for quantifying the value of insider information base on the theoretical model of \citet{back2000imperfect}.

I employ the measure for two primary reasons. First, the framework is rooted in a general setting of imperfect competition among informed traders, which closely aligns with the reality of multiple investors with asymmetric yet overlapping access to analyst information. Second, this measure provides a conservative lower bound for the value estimation, allowing quantification of the minimum potential value of analyst information without overstating its importance.

The economic value of information ($\widehat{\Omega}_{it}$) for stock $i$ on day $t$ is defined as the ratio of the explainable component of return variance to the standardized price impact of trading:
\begin{equation}
\widehat{\Omega}_{it} = \frac{\text{Explainable Return Variance}}{\text{Standardized Price Impact}} = \frac{r_{it}^2 - \left(r_{it} - \widehat{r}_{it}\right)^2}{\widehat{\lambda}_{it} / p_{it_-}}.\label{equationiv}
\end{equation}
Here, $\widehat{r}_{it}$ represents the estimated cumulative abnormal return $\widehat{CAR}_{[-1,+1]}$ for stock $i$ on day $t$, and $r_{it}$ denotes the realized $CAR_{[-1,+1]}$ (where $\hat{r}_{it}=\frac{1}{N} \sum_{j=1}^N \hat{r}_{ijt}$ if multiple reports exist). The numerator thus captures the portion of return variance explained by the analyst reports. The denominator consists of the estimated price impact, $\widehat{\lambda}_{it}$, scaled by the stock's closing price at day $t-2$, $p_{it\_}$.

This measure provides an estimate of the potential trading profits available to strategic investors who can trade on the information contained in analyst reports. By construction, the value of information ($\widehat{\Omega}_{it}$) increases with the model's explanatory power (a higher numerator) and decreases with market liquidity (a lower price impact, $\widehat{\lambda}_{it}$, in the denominator), linking statistical informativeness with economic relevance.

To estimate the price impact $\widehat{\lambda}_{it}$, I regress 1-minute log returns $r_{itk}$ on contemporaneous order flow $y_{itk}$: 
\begin{equation}
r_{itk} = \widehat{\lambda}_{it} y_{itk} + \epsilon_{it},
\end{equation}
where $r_{itk} = p_{it\tau_k} - p_{it\tau_{k-1}}$ is the log price change, and $y_{itk} = Y_{it\tau_k} - Y_{it\tau_{k-1}}$ denotes signed order flow over interval $[\tau_{k-1}, \tau_k]$. $\widehat{\lambda}_{it}$ (Kyle's lambda) represents how sensitive the price is to trading volume. Buy and sell trades are classified using the algorithm of \citet{lee1991inferring}. Robustness checks also consider methods from \citet{ellis2000underwriter} and \citet{chakrabarty2007trade}. The three-day interval $T$ is partitioned into one-minute segments $\tau_0, \tau_1, \ldots, \tau_K$, yielding 1,170 observations (390 × 3) per regression.


Given that $\widehat{\Omega}_{it}$ is a ratio of two random variables measured with error, the interpretation of the information value estimate requires care. To address this challenge, I apply the delta method to estimate the mean and variance of the information value. The derivation of first-order estimations using the delta method is detailed in the Online Appendix E.

Table \ref{tab:iv_ss} reports the average analyst information value estimates. The average economic value of analyst reports for an S\&P 100 stock is \$0.47 million over a three-day event window (SE = \$0.05 million). This figure represents the potential profits available to strategic investors who can trade on the combined numerical and textual information.

To understand the sources of this value, I estimate the information value derived from textual and numerical content separately. A model based solely on text embeddings yields information value of \$0.38 million, while a model based only on numerical forecast revisions yields \$0.34 million. The fact that the combined model's value (\$0.47M) is less than the sum of the individual components suggests a significant overlap or interaction between these information sources. Extrapolating from an average of 15 report days per year, the annualized lower-bound estimate of investor profits from early access to analyst information is \$6.89 million for a typical S\&P 100 stock.

A key input to this calculation is the price impact of trading. The average standardized price impact ($\widehat{\lambda}_{it} / p_{it\_}$) is 0.34, which implies that a \$1 million trade moves the stock's price by approximately 3.4 basis points. The modest price impact of these highly liquid, large-cap stocks helps explain the substantial economic value.

\begin{figure}[H]
    \begin{center}
        \includegraphics[width=0.9\textwidth]{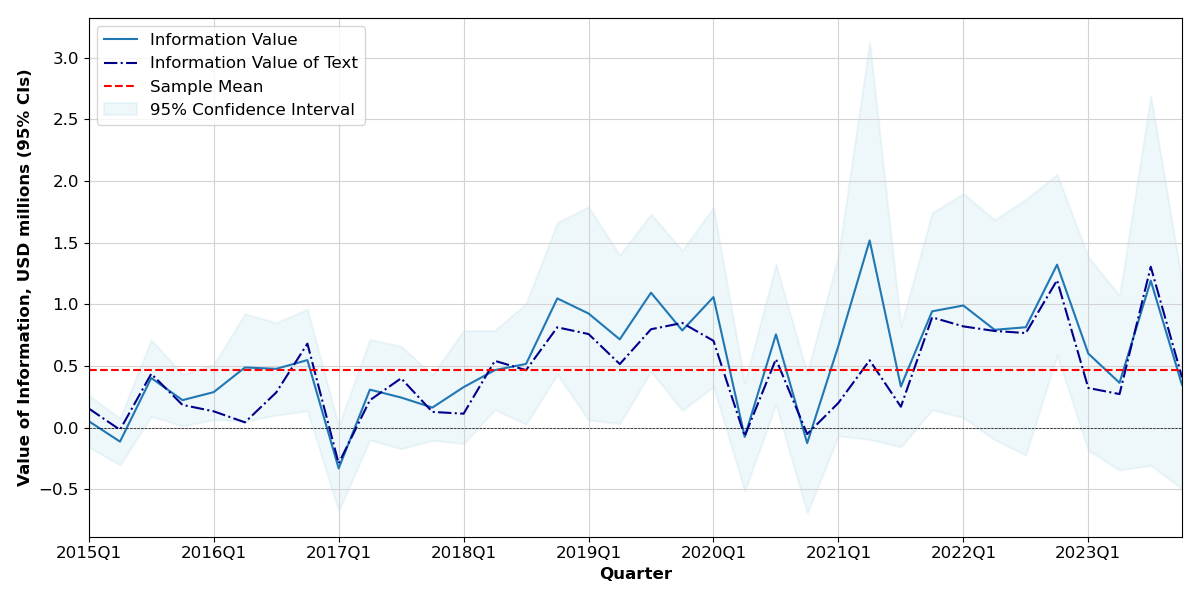}
    \end{center}
    \caption{Information Value of Analyst Reports over Time}\label{fig:iv_ts}
    {\footnotesize 
    This figure presents the estimated analyst information value between 2015Q1 and 2023Q4. The value is measured in millions of dollars, adjusted to constant 2020 dollars, with quarterly means estimated via the delta method. The solid line represents the combined strategic value of numerical and textual information, while the dashed line shows the value of textual information alone. The shaded area denotes the 95\% confidence interval for the total value. For reference, the horizontal line marks the average value over the entire sample period.}
\end{figure}

Figure \ref{fig:iv_ts} plots the time series of information value estimates from 2015Q1 to 2023Q4. The solid line, representing the combined value from text and numbers, exhibits a mild upward trend, suggesting that the economic value of analyst reports has, on average, increased over the past decade. The persistent gap between the combined value and the value from text alone confirms the complementary nature of quantitative and qualitative information. Notably, the shaded 95\% confidence band widens post-2020, signaling greater uncertainty and variation in information value during and after the COVID-19 pandemic. The fluctuations demonstrate the dynamic nature of analyst information value and its sensitivity to market conditions.

Further robustness checks are detailed in the Online Appendix. Table \ref{tab:iv_ss_robust} shows that using alternative high-frequency volatility measures and different trade-signing algorithms (e.g., \citeay{ellis2000underwriter}; \citeay{chakrabarty2007trade}) yields consistent estimates, with average information values ranging from \$0.42 million to \$0.64 million. Furthermore, Figure \ref{fig:iv_ts_robust} demonstrates that the temporal trends in information value are consistent across these alternative specifications.

\subsubsection{The Value of Analyst Information in the Cross-section of Stocks}

The economic value of analyst information may vary across the cross-section of stocks. Prior literature offers competing hypotheses regarding firm size. On the one hand, information acquisition and analysis are typically more costly for smaller, less-followed firms, suggesting analysts might focus on larger firms \citep{zeghal1984firm}. On the other hand, the relative scarcity of public information for smaller firms could make analyst opinions more influential and thus more valuable to the market \citep{stickel1995anatomy}. This creates an empirical question as to whether information value is positively or negatively related to firm size.

Figure \ref{fig:iv_stock} illustrates this relationship through a scatter plot of information value against stock market capitalization. While the plot reveals a positive association between market equity and information value, the relationship exhibits notable nonlinearities. To formally test this relationship, I estimate regressions of the information value measure on firm size (log market equity), and controlling for the book-to-market ratio, firm, and year fixed effects.

The results in Table \ref{tab:iv_size} confirm a strong positive relationship. The coefficient estimates in Column (4) imply an elasticity of 0.864, meaning a 1\% increase in market equity is associated with a 0.864\% increase in analyst information value. Notably, Columns (5) through (8) show that this elasticity is even stronger for text information value, suggesting that the interpretive role of analysts is particularly important for larger, often more complex firms.

To better understand the mechanisms behind this relationship, I decompose information value into two components: the explanatory power of the report and the price impact of trading. Expressed in logarithmic form, the decomposition is given by:
\begin{equation}
\log \widehat{\Omega}_{i t} = 
\underbrace{\log \left[ r_{i t}^2 - \left( r_{i t} - \hat{r}_{it}\right)^2 \right]}_{\text{log explained return variance}} 
- 
\underbrace{\log \frac{\widehat{\lambda}_{i t}} {P_{i t_{-}}}}_{\text{log price impact}}.
\end{equation}

This allows for testing whether the higher information value for large firms stems from an information channel (analysts providing better or more impactful information) or a liquidity channel (trading on the information is cheaper due to lower price impact).

Panel B of Table \ref{tab:iv_size} shows that the higher information value for large-cap stocks arises mainly from lower price impact, not greater explanatory power. That is, while analyst reports do not better explain return variance for larger firms, their improved liquidity reduces transaction costs, thereby enhancing the value of analyst information from a trading perspective.

Panel B of Table \ref{tab:iv_size} presents results of the decomposition analysis. The findings are striking: the positive relationship between firm size and information value arises mainly from the liquidity channel. While analyst reports do not exhibit significantly greater explanatory power for larger firms, the price impact of trading is substantially lower for these high-liquidity stocks. This reduced transaction cost enhances the value of analyst information. As my sample is composed of large stocks, it is crucial to recognize that the economic value of analyst information likely decreases for smaller, less liquid firms.

\subsubsection{The Value of Analyst Information in the Cross-section of Analysts}

This section explores how the economic value of analyst reports varies with analyst characteristics, focusing on bold and herding forecast behavior. Prior research suggests that bold forecasts are more likely to be based on an analyst's private information, whereas herding behavior often reflects a reliance on public signals or consensus views \citep{clement2005financial}. I therefore test the hypothesis that bold reports generate greater strategic information value.

To examine this cross-sectional variation, I regress the stock-analyst-day level information value on an indicator for forecast boldness, controlling for brokerage size as well as firm, analyst, and year fixed effects. The analyst-specific information value measure is defined as:
\begin{equation}
\widehat{\Omega}_{ijt}=\frac{r_{i t}^2-\left(r_{i t}- \widehat{r}_{i j t}\right)^2}{\lambda_{i t}} \cdot p_{i t_{-}},
\end{equation}
where $\hat{r}_{ijt}$ is the out-of-sample CAR estimation for report $j$ of stock $i$ on day $t$.

The regression results are presented in Table \ref{tab:iv_bold}. Panel A shows that the coefficient on the boldness indicator (Bold) is positive and highly significant across all specifications. The estimated magnitude implies that, on average, bold analyst reports are 29\% more valuable than their herding counterparts. This pattern holds when analyzing the value derived from textual content alone. In contrast, the coefficient on brokerage size is consistently insignificant, suggesting that after controlling for other factors, the size of brokerage firms does not materially influence the economic value of analysts' reports.

To understand the channel through which boldness creates value, Panel B decomposes the information value into its two components: explained return variance (the information channel) and price impact (the liquidity channel). The results clearly indicate that the effect is driven by the information channel. Columns (1)–(4) show that bold reports are associated with significantly larger explained return variance, reinforcing the view that they contain meaningful signals. Conversely, Columns (5)–(8) reveal no systematic relationship between boldness and price impact, going against the possibility that the higher value of bold reports stems from differential market liquidity or trading frictions.

\subsubsection{The Value of Analyst Information around Earnings Announcements} \label{iv_ead}

Building on earlier findings that the statistical information content of analyst reports peaks shortly after earnings announcements, this section examines whether the economic value of these reports follows a similar pattern. I investigate how the value of analyst information varies based on its timeliness relative to corporate earnings disclosures.

\begin{figure}[H]
    \begin{center}
        \includegraphics[width=0.9\textwidth]{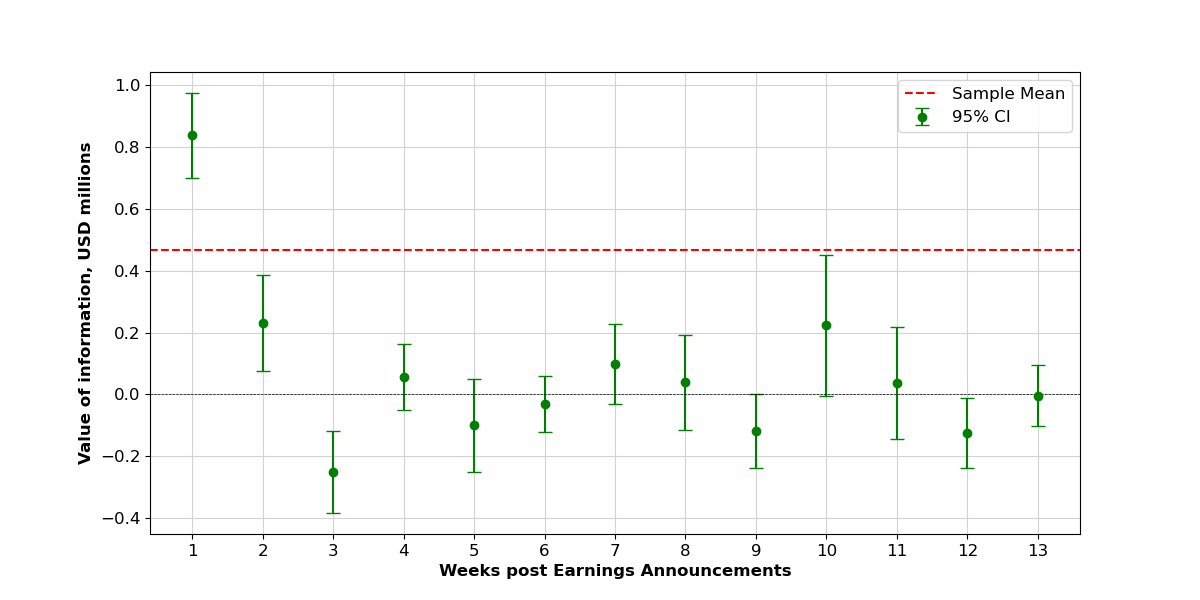}
    \end{center}
    \caption{Information Value of Analyst Reports Post-Earnings Announcement}\label{fig:iv_ead}
    {\footnotesize 
    This figure presents the estimated information value of analyst reports released 1 to 13 weeks after an earnings announcement. All values are in millions of 2020 dollars. The sample spans 2015Q1 to 2023Q4. The vertical axis shows the estimated value of information, and the horizontal axis indicates the number of weeks post-earnings announcement. Each point is the average information value for reports in that week, with the green bars denoting the 95\% confidence interval. The red dashed line is the average value across the full sample, regardless of the week.}
\end{figure}

Figure \ref{fig:iv_ead} plots the average information value of analyst report subsamples across 13 weekly bins following earnings announcements. Reports issued in the first week show the highest average value, at approximately \$0.84 million. This value declines sharply in subsequent weeks, suggesting that the market places a significant premium on timely analyst reports that digest and interpret corporate news.

To formally test this observation, I estimate panel regressions of log information value on a post-earnings week indicator ($Week$), which equals one if the report is issued within one week of the earnings announcement. Following prior literature that suggests analyst interpretation is most valuable amidst uncertainty \citep{chen2010relationship}, I also test if this effect is amplified during periods of high investor disagreement. I proxy for disagreement using trading volume on the earnings announcement day ($TradingVolume$) and include an interaction term ($Week \times TradingVolume$), following \citet{banerjee2011learning} and \citet{beckmann2024unusual}. All specifications include stock and year fixed effects.

The regression results are presented in Table \ref{tab:iv_ead}. Panel A documents significantly higher information value for reports issued within the first post-earnings week, as evidenced by positive and significant coefficients on $Week$ across all models. Furthermore, the positive and significant coefficient on the interaction term indicates that this timeliness premium is even greater when there is high investor disagreement surrounding earnings announcements. This is consistent with investors placing the greatest reliance on analyst sense-making during periods of heightened uncertainty.

Panel B decomposes this finding to identify the channel through which timeliness creates value. By regressing the two components of the information value measure on the same set of variables, I can distinguish between an information channel and a liquidity channel. The results show that the timeliness effect is driven primarily by the information channel: reports issued shortly in the post-earnings announcement window are associated with a significant increase in explained return variance, while changes in price impact are relatively modest.

\section{Conclusion}\label{conclu}

This paper examines both the quantitative and qualitative information content of analyst reports. I find that the textual content of analyst reports carries more economically significant, market-valued information than quantifiable forecast revisions. This effect is especially pronounced in the period immediately following earnings announcements. A Shapley value decomposition identifies analysts' interpretation of income statements as the most valuable topic within written reports.

To translate these findings into economic terms, I adopt a measure of information value accounting for both the explained return variance and the price impact. The results suggest that early acquisition of analyst reports for an average S\&P 100 constituent stock yields annualized profits exceeding \$6.89 million. The value of analyst information is higher for large-cap firms, bold analyst forecasts, and when investors' disagreement is high.

These findings have important implications for regulators. By quantifying the substantial economic value of privileged access to analyst reports, I reveal strong incentives for selective distribution, a practice prohibited by FINRA Rule 2241. It is important to note that the analysis provides a hypothetical framework for understanding the potential value of such information, not an empirical assessment of market practices. Nevertheless, the magnitude of these potential profits reinforces the fundamental principle that equal access to value-relevant information is essential for market integrity.

\clearpage
    \newpage
     \onehalfspacing
    \bibliographystyle{jfe}
    \bibliography{references}

\newpage

\newpage

\newpage

\newpage

\newpage

\newpage
\setcounter{table}{0}

\scriptsize
\renewcommand{\arraystretch}{1.2}
\begin{longtable}{>{\raggedright\arraybackslash}p{4cm}>
{\raggedright\arraybackslash}p{12cm}}
\caption{Definition of Topic Categories}\label{tab: topic} \\
\toprule
\textbf{Topic} & \textbf{Descriptions} \\
\midrule
\endfirsthead
\multicolumn{2}{c}{{\tablename\ \thetable{} -- continued from previous page}} \\
\toprule
\textbf{Topic} & \textbf{Descriptions} \\
\midrule
\endhead
\endfoot
\bottomrule
\endlastfoot
\multicolumn{2}{l}{\textbf{Company and Industry Overview}} \\
Executive Summary & Provides a high-level overview of the report's key findings and conclusions; includes a brief description of the company, its industry, and the purpose of the report; highlights the most important points from the analysis, such as the company's financial performance, competitive position, and growth prospects. \\
Company Overview & Offers a comprehensive description of the company, including its history, business model, and key products or services; discusses the company's organizational structure, management team, and corporate governance; analyzes the company's mission, vision, and strategic objectives. \\
Industry Analysis & Provides an in-depth analysis of the industry in which the company operates; includes information on market size, growth trends, and key drivers; discusses the regulatory environment, technological advancements, and other external factors affecting the industry; analyzes the industry's competitive dynamics and the company's position within the industry. \\
Competitive Landscape & Identifies the company's main competitors and their market share; compares the company's products, services, and pricing strategies with those of its competitors; analyzes the strengths and weaknesses of the company and its competitors; discusses potential new entrants and substitutes that could disrupt the competitive landscape. \\
Business Segments & Provides a detailed analysis of the company's various business segments or divisions; discusses the financial performance, growth prospects, and challenges of each segment; analyzes the contribution of each segment to the company's overall revenue and profitability. \\
Growth Strategies & Discusses the company's strategies for driving future growth, such as organic growth initiatives, product innovations, and geographic expansions; analyzes the company's mergers and acquisitions (M\&A) strategy and potential targets; examines the company's investments in research and development (R\&D) and marketing. \\
\midrule
\multicolumn{2}{l}{\textbf{Financial Analysis}} \\
Income Statement Analysis & Analyzes the company's revenue, expenses, and profitability. \\
Balance Sheet Analysis & Examines the company's assets, liabilities, and shareholders' equity. \\
Cash Flow Analysis & Analyzes the company's cash inflows and outflows to evaluate liquidity. \\
Financial Ratios & Discusses key ratios like profitability, liquidity, and solvency ratios. \\
\midrule
\multicolumn{2}{l}{\textbf{Strategic Outlook}} \\
Investment Thesis & Summarizes the key reasons for investing (or not investing) in the company's shares; discusses the potential catalysts and risks that could impact the company's valuation and stock price performance; provides a target price or price range for the company's shares based on the valuation analyses and investment thesis. \\
Valuation & Estimates the intrinsic value of the company's shares using various valuation methodologies, such as discounted cash flow (DCF) analysis, relative valuation multiples, and sum-of-the-parts analysis; compares the company's valuation with that of its peers and historical benchmarks; discusses the key assumptions and sensitivities underlying the valuation analyses. \\
\midrule
\multicolumn{2}{l}{\textbf{Risk and Governance}} \\
Risk Factors & Identifies and analyzes the key risks facing the company, such as market risks, operational risks, financial risks, and legal/regulatory risks; discusses the potential impact of these risks on the company's financial performance and growth prospects; examines the company's risk management strategies and mitigation measures. \\
Management and Governance & Provides an overview of the company's management team, including their experience, expertise, and track record; analyzes the company's corporate governance practices, such as board composition, executive compensation, and shareholder rights; discusses the company's succession planning and key person risks. \\
\bottomrule
ESG & Analyzes the company's performance and initiatives related to environmental sustainability, social responsibility, and corporate governance; discusses the potential impact of ESG factors on the company's reputation, risk profile, and financial performance; examines the company's compliance with ESG regulations and industry standards. \\
\midrule
\multicolumn{2}{l}{\textbf{Additional Content}} \\
Appendices and Disclosures & Includes additional supporting materials, such as financial statements, ratio calculations, and detailed segment data; provides important disclosures, such as the analyst's rating system, potential conflicts of interest, and disclaimers; discusses the limitations and uncertainties of the analysis and the need for further due diligence by investors. \\
None of the Above & Covers any content that does not fall into the specified topics. \\
\end{longtable}

\newpage
\begin{table}[!htb]
    \caption{Summary Statistics of Analyst Reports}\label{tab:sum1}
    {\footnotesize This table presents summary statistics for analyst reports on S\&P 100 firms from 2000 to 2023. Panel A provides annual statistics, including the number of reports, unique brokerage firms and analysts, and the average report length in pages and tokens. Panel B presents the same statistics aggregated by Fama-French 12 (FF12) industry. FF12 industry definitions are available on Kenneth French's data library.}
    \begin{center}
    \scriptsize
    \tabcolsep = 0.23cm
    \renewcommand{\arraystretch}{1.2}
    \begin{tabularx}{\textwidth}{lYYYYY}
\toprule
\multicolumn{6}{l}{Panel A: Sell-side analyst reports by year (2000-2023)}\\
\midrule
 Year &  Reports &  Brokerage firms &  Analysts &  Pages &  Tokens \\
\midrule
 2000 &      582 &               28 &        78 &      5 &    1992 \\
 2001 &     1080 &               34 &       116 &      5 &    2043 \\
 2002 &     1973 &               42 &       167 &      5 &    1806 \\
 2003 &     2735 &               45 &       192 &      6 &    2006 \\
 2004 &     3622 &               49 &       262 &      7 &    1910 \\
 2005 &     4434 &               50 &       269 &      7 &    2154 \\
 2006 &     4716 &               45 &       234 &      7 &    2065 \\
 2007 &     4872 &               44 &       243 &      7 &    2259 \\
 2008 &     5834 &               51 &       279 &      8 &    2468 \\
 2009 &     5745 &               62 &       326 &      8 &    2322 \\
 2010 &     4957 &               67 &       313 &      7 &    2171 \\
 2011 &     7327 &               56 &       381 &      8 &    2164 \\
 2012 &     7534 &               54 &       379 &      8 &    1943 \\
 2013 &     7936 &               52 &       403 &      8 &    1941 \\
 2014 &     7350 &               50 &       407 &      8 &    1824 \\
 2015 &     7534 &               50 &       380 &      8 &    1866 \\
 2016 &     7009 &               50 &       374 &      8 &    1987 \\
 2017 &     6628 &               48 &       344 &      9 &    2039 \\
 2018 &     5481 &               37 &       282 &      8 &    2034 \\
 2019 &     5958 &               38 &       305 &      8 &    1992 \\
 2020 &     6049 &               40 &       299 &      8 &    2017 \\
 2021 &     3967 &               27 &       196 &      9 &    2047 \\
 2022 &     4536 &               25 &       210 &      9 &    2146 \\
 2023 &     4393 &               34 &       212 &      9 &    2094 \\
\midrule
\multicolumn{6}{l}{Panel B: Sell-side analyst reports across Fama-French 12 industries}\\
\midrule
Industry &  Reports &  Brokerage firms &  Analysts &  Pages &  Tokens \\
\midrule
   BusEq &    25258 &              101 &       411 &      8 &    2133 \\
    Hlth &    20257 &               66 &       191 &      8 &    2086 \\
   Money &    19689 &               63 &       230 &      7 &    1955 \\
   Shops &    11602 &               75 &       186 &      7 &    1968 \\
   Manuf &    10531 &               57 &       170 &      8 &    1954 \\
   Other &     9763 &               79 &       259 &      9 &    2303 \\
   Telcm &     6022 &               60 &        89 &      9 &    2149 \\
   Utils &     5068 &               30 &        54 &      6 &    1827 \\
   Enrgy &     5049 &               41 &        67 &      8 &    1927 \\
   NoDur &     3862 &               36 &        65 &      8 &    2209 \\
   Durbl &     2915 &               29 &        40 &      8 &    1867 \\
   Chems &     2236 &               33 &        53 &      8 &    2141 \\
\bottomrule
    \end{tabularx}
    \end{center}
\end{table}

\newpage
\begin{table}[!htb]
    \caption{Analyst Quantitative and Qualitative Information Content}\label{tab:base}
    {\footnotesize This table presents the information content of quantitative and qualitative measures. The analysis employs four distinct input configurations: (1) `Revision only', incorporating three analyst forecast revision measures; (2) `Numerical only', augmenting the forecast revisions with additional numerical measures as described in Section \ref{data}; (3) `Text only', using analyst report text embeddings exclusively; and (4) `Rev + text', combining forecast revision measures with text embeddings. The DM t-statistics for the $R^2_{\mathrm{OOS}}$ is calculated using the procedure outlined by \citet{gu2020empirical}. In Panel A, the benchmark estimation is set to zero. Differences such as (5)-(1) represent the gain from replacing Revision-only input with Text-only, and (7)-(5) compares the additive value of combining revisions and text. In Panel B, the reported t-statistics compare the estimation of alternative large language models with those of LLaMA-2-13B. BERT denotes the \texttt{bert-base-uncased} model. OpenAI denotes the \texttt{text-embedding-3-small} model. LLaMA-3 denotes the \texttt{LLaMA-3-8B} model.}
    \begin{center}
    \scriptsize
    \tabcolsep = 0.23cm
    \renewcommand{\arraystretch}{1.2}
    \begin{tabularx}{\textwidth}{lYYYYYYYYYYY}
    \toprule
    \multicolumn{12}{l}{Panel A: Information content of quantitative and qualitative information}\\
    \midrule
   Year & Revision only &  t-stat & Numerical only &  t-stat & Text only &  t-stat & Rev + text &  t-stat &  t-stat &  t-stat &  t-stat \\
    \midrule
   {} &   (1) &  (2) &   (3) &  (4) &  (5) &  (6) &  (7) &  (8) & (5)-(1) &  (7)-(1) &  (7)-(5) \\
    \midrule
    2015 & 10.31\% & 4.27 & 7.75\% & 2.59 & 12.63\% & 6.39 & 10.63\% & 2.98 & 3.63 & 0.16 & -1.25 \\
    2016 & 14.61\% & 11.26 & 15.35\% & 11.26 & 11.98\% & 3.93 & 17.08\% & 9.67 & -1.15 & 3.82 & 2.93 \\
    2017 & 8.99\% & 6.23 & 9.99\% & 6.41 & 11.11\% & 6.40 & 11.98\% & 7.09 & 4.99 & 5.96 & 2.38 \\
    2018 & 10.05\% & 3.28 & 10.52\% & 3.60 & 10.87\% & 5.85 & 13.64\% & 5.95 & 0.87 & 5.77 & 4.47 \\
    2019 & 9.94\% & 20.14 & 9.97\% & 14.76 & 12.16\% & 17.68 & 14.44\% & 26.17 & 2.68 & 19.77 & 4.83 \\
    2020 & 5.52\% & 3.88 & 6.10\% & 4.55 & 3.82\% & 5.18 & 6.34\% & 6.16 & -1.75 & 1.57 & 7.11 \\
    2021 & 5.43\% & 5.83 & 5.96\% & 6.82 & 8.50\% & 6.21 & 11.94\% & 27.16 & 1.48 & 7.48 & 3.28 \\
    2022 & 9.78\% & 6.03 & 9.16\% & 4.97 & 14.88\% & 10.34 & 16.95\% & 10.09 & 9.21 & 8.01 & 5.30 \\
    2023 & 6.68\% & 5.48 & 7.19\% & 4.00 & 9.30\% & 4.17 & 10.76\% & 6.09 & 2.13 & 4.87 & 3.08 \\
    Overall & 9.01\% & 9.45 & 9.08\% & 9.44 & 10.19\% & 8.20 & 12.28\% & 8.87 & 1.66 & 3.95 & 3.77 \\
    \midrule
    \end{tabularx}
    
    \begin{tabularx}{\textwidth}{lYYYYYY}
    \multicolumn{6}{l}{Panel B: Information content of qualitative information with alternative LLMs}\\
    \midrule
       Year &   BERT &  t-stat & OpenAI &  t-stat & LLaMA-3 &  t-stat \\
    \midrule
       {} &   (1) &  (2) &   (3) &  (4) &  (5) &  (6) \\
    \midrule
   2015 & 7.24\% &  -13.45 & 6.76\% &   -9.97 & 11.28\% &   -8.04 \\
   2016 & 6.50\% &   -2.22 & 5.94\% &   -4.05 & 10.48\% &    2.23 \\
   2017 & 5.48\% &   -5.95 & 5.88\% &   -7.72 & 10.25\% &    0.11 \\
   2018 & 6.94\% &  -14.05 & 6.48\% &   -8.15 & 10.21\% &    1.29 \\
   2019 & 6.16\% &  -13.80 & 5.41\% &  -16.94 & 10.48\% &   -6.19 \\
   2020 & 3.92\% &   -0.33 & 4.10\% &    0.09 &  6.07\% &    6.27 \\
   2021 & 2.63\% &  -18.53 & 3.07\% &   -4.11 &  6.98\% &   -1.26 \\
   2022 & 7.75\% &   -9.93 & 7.89\% &   -7.09 & 13.02\% &   -2.96 \\
   2023 & 4.18\% &   -5.65 & 4.03\% &   -8.04 &  8.68\% &   12.05 \\
Overall & 5.72\% &   -5.54 & 5.57\% &   -5.66 &  9.66\% &    0.19 \\    \bottomrule
    \end{tabularx}
    
    \end{center}
\end{table}

\newpage
\begin{table}[!htb]
\caption{OLS Regressions of Market Reactions on Analyst Report Content}\label{tab:ols}
    {\footnotesize This table presents OLS regression results analyzing the information content of analyst reports. Panel A shows summary statistics, and Panel B reports regression coefficients. The dependent variable is $CAR_{[-1,+1]}$, the three-day cumulative abnormal return around the report release, computed as the raw return minus the NYSE/AMEX/NASDAQ value-weighted market return. $REC_{REV}$ denotes recommendation revision, calculated as the current report's recommendation minus the last recommendation in I/B/E/S issued by the same analyst for the same stock. $EF_{REV}$ represents earnings forecast revision, calculated as the current report's EPS forecast minus the last EPS forecast in I/B/E/S issued by the same analyst for the same stock, scaled by the stock price 50 days before the report release. $TP_{REV}$ indicates target price revision, calculated as the current report's target price minus the last target price in I/B/E/S issued by the same analyst for the same stock, scaled by the stock price 50 days before the report release. $PriorCAR$ represents the cumulative abnormal return over the period [-10, -2]. $SUE$ denotes the earnings surprise. $\widehat{CAR}_{txt}$ and $\widehat{CAR}_{rev}$ are the out-of-sample fitted values from Ridge regressions using text embeddings and the three forecast revision measures, respectively. All explanatory variables are standardized. The t-statistics (in parentheses) are clustered two-way by stock and year. $^*p<.1$, $^{**}p<.05$, $^{***}p<.01$.}
\begin{center}
\scriptsize
\begin{tabularx}{\textwidth}{l*{6}{Y}}
\toprule
\multicolumn{7}{l}{Panel A: Summary statistics} \\
\midrule
{} &   Mean &    Std. Dev. &    P25 &    P50 &    P75 &      N \\
\midrule
$CAR_{[-1,+1]}$     &  0.002 &  0.048 & -0.019 &  0.001 &  0.021 &  28837 \\
$REC_{REV}$            &  0.002 &  0.151 &  0.000 &  0.000 &  0.000 &  28837 \\
$EF_{REV}$             &  0.000 &  0.005 &  0.000 &  0.000 &  0.001 &  28837 \\
$TP_{REV}$             &  0.010 &  0.067 &  0.000 &  0.000 &  0.000 &  28837 \\
$\widehat{CAR_{rev}}$ &  0.002 &  0.014 & -0.001 & -0.000 &  0.003 &  28837 \\
$\widehat{CAR_{txt}}$ &  0.002 &  0.018 & -0.009 &  0.002 &  0.013 &  28837 \\
\midrule
\end{tabularx}
\begin{tabularx}{\textwidth}{l*{7}{Y}}
\multicolumn{8}{l}{Panel B: Market reaction to forecast revisions and report text} \\
\midrule
            &\multicolumn{1}{c}{(1)}&\multicolumn{1}{c}{(2)}&\multicolumn{1}{c}{(3)}&\multicolumn{1}{c}{(4)}&\multicolumn{1}{c}{(5)}&\multicolumn{1}{c}{(6)}&\multicolumn{1}{c}{(7)}\\
\midrule
$REC_{REV}$             &       0.001\sym{***}&                     &       0.001\sym{**} &                     &                     &                     &                     \\
                    &      (3.26)         &                     &      (2.06)         &                     &                     &                     &                     \\
$EF_{REV}$              &       0.005\sym{***}&                     &       0.004\sym{***}&                     &                     &                     &                     \\
                    &      (7.72)         &                     &      (5.39)         &                     &                     &                     &                     \\
$TP_{REV}$              &       0.012\sym{***}&                     &       0.009\sym{***}&                     &                     &                     &                     \\
                    &     (14.95)         &                     &     (11.71)         &                     &                     &                     &                     \\
$\widehat{CAR_{txt}}$&                     &       0.015\sym{***}&       0.012\sym{***}&                     &       0.012\sym{***}&       0.012\sym{***}&       0.012\sym{***}\\
                    &                     &     (24.23)         &     (22.40)         &                     &     (22.30)         &     (19.13)         &     (22.62)         \\
$\widehat{CAR_{rev}}$&                     &                     &                     &       0.014\sym{***}&       0.010\sym{***}&       0.010\sym{***}&       0.010\sym{***}\\
                    &                     &                     &                     &     (18.29)         &     (14.00)         &     (13.44)         &     (13.35)         \\
$SUE$                 &                     &                     &                     &                     &                     &       0.001\sym{***}&       0.001\sym{**} \\
                    &                     &                     &                     &                     &                     &      (3.97)         &      (2.36)         \\
$PriorCAR$           &                     &                     &                     &                     &                     &      -0.003\sym{***}&      -0.003\sym{***}\\
                    &                     &                     &                     &                     &                     &     (-4.45)         &     (-5.01)         \\
Constant            &       0.002\sym{***}&       0.002\sym{***}&       0.002\sym{***}&       0.002\sym{***}&       0.002\sym{***}&       0.002\sym{***}&       0.002\sym{***}\\
                    &      (3.14)         &      (2.82)         &      (3.09)         &      (3.11)         &      (3.07)         &      (3.41)         &      (3.83)         \\
\midrule
Year FE & No & No & No & No & No & No & Yes \\
Stock FE & No & No & No & No & No & No & Yes \\
\(N\)        &       28837         &       28837         &       28837         &       28837         &       28837         &       23314         &       20987         \\
Adjusted \(R^{2}\)  &       0.091         &       0.105         &       0.149         &       0.089         &       0.147         &       0.137         &       0.146         \\
\bottomrule
\end{tabularx}
\end{center}
\end{table}

\clearpage
\newpage
\begin{table}[!htb]
\caption{Information Content around Earnings Announcements}\label{tab:ea}
    {\footnotesize This table analyzes the information content ($R^2_{\mathrm{OOS}}$) of analyst reports and earnings announcement (EA) transcripts based on their timing relative to the EA. The out-of-sample period is 2015-2023. Panel A documents the decay in the information content of analyst reports over time. It presents the $R^2_{\mathrm{OOS}}$ for reports grouped into 13 separate weekly bins following the EA date. Panel B focuses on the short-term window, comparing the information content of reports issued within 1, 2, 3, and 7 days of the EA to those issued beyond those cutoff points. Panel C assesses the incremental value of analyst reports over EA transcripts. This analysis uses only reports released one day after the EA. `Transcripts' and `Reports' denote models using only the respective text embeddings, while `Reports + Transcripts' combines both. The `Diff' column reports the incremental $R^2_{\mathrm{OOS}}$ from adding report text to the transcript embeddings, i.e., `(Reports + Transcripts) - (Transcripts)'. The t-statistics for the $R^2_{\mathrm{OOS}}$ are calculated using the procedure outlined by \citet{gu2020empirical}, with a benchmark prediction of zero.}
    \begin{center}
    \scriptsize
    \tabcolsep = 0.23cm
    \renewcommand{\arraystretch}{1.2}
\begin{tabularx}{\textwidth}{YYYY}
\toprule
\multicolumn{4}{l}{Panel A: Weekly bins}\\
\midrule
 Weeks &  $R^2_{\mathrm{OOS}}$ &  t-stat &     N \\
\midrule
    1 &  9.80\% &   10.17 & 26490 \\
    2 &  0.60\% &   -0.08 &  4586 \\
    3 & -8.71\% &   -2.95 &  4144 \\
    4 & -3.61\% &   -1.36 &  4320 \\
    5 &  0.23\% &    0.53 &  3796 \\
    6 & -4.07\% &   -1.53 &  4230 \\
    7 & -4.26\% &   -1.62 &  4404 \\
    8 &  2.88\% &    1.57 &  4286 \\
    9 & -4.89\% &   -3.50 &  3722 \\
   10 &  2.06\% &    0.83 &  3528 \\
   11 &  2.74\% &    0.83 &  4116 \\
   12 & -0.02\% &   -0.42 &  4764 \\
   13 & -0.11\% &   -0.76 &  5642 \\
\midrule
\end{tabularx}

\begin{tabularx}{\textwidth}{YYYYY}
\multicolumn{5}{l}{Panel B: Sub-sample analyses of earnings announcement periods}\\
\midrule
Window & $R^2_{\mathrm{OOS}}$ (Within EA) &  t-stat & $R^2_{\mathrm{OOS}}$ (Beyond EA) &  t-stat \\
\midrule
1 day &               11.84\% &    3.50 &                    4.59\% &    2.55 \\
2 days &               11.97\% &    3.38 &                    4.38\% &    2.47 \\
3 days &               11.69\% &    3.38 &                    4.13\% &    2.38 \\
7 days &                7.29\% &    2.57 &                    5.08\% &    2.59 \\\midrule
\end{tabularx}

\begin{tabularx}{\textwidth}{YYYYY}
\multicolumn{5}{l}{Panel C: Information content of earnings call transcripts}\\
\midrule
{} &  Transcripts & Reports & Reports + Transcripts &  Diff \\
{} &  (1) &  (2) &  (3) &  (3) - (1) \\
\midrule
$R^2_{\mathrm{OOS}}$  &                 4.20\% &          9.72\% &                                  11.96\% &      7.76\% \\
t-stat  &                   5.16 &            3.24 &                                     6.42 &        5.12 \\
\bottomrule
\end{tabularx}

\end{center}
\end{table}

\clearpage
\newpage
\begin{table}[!htb]
\caption{Information Content: Forecast Revisions vs. Reiterations}\label{tab:reiteration}
    {\footnotesize This table compares the information content ($R^2_{\mathrm{OOS}}$) of analyst reports that revise prior forecasts versus those that reiterate them. The sample period is 2015–2023. The analysis compares three input types: forecast revisions only (`Revision only'), text embeddings only (`Text only'), and a combination (`Rev + text'). Panel A presents results for reports that reiterate prior \{Target\} predictions. Panel B presents the $R^2_{\mathrm{OOS}}$ for reports that revise \{Target\} forecasts. The t-statistics for $R^2_{\mathrm{OOS}}$ are calculated using zero benchmarks estimation following \citet{gu2020empirical}.}
    \begin{center}
    \scriptsize
    \tabcolsep = 0.23cm
    \renewcommand{\arraystretch}{1.2}
\begin{tabularx}{\textwidth}{lYYYYYY}
\toprule
\multicolumn{7}{l}{Panel A: Reiterating reports}\\
\midrule
 Target & Revision only &  t-stat & Text only &  t-stat & Rev + text &  t-stat \\
\midrule
Stock Recommendation &        8.71\% &    9.86 &    9.85\% &    8.21 &           11.76\% &    9.12 \\
Target Price &        1.20\% &    1.63 &    2.57\% &    2.26 &            3.58\% &    3.40 \\
Earnings Forecast &        2.89\% &    2.96 &   -2.71\% &   -2.05 &           -0.66\% &    0.06 \\
\midrule
\end{tabularx}
\begin{tabularx}{\textwidth}{lYYYYYY}
\multicolumn{7}{l}{Panel B: Revising reports}\\
\midrule
 Target & Revision only &  t-stat & Text only &  t-stat & Rev + text &  t-stat \\
\midrule
Stock Recommendation &       14.99\% &    2.92 &   16.80\% &     4.8 &           22.63\% &    4.83 \\
Target Price &       16.72\% &   11.14 &   17.72\% &    11.1 &           20.88\% &   11.18 \\
Earnings Forecast &       10.94\% &    7.48 &   14.27\% &    10.9 &           16.37\% &   11.42 \\
\bottomrule
\end{tabularx}
\end{center}
\end{table}

\newpage
\begin{table}[!htb]
\caption{Information Content of Text Embeddings vs. Text Tones}\label{tab:tone}
    {\footnotesize This table compares the information content of text embeddings against various text tone measures. The tone measures are calculated for the whole report ($Tone_{NB}$, $Tone_{BERT}$), for income statement topics ($Tone_{Income, NB/BERT}$), and for non-income statement topics ($Tone_{NonIncome, NB/BERT}$), using both Naive Bayes (NB) and BERT-based methods. Panel A presents OLS regression results of cumulative abnormal returns ($CAR[-1,+1]$) on the various tone measures. T-statistics (in parentheses) are calculated using standard errors clustered two-way by stock and year. Panel B presents the out-of-sample performance of Ridge regression models estimating $CAR[-1,+1]$ using distinct inputs. `Rev' denotes three analyst forecast revision measures, and `Emb' denotes report text embeddings. The t-statistics for the $R^2_{\mathrm{OOS}}$ are calculated following \citet{gu2020empirical}. $^*p<.1$, $^{**}p<.05$, $^{***}p<.01$.}
\begin{center}
\scriptsize
\begin{tabularx}{\textwidth}{l*{7}{Y}}

\toprule
\multicolumn{8}{l}{Panel A: Market reaction to tone measures} \\
\midrule
            &\multicolumn{1}{c}{(1)}&\multicolumn{1}{c}{(2)}&\multicolumn{1}{c}{(3)}&\multicolumn{1}{c}{(4)}&\multicolumn{1}{c}{(5)}&\multicolumn{1}{c}{(6)}&\multicolumn{1}{c}{(7)}\\
\midrule
$Tone_{NB}$             &       0.022\sym{***}&                     &                     &                     &                     &                     &                     \\
                    &     (15.86)         &                     &                     &                     &                     &                     &                     \\
$Tone_{Income, NB}$      &                     &       0.006\sym{***}&       0.006\sym{***}&                     &                     &                     &       0.000         \\
                    &                     &     (15.02)         &     (12.05)         &                     &                     &                     &      (1.12)         \\
$Tone_{NonIncome, NB}$   &                     &       0.014\sym{***}&       0.015\sym{***}&                     &                     &                     &      -0.001         \\
                    &                     &     (13.10)         &     (11.35)         &                     &                     &                     &     (-1.47)         \\
$Tone_{BERT}$           &                     &                     &                     &       0.035\sym{***}&                     &                     &                     \\
                    &                     &                     &                     &     (17.71)         &                     &                     &                     \\
$Tone_{Income, BERT}$    &                     &                     &                     &                     &       0.011\sym{***}&       0.011\sym{***}&       0.010\sym{***}\\
                    &                     &                     &                     &                     &     (15.44)         &     (13.99)         &     (13.57)         \\
$Tone_{NonIncome, BERT}$ &                     &                     &                     &                     &       0.020\sym{***}&       0.021\sym{***}&       0.022\sym{***}\\
                    &                     &                     &                     &                     &     (15.90)         &     (14.94)         &     (15.57)         \\
Constant            &      -0.008\sym{***}&      -0.007\sym{***}&      -0.008\sym{***}&      -0.008\sym{***}&      -0.007\sym{***}&      -0.007\sym{***}&      -0.007\sym{***}\\
                    &    (-13.25)         &    (-13.42)         &    (-16.31)         &    (-12.18)         &    (-10.98)         &    (-18.23)         &    (-17.36)         \\
\midrule
Firm FE  &      No         &      No        &      Yes         &      No         &      No         &      Yes         &      Yes         \\
Year FE  &      No         &      No        &      Yes         &      No         &      No         &      Yes         &      Yes         \\
\(N\)&      99705         &       99705         &       99705         &       99705         &       99705         &       99705         &       99705         \\
Adjusted \(R^{2}\)  &       0.012         &       0.011         &       0.018         &       0.036         &       0.034         &       0.041         &       0.041         \\
\midrule
\end{tabularx}

\begin{tabularx}{\textwidth}{lYY}
\multicolumn{3}{l}{Panel B: Information content of text tones and text embeddings} \\
\midrule
Model Input & $R^2_{\mathrm{OOS}}$ & t-stat \\
\midrule
                              $Tone_{NB}$ &  0.05\% &   -0.30 \\
                                       $Tone_{BERT}$ &  3.78\% &   10.55 \\
                  $Tone_{NB}$ + Rev &  9.49\% &   10.19 \\
                           $Tone_{BERT}$ + Rev & 10.58\% &   11.11 \\
          $Tone_{NB}$ + Rev + Emb  & 12.28\% &    8.87 \\
                    $Tone_{BERT}$ + Rev + Emb & 12.27\% &    8.91 \\
``$Tone_{NB}$ + Rev + Emb" - ``Rev + Emb" & 0.00\% &    0.67 \\
``$Tone_{BERT}$ + Rev + Emb" - ``Rev + Emb" & -0.01\% &   -0.79 \\
\bottomrule
\end{tabularx}
\end{center}
\end{table}

\newpage
\begin{table}[!htb]
\caption{Summary Statistics of Analyst Information Value}\label{tab:iv_ss}
    {\footnotesize 
    This table presents summary statistics for the information value of analyst reports on S\&P 100 stocks from 2015Q1 to 2023Q4. All monetary values are adjusted to 2020 dollars using the CPI. The value of information is defined as explained return volatility divided by price impact. Panel A details this information value, breaking it down into components from text embeddings and forecast revisions. Means, standard errors, and confidence intervals are estimated using the delta method. Panel B provides summary statistics for market conditions, including the stock price at day t-2 and the price impact of trading. Price impact is calculated from a regression of one-minute returns on signed order flow, scaled by the stock price at t-2, and is reported in billions of dollars.}
    \begin{center}
    \scriptsize
    \tabcolsep = 0.23cm
    \renewcommand{\arraystretch}{1.2}
    \begin{tabularx}{\textwidth}{lYYYYY}
    \toprule
    \multicolumn{6}{l}{Panel A: Dollar value of analyst information} \\
    \midrule
    {} & Mean & SE & 95\%CI & 99\%CI & N \\
    \midrule
    Information value, \$M                & 0.47 & 0.05 & [0.38, 0.56] & [0.35, 0.58] & 17672 \\
    Information value of text embeddings, \$M        & 0.38 & 0.04 & [0.30, 0.46] & [0.28, 0.48] & 17672 \\
    Information value of forecast revisions, \$M   & 0.34 & 0.04 & [0.26, 0.43] & [0.23, 0.46] & 17672 \\
    \midrule
    \end{tabularx}
    
    \begin{tabularx}{\textwidth}{lYYYYYYY}
    \multicolumn{8}{l}{Panel B: Price impact and stock price} \\
    \midrule
    {} &{} & Mean & Std. Dev. & p25 & P50 & P75 & N \\
    \midrule
    Price impact per \$B   & {} & 0.34 & 1.29 & 0.05 & 0.13 & 0.31 & 17672 \\
    Stock price            & {} & 118.49 & 130.34 & 51.15 & 82.0 & 143.58 & 17672 \\
    \bottomrule
    \end{tabularx}
    
    \end{center}
\end{table}

\newpage
\begin{table}[!htb]
    \caption{Analyst Information Value and Stock Characteristics}
    {\footnotesize
    This table presents OLS regressions of the logarithm of analyst information value on stock characteristics. Log information value (text) specifically measures the value of information from the analyst report text. The overall log information value is computed as the difference between log explained return variance and log price impact, which are separately analyzed in Panel B. Observations with negative price impact and negative explained return variance are excluded. Variable definitions are presented in Table \ref{tab:numdef}. T-statistics (in parentheses) are clustered by stock and year. $^*p<.1$, $^{**}p<.05$, $^{***}p<.01$.}
    \begin{center}
    \scriptsize
    \label{tab:iv_size}
    \begin{tabularx}{\textwidth}{lYYYYYYYY}
        \toprule
        \multicolumn{9}{l}{Panel A: Analyst information value} \\
        \midrule
        & \multicolumn{4}{c}{log information value} & \multicolumn{4}{c}{log information value (text)} \\
        \cmidrule(lr){2-5} \cmidrule(lr){6-9}
        & (1) & (2) & (3) & (4) & (5) & (6) & (7) & (8) \\
        \midrule
        Size & 0.574*** & 0.853*** & 0.573*** & 0.864*** & 0.667*** & 0.922*** & 0.669*** & 0.934*** \\ 
        & (8.01) & (9.25) & (8.33) & (9.62) & (9.20) & (8.45) & (9.41) & (8.38) \\ 
        BtoM & -0.222 & 0.416 & -0.139 & 0.457 & -0.218 & 0.326 & -0.161 & 0.335 \\ 
        & (-1.38) & (1.27) & (-0.94) & (1.36) & (-1.34) & (0.93) & (-1.05) & (0.92) \\ 
        Year FE & No & No & Yes & Yes & No & No & Yes & Yes \\ 
        Stock FE & No & Yes & No & Yes & No & Yes & No & Yes \\ 
        \midrule 
        $N$ & 8030 & 8030 & 8030 & 8030 & 8509 & 8509 & 8509 & 8509 \\ 
        Adjusted $R^2$ & 0.072 & 0.155 & 0.088 & 0.168 & 0.093 & 0.167 & 0.103 & 0.176 \\ 
        \midrule 
        \multicolumn{9}{l}{Panel B: Explained return variance and price impact} \\
        \midrule
        & \multicolumn{4}{c}{log explained return variance} & \multicolumn{4}{c}{log price impact} \\ 
        \cmidrule(lr){2-5} \cmidrule(lr){6-9} 
        & (1) & (2) & (3) & (4) & (5) & (6) & (7) & (8) \\ 
        \midrule 
        Size & -0.229*** & 0.028 & -0.228*** & 0.044 & -0.801*** & -0.831*** & -0.798*** & -0.826*** \\ 
        & (-5.58) & (0.53) & (-6.07) & (0.78) & (-13.15) & (-9.24) & (-13.36) & (-9.46) \\ 
        BtoM & -0.273** & 0.153 & -0.192* & 0.180 & -0.059 & -0.260 & -0.060 & -0.274 \\ 
        & (-2.31) & (0.62) & (-1.82) & (0.70) & (-0.51) & (-1.26) & (-0.56) & (-1.33) \\ 
        Year FE & No & No & Yes & Yes & No & No & Yes & Yes \\ 
        Stock FE & No & Yes & No & Yes & No & Yes & No & Yes \\ 
        \midrule 
        $N$ & 8030 & 8030 & 8030 & 8030 & 8030 & 8030 & 8030 & 8030 \\ 
        Adjusted $R^2$ & 0.015 & 0.076 & 0.036 & 0.092 & 0.318 & 0.449 & 0.318 & 0.450 \\ 
        \bottomrule
    \end{tabularx}
    \end{center}
\end{table}

\newpage
\begin{table}[!htb]
    \caption{Analyst Information Value and Analyst Characteristics}
    {\footnotesize 
    This table reports regression results examining the relationship between the value of analyst information and analyst boldness. Log information value (text) specifically measures the value of information from the analyst report text. The overall log information value is computed as the difference between log explained return variance and log price impact, which are separately analyzed in Panel B. Observations with negative price impact and negative explained return variance are excluded. Variable definitions are presented in Table \ref{tab:numdef}. T-statistics (in parentheses) are clustered by stock and year. $^*p<.1$, $^{**}p<.05$, $^{***}p<.01$.}    
    \begin{center}
    \scriptsize
    \label{tab:iv_bold}
    \begin{tabularx}{\textwidth}{lYYYYYYYY}
        \toprule
        \multicolumn{9}{l}{Panel A: Analyst information value} \\
        \midrule
        & \multicolumn{4}{c}{log information value} & \multicolumn{4}{c}{log information value (text)} \\
        \cmidrule(lr){2-5} \cmidrule(lr){6-9}
        & (1) & (2) & (3) & (4) & (5) & (6) & (7) & (8) \\
        \midrule
        Bold & 0.300** & 0.305*** & 0.275*** & 0.290*** & 0.240** & 0.249** & 0.230** & 0.248** \\
             & (3.000) & (4.150)  & (3.420)  & (4.090)  & (2.360) & (3.120) & (2.810) & (3.100) \\
        Brokersize & -0.000 & 0.000 & 0.001 & 0.001 & -0.000 & 0.000 & 0.002 & 0.002 \\
             & (-0.340) & (0.760) & (0.890) & (0.880) & (-0.570) & (1.180) & (1.010) & (1.140) \\
        Year FE & No & Yes & Yes & Yes & No & Yes & Yes & Yes \\
        Stock FE & No & Yes & No & Yes & No & Yes & No & Yes \\
        Analyst FE & No & No & Yes & Yes & No & No & Yes & Yes \\
        \midrule
        $N$ & 13652 & 13650 & 13607 & 13606 & 13520 & 13518 & 13477 & 13476 \\
        Adjusted $R^2$ & 0.003 & 0.182 & 0.141 & 0.201 & 0.002 & 0.181 & 0.141 & 0.204 \\
        \midrule
        \multicolumn{9}{l}{Panel B: Explained return variance and price impact} \\
        \midrule
        & \multicolumn{4}{c}{log explained return variance} & \multicolumn{4}{c}{log price impact} \\
        \cmidrule(lr){2-5} \cmidrule(lr){6-9}
        & (1) & (2) & (3) & (4) & (5) & (6) & (7) & (8) \\
        \midrule
        Bold & 0.290*** & 0.241*** & 0.251*** & 0.231*** & -0.017 & -0.065 & -0.025 & -0.059 \\
             & (5.770)  & (5.960)  & (5.910)  & (5.350)  & (-0.280) & (-1.350) & (-0.420) & (-1.340) \\
        Brokersize & -0.000 & -0.000 & -0.000 & -0.000 & 0.000 & -0.000 & -0.002 & -0.002 \\
             & (-0.700) & (-0.050) & (-0.570) & (-0.220) & (0.180) & (-1.130) & (-1.730) & (-1.680) \\
        Year FE & No & Yes & Yes & Yes & No & Yes & Yes & Yes \\
        Stock FE & No & Yes & No & Yes & No & Yes & No & Yes \\
        Analyst FE & No & No & Yes & Yes & No & No & Yes & Yes \\
        \midrule
        $N$ & 13642 & 13640 & 13588 & 13587 & 12867 & 12865 & 12813 & 12812 \\
        Adjusted $R^2$ & 0.004 & 0.095 & 0.091 & 0.122 & -0.000 & 0.416 & 0.228 & 0.442 \\
        \bottomrule
    \end{tabularx}
    \end{center}
\end{table}

\newpage
\begin{table}[!htb]
    \caption{Analyst Information Value and Earnings Announcements}
    {\footnotesize 
    This table presents regression results examining how analyst information value varies around earnings announcements and with trading volume. Log information value (text) specifically measures the value of information from the analyst report text. The overall log information value is computed as the difference between log explained return variance and log price impact, which are separately analyzed in Panel B. Observations with negative price impact and negative explained return variance are excluded. $Week$ is an indicator equal to 1 if the report is released within one week after the earnings announcement date, and 0 otherwise. $TradingVolume$ is calculated as trading volume on the earnings announcement day, scaled by shares outstanding. T-statistics (in parentheses) are clustered by stock and year. $^*p<.1$, $^{**}p<.05$, $^{***}p<.01$.}
    \begin{center}
    \scriptsize
    \label{tab:iv_ead}
    \begin{tabularx}{\textwidth}{lYYYYYYYY}
        \toprule
        \multicolumn{9}{l}{Panel A: Analyst information value} \\
        \midrule
        & \multicolumn{4}{c}{log information value} & \multicolumn{4}{c}{log information value (text)} \\
        \cmidrule(lr){2-5} \cmidrule(lr){6-9}
        & (1) & (2) & (3) & (4) & (5) & (6) & (7) & (8) \\
        \midrule
        Week & 0.246* & 0.392*** & 0.225* & 0.380*** & 0.199 & 0.354*** & 0.173 & 0.335*** \\
             & (1.910) & (3.560) & (1.880) & (3.540) & (1.550) & (3.200) & (1.450) & (3.190) \\
        TradingVolume & 0.003 & -0.005 & 0.003 & -0.004 & 0.000 & -0.005 & -0.000 & -0.004 \\
             & (0.770) & (-1.050) & (0.600) & (-0.960) & (0.110) & (-0.940) & (-0.020) & (-0.890) \\
        Week $\times$ TradingVolume & 0.018*** & 0.019*** & 0.015*** & 0.017*** & 0.013*** & 0.015** & 0.010* & 0.013** \\
             & (3.390) & (3.540) & (2.740) & (2.950) & (2.720) & (2.510) & (1.970) & (2.130) \\
        Year FE & No & No & Yes & Yes & No & No & Yes & Yes \\
        Stock FE & No & Yes & No & Yes & No & Yes & No & Yes \\
        \midrule
        $N$ & 8030 & 8030 & 8030 & 8030 & 8509 & 8509 & 8509 & 8509 \\
        Adjusted $R^2$ & 0.013 & 0.149 & 0.029 & 0.158 & 0.006 & 0.152 & 0.017 & 0.158 \\
        \midrule
        \multicolumn{9}{l}{Panel B: Explained return variance and price impact} \\
        \midrule
        & \multicolumn{4}{c}{log explained return variance} & \multicolumn{4}{c}{log price impact} \\
        \cmidrule(lr){2-5} \cmidrule(lr){6-9}
        & (1) & (2) & (3) & (4) & (5) & (6) & (7) & (8) \\
        \midrule
        Week & 0.501*** & 0.491*** & 0.477*** & 0.473*** & 0.243*** & 0.088** & 0.241*** & 0.083** \\
             & (5.090) & (5.280) & (5.120) & (5.380) & (3.430) & (2.110) & (3.640) & (2.000) \\
        TradingVolume & 0.020*** & 0.004 & 0.019*** & 0.005 & 0.016*** & 0.009** & 0.016*** & 0.009** \\
             & (5.910) & (1.170) & (5.740) & (1.410) & (3.080) & (2.020) & (3.090) & (2.030) \\
        Week $\times$ TradingVolume & 0.014*** & 0.013*** & 0.012** & 0.011** & -0.003** & -0.005** & -0.002 & -0.005** \\
             & (3.120) & (3.160) & (2.530) & (2.410) & (-2.050) & (-2.290) & (-1.350) & (-2.370) \\
        Year FE & No & No & Yes & Yes & No & No & Yes & Yes \\
        Stock FE & No & Yes & No & Yes & No & Yes & No & Yes \\
        \midrule
        $N$ & 8030 & 8030 & 8030 & 8030 & 8030 & 8030 & 8030 & 8030 \\
        Adjusted $R^2$ & 0.059 & 0.104 & 0.073 & 0.115 & 0.025 & 0.393 & 0.033 & 0.396 \\
        \bottomrule
    \end{tabularx}
    \end{center}
\end{table}

\newpage
\begin{center}{\bf{\LARGE Internet Appendix for \\[0.5cm] ``The Value of Information from Sell-side Analysts''}} \\[0.9cm]

\textit{Not for Publication} \\

\end{center}

\setcounter{page}{1}
\setcounter{table}{0}
\setcounter{figure}{0}

\renewcommand{\thetable}{A\arabic{table}}
\renewcommand{\thefigure}{A\arabic{figure}}

\newpage
\section*{A. Additional Figures}
\captionsetup[subfigure]{labelformat=empty, justification=centering}

\begin{figure}[H]
    \begin{center}
        \includegraphics[width=0.9\textwidth]{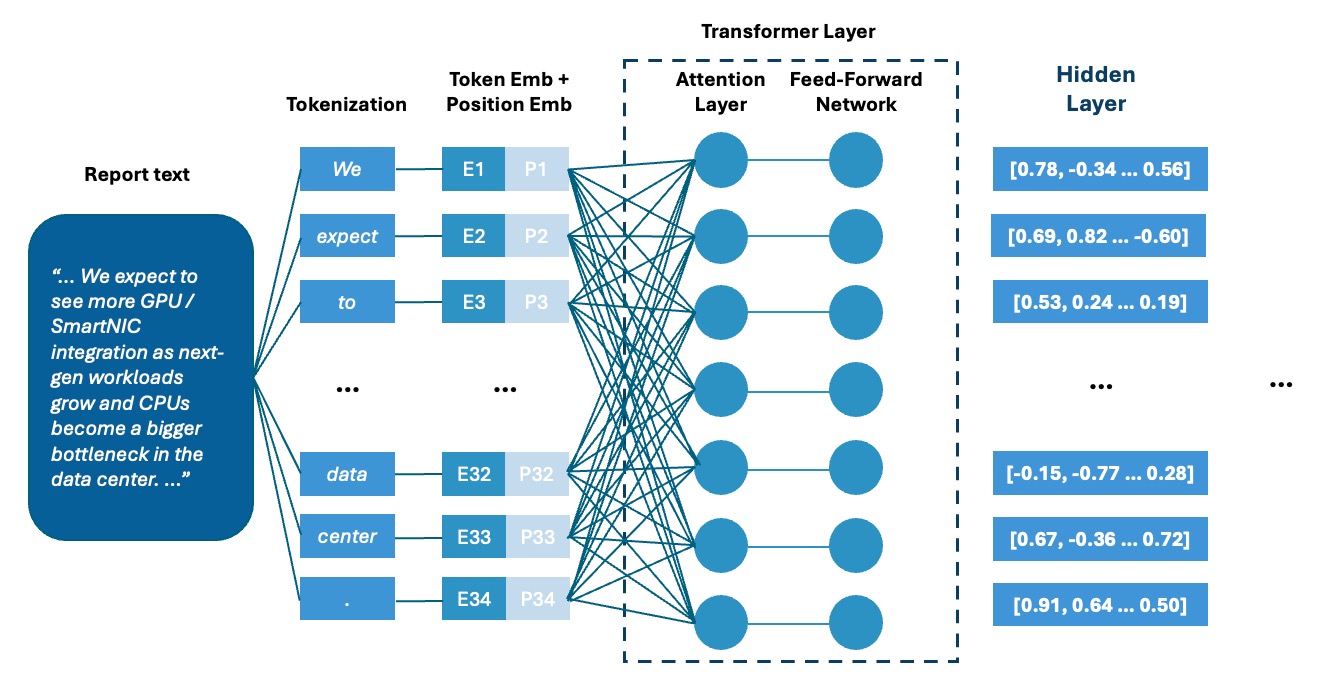}
    \end{center}
    \caption{Large Language Model Architecture}\label{fig:transformer}
    {\footnotesize 
    This figure demonstrates how a large language model generates context-aware representations using a sample sentence from an analyst report on NVIDIA. The model first tokenizes the sentence and generates two types of embeddings: token embeddings and positional embeddings. These are then passed through a stack of transformer layers, the core of the architecture. The output of each transformer layer is a sequence of context-aware vectors, one for each token, capturing its meaning concerning the full context.}
\end{figure}

\begin{figure}[H] \label{fig:word_clould}
\begin{center}
\begin{subfigure}[b]{0.48\textwidth}
\includegraphics[width=\textwidth]{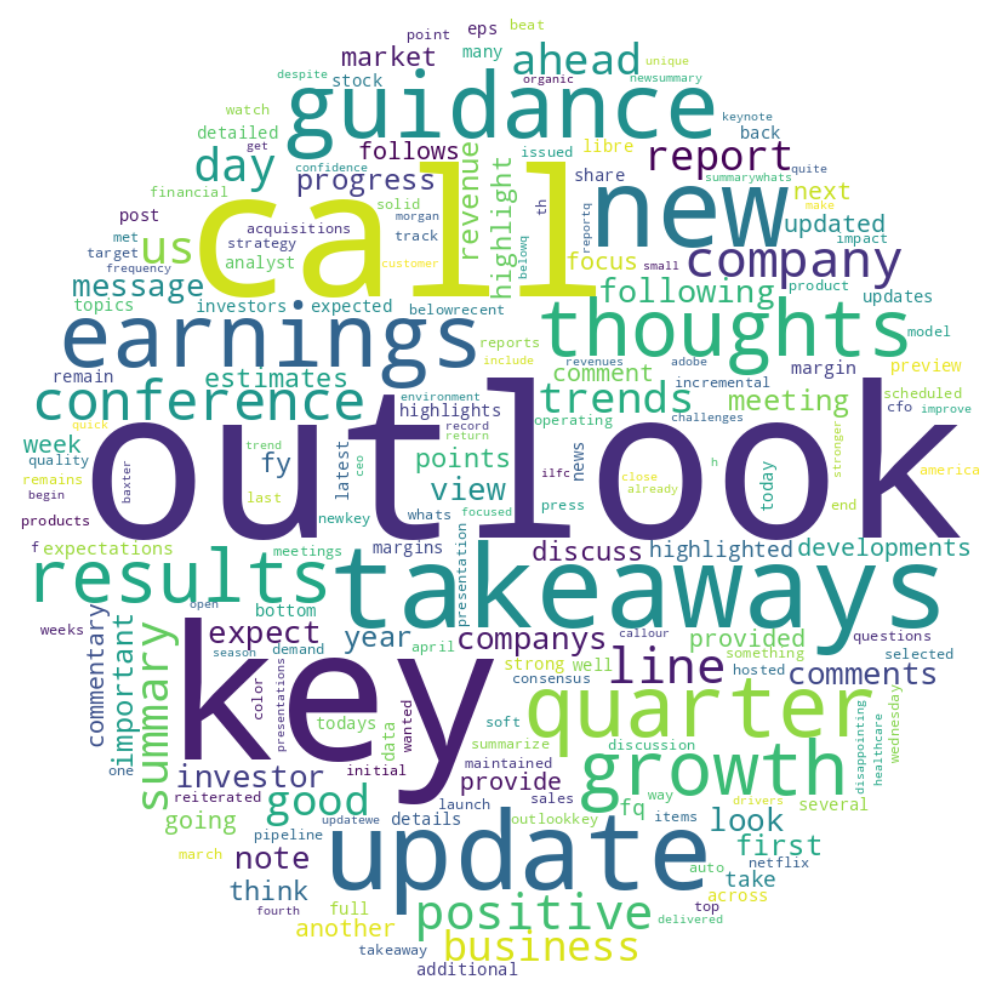}
\caption{(1) Executive Summary}
\end{subfigure}
\hfill
\begin{subfigure}[b]{0.48\textwidth}
\includegraphics[width=\textwidth]{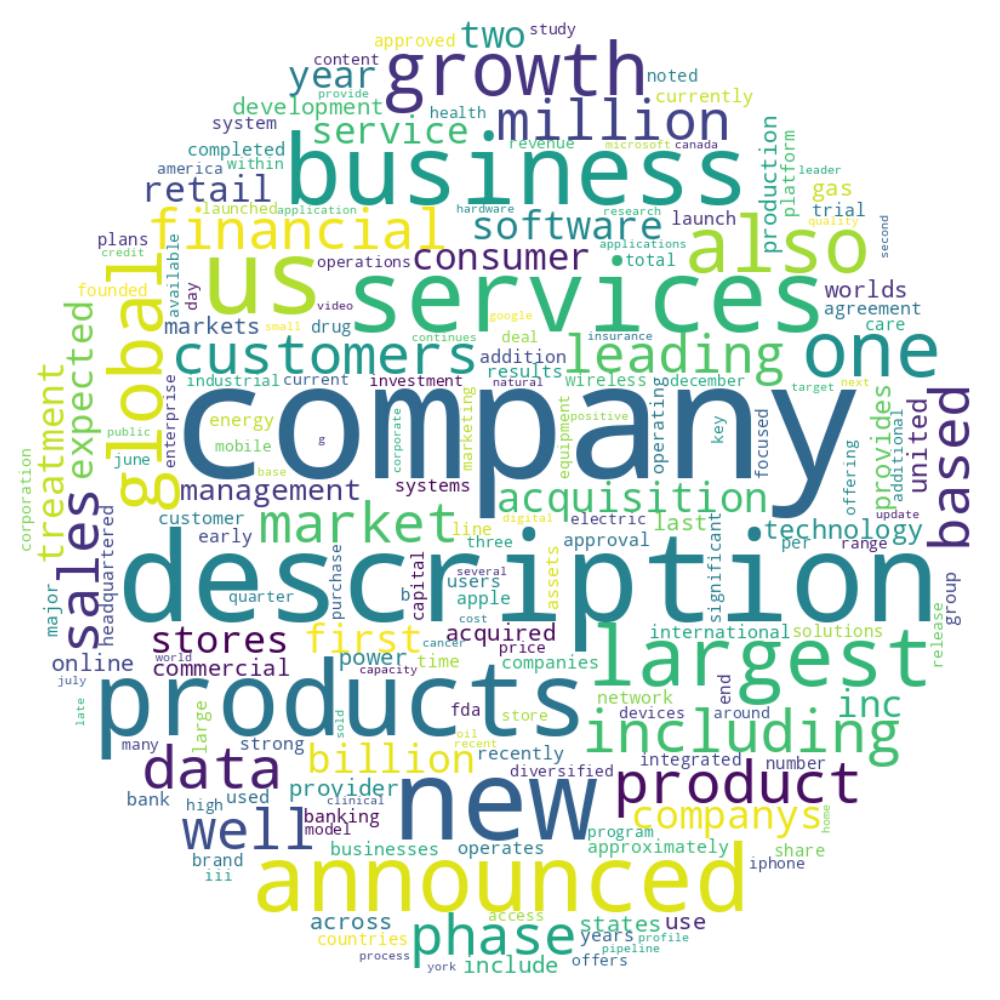}
\caption{(2) Company Overview}
\end{subfigure}
\vfill
\begin{subfigure}[b]{0.48\textwidth}
\includegraphics[width=\textwidth]{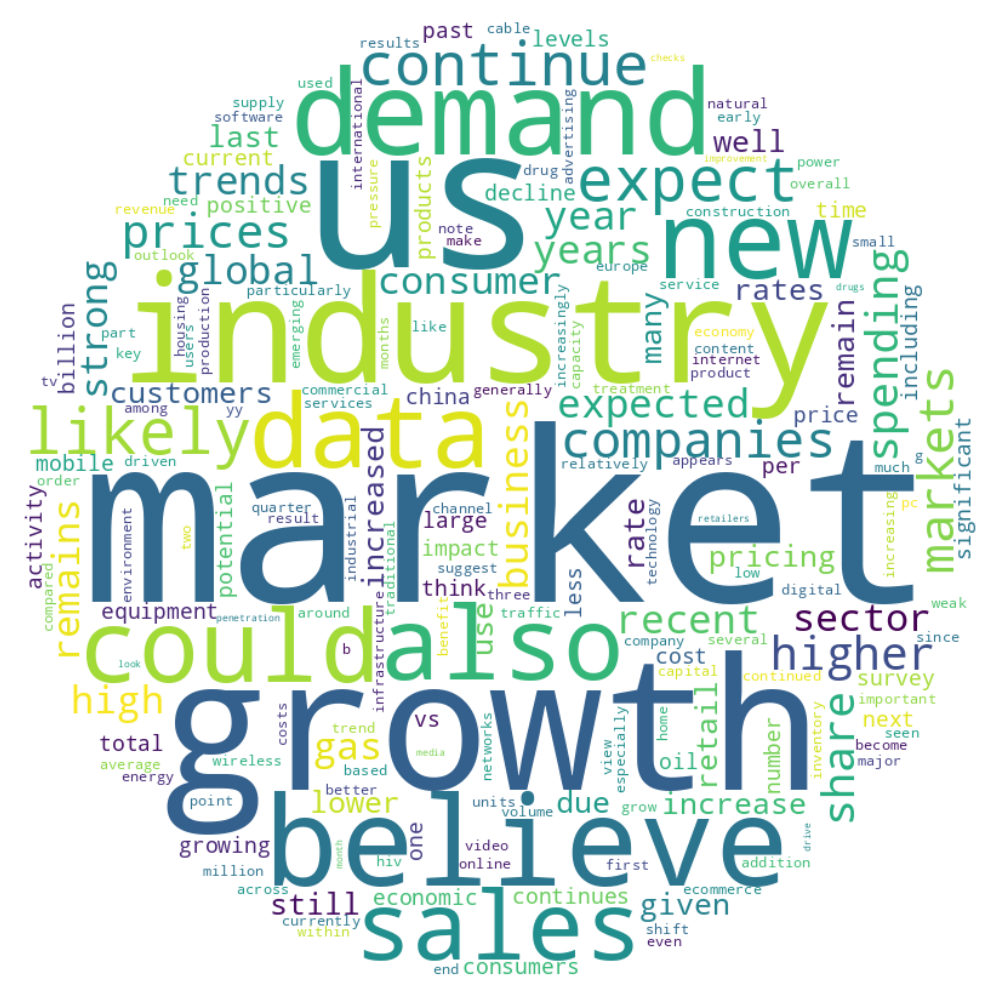}
\caption{(3) Industry Analysis}
\end{subfigure}
\hfill
\begin{subfigure}[b]{0.48\textwidth}
\includegraphics[width=\textwidth]{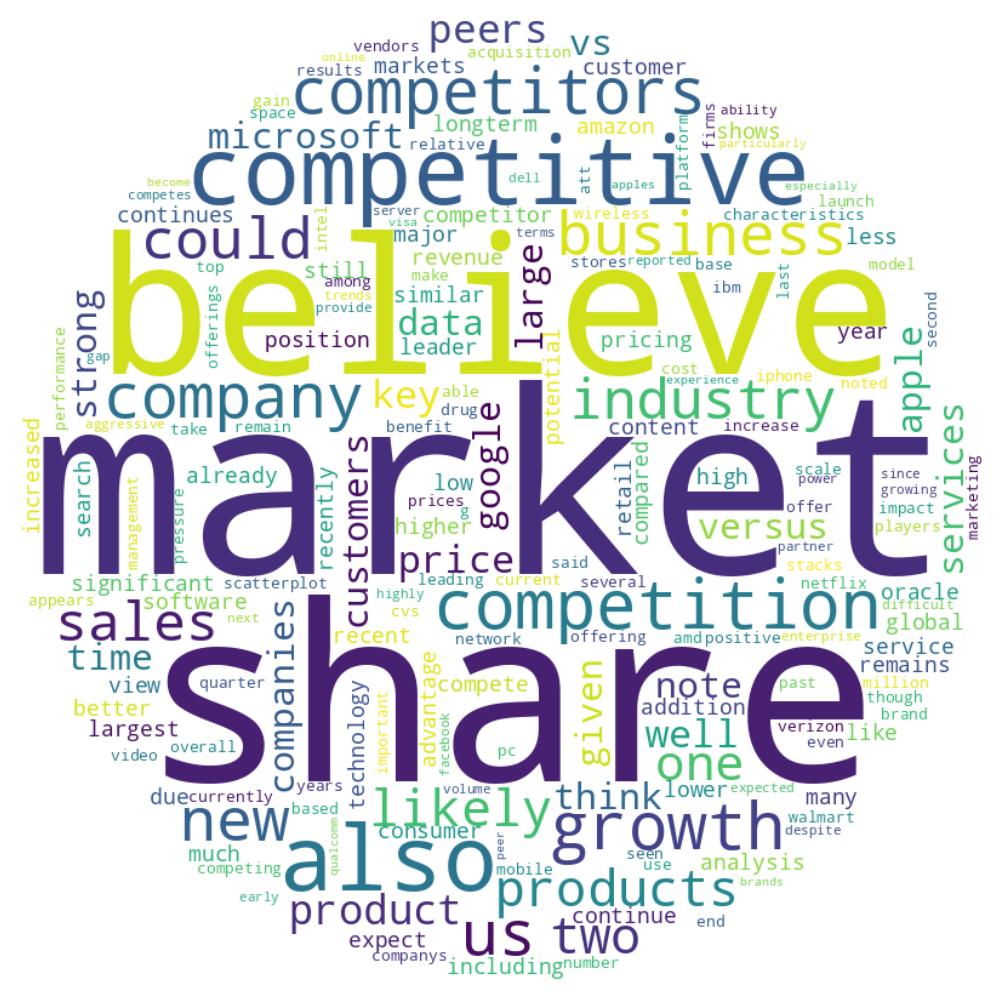}
\caption{(4) Competitive Landscape}
\end{subfigure}
\end{center}
\caption{Word Clouds of Analyst Report Topics}\label{fig: a1}
    {\begin{onehalfspace}
    \footnotesize This figure displays word clouds for 16 key topics commonly discussed in analyst reports. Each word cloud highlights the most frequently occurring terms within each topic, with word size proportional to term frequency. The topics include Executive Summary, Company Overview, Industry Analysis, Competitive Landscape, Income Statement Analysis, Balance Sheet Analysis, Cash Flow Analysis, Financial Ratios, Business Segments, Growth Strategies, Risk Factors, Management and Governance, ESG Factors, Valuation, Investment Thesis, and Appendices and Disclosures. The ``None of the Above" category is excluded from the visualization.
    \end{onehalfspace}}
\end{figure}

\newpage
\addtocounter{figure}{-1} 

\begin{figure}[H]
\begin{center}
\begin{subfigure}[b]{0.48\textwidth}
\includegraphics[width=\textwidth]{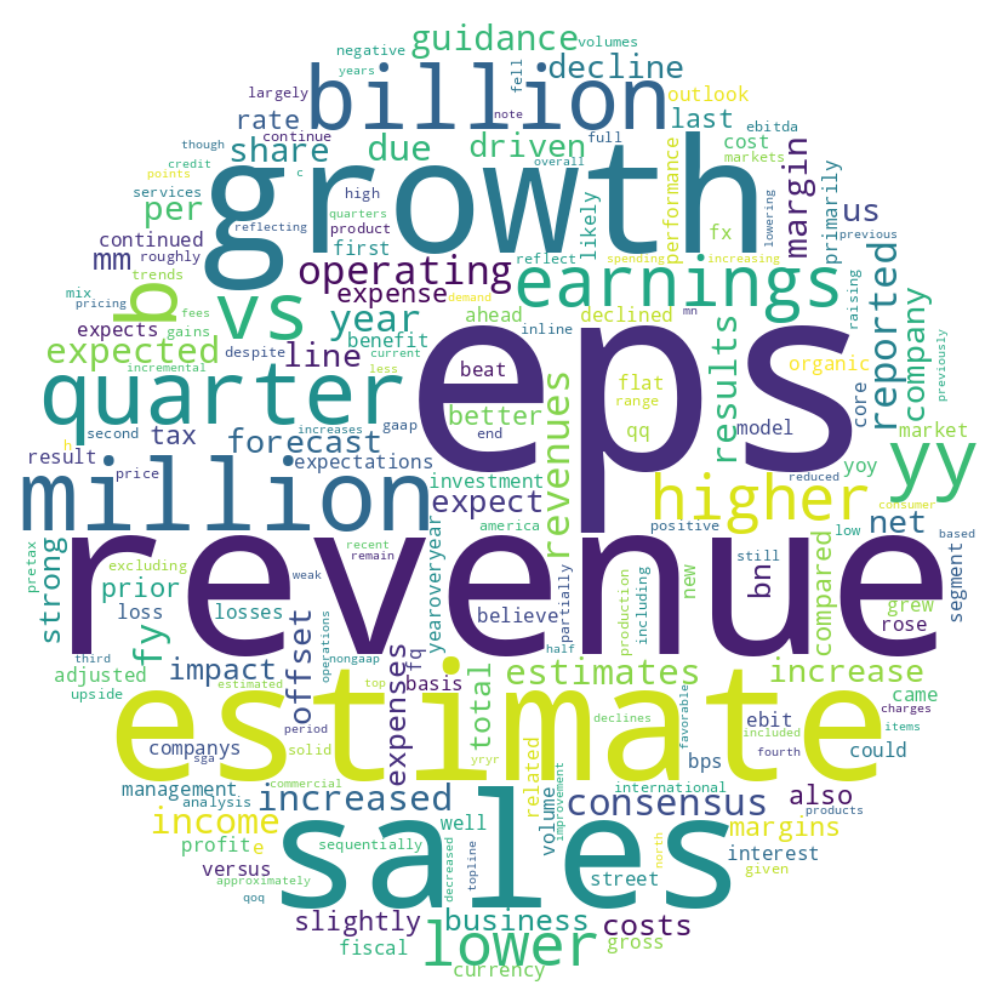}
\caption{(5) Income Statement Analysis}
\end{subfigure}
\hfill
\begin{subfigure}[b]{0.48\textwidth}
\includegraphics[width=\textwidth]{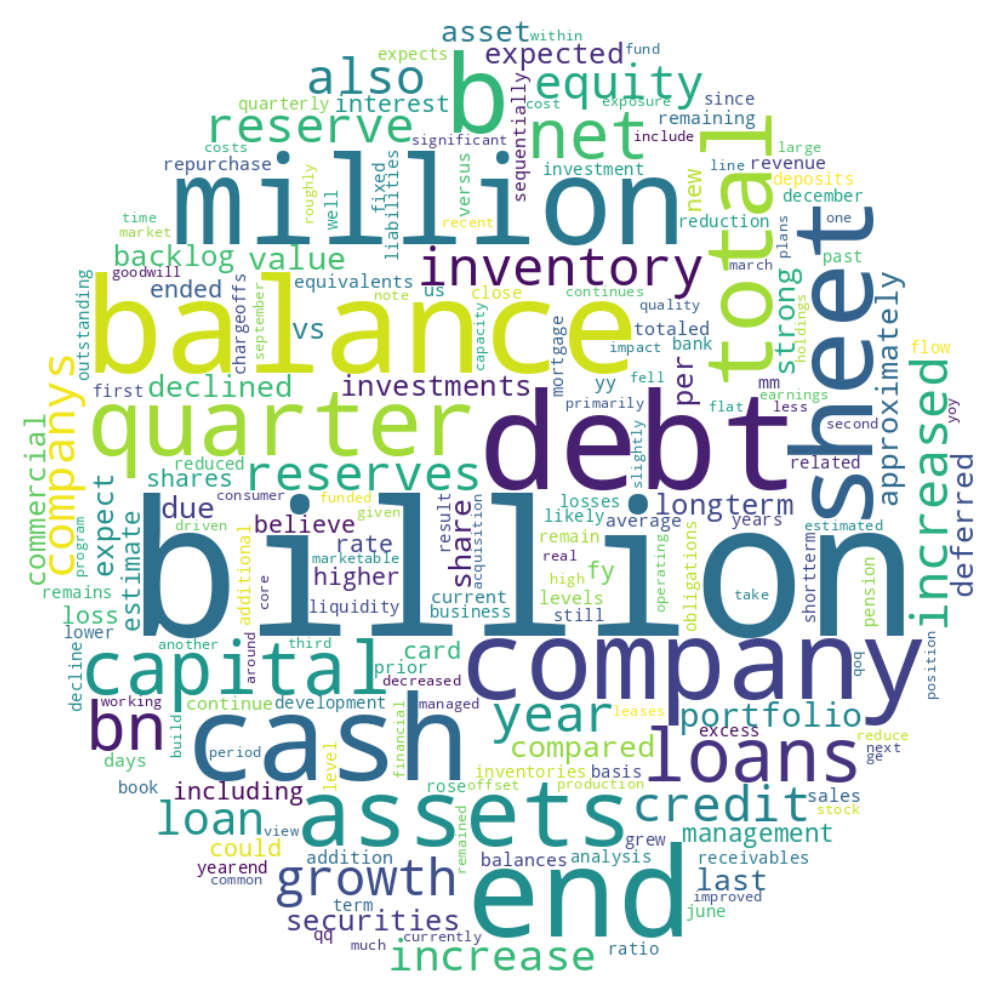}
\caption{(6) Balance Sheet Analysis}
\end{subfigure}
\vfill
\begin{subfigure}[b]{0.48\textwidth}
\includegraphics[width=\textwidth]{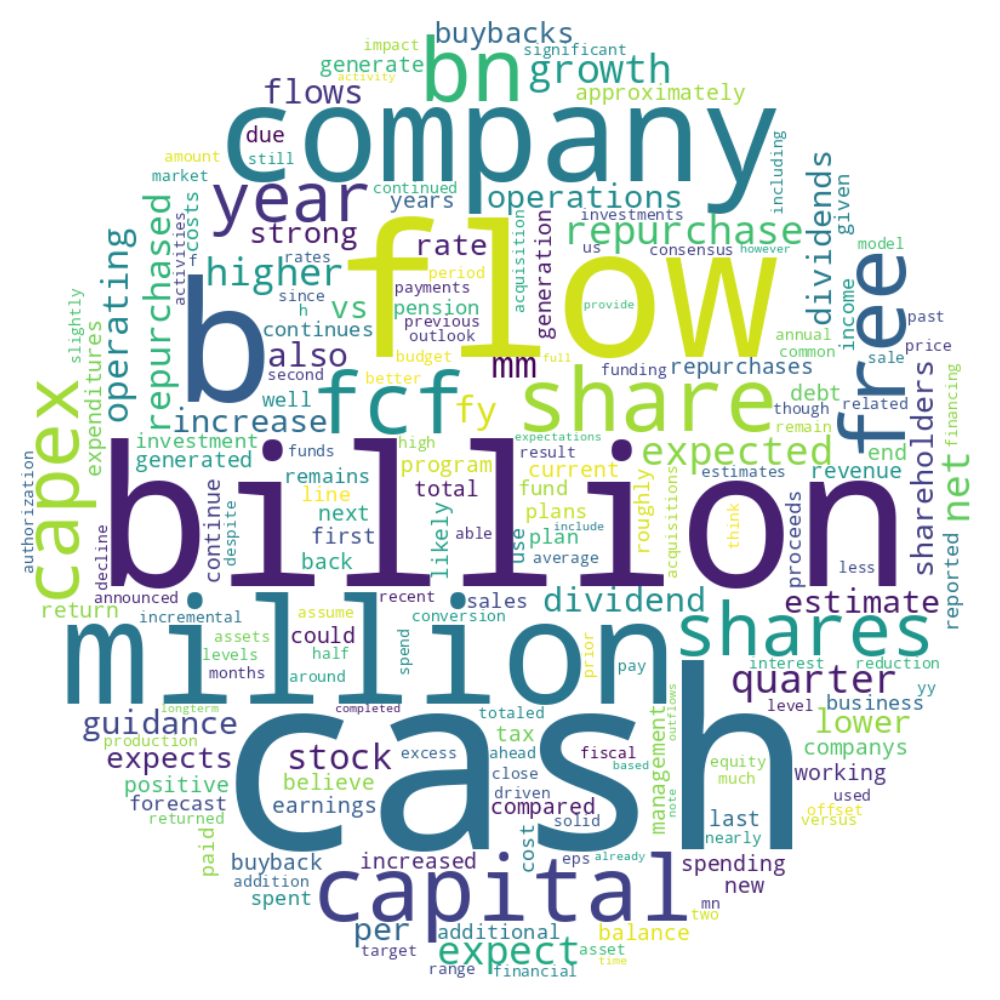}
\caption{(7) Cash Flow Analysis}
\end{subfigure}
\hfill
\begin{subfigure}[b]{0.48\textwidth}
\includegraphics[width=\textwidth]{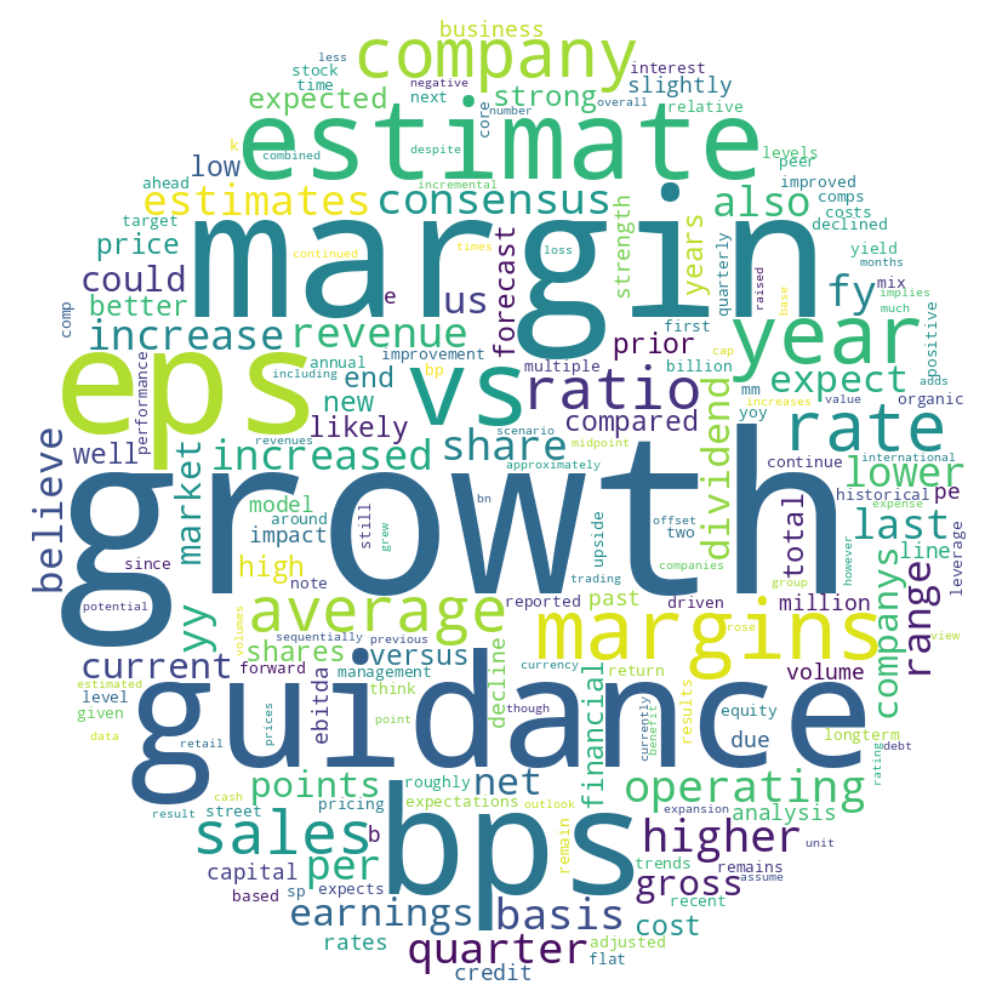}
\caption{(8) Financial Ratios}
\end{subfigure}
\end{center}
\caption{Word Clouds of Analyst Report Topics}
    {\footnotesize This figure displays word clouds for 16 key topics commonly discussed in analyst reports. Each word cloud highlights the most frequently occurring terms within each topic, with word size proportional to term frequency. The topics include Executive Summary, Company Overview, Industry Analysis, Competitive Landscape, Income Statement Analysis, Balance Sheet Analysis, Cash Flow Analysis, Financial Ratios, Business Segments, Growth Strategies, Risk Factors, Management and Governance, ESG Factors, Valuation, Investment Thesis, and Appendices and Disclosures. The ``None of the Above" category is excluded from the visualization.}
\end{figure}

\newpage
\addtocounter{figure}{-1} 

\begin{figure}[H]
\begin{center}
\begin{subfigure}[b]{0.48\textwidth}
\includegraphics[width=\textwidth]{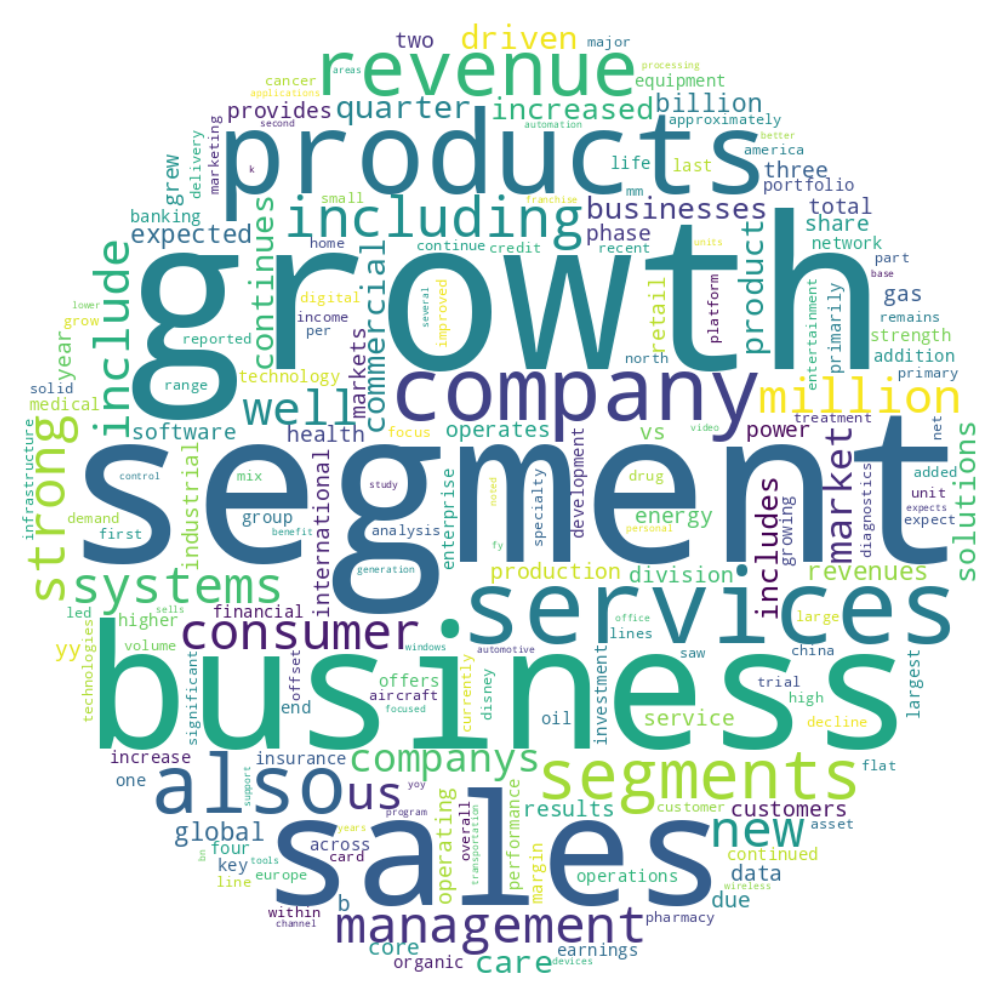}
\caption{(9) Business Segments}
\end{subfigure}
\hfill
\begin{subfigure}[b]{0.48\textwidth}
\includegraphics[width=\textwidth]{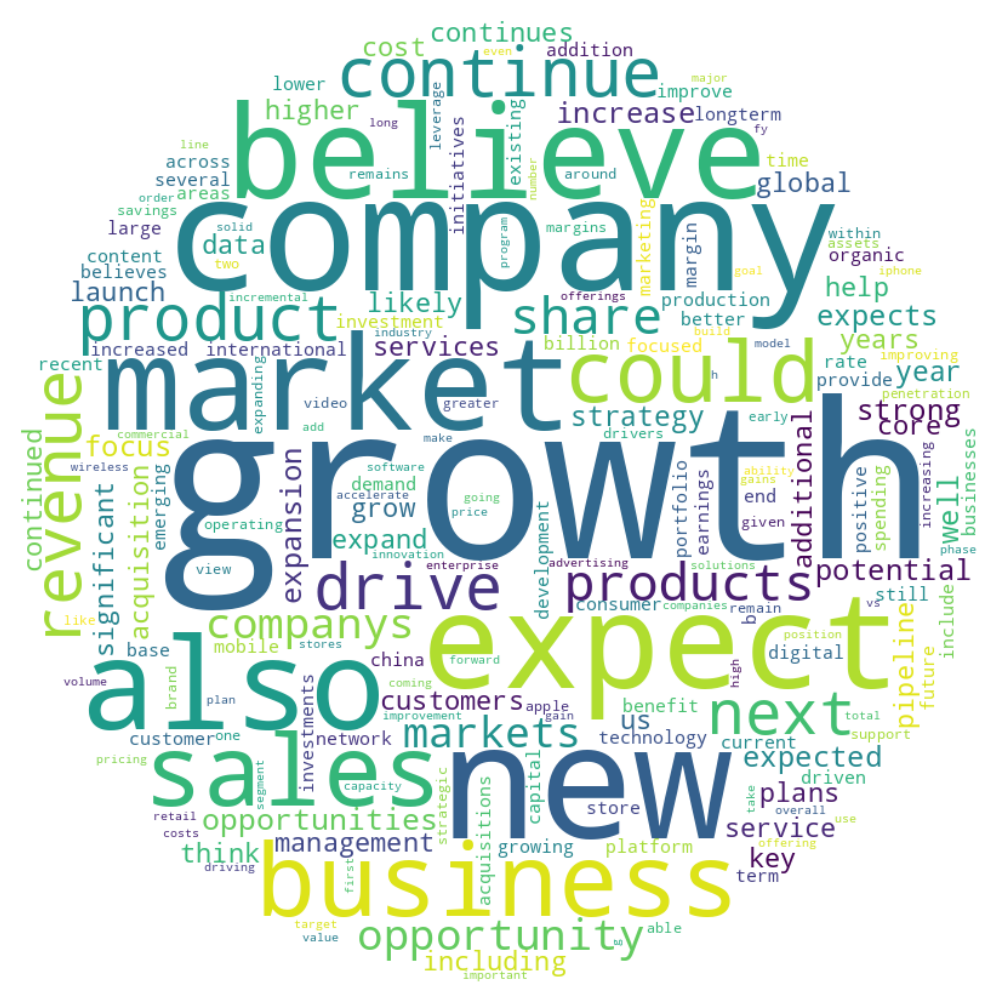}
\caption{(10) Growth Strategies}
\end{subfigure}
\vfill
\begin{subfigure}[b]{0.48\textwidth}
\includegraphics[width=\textwidth]{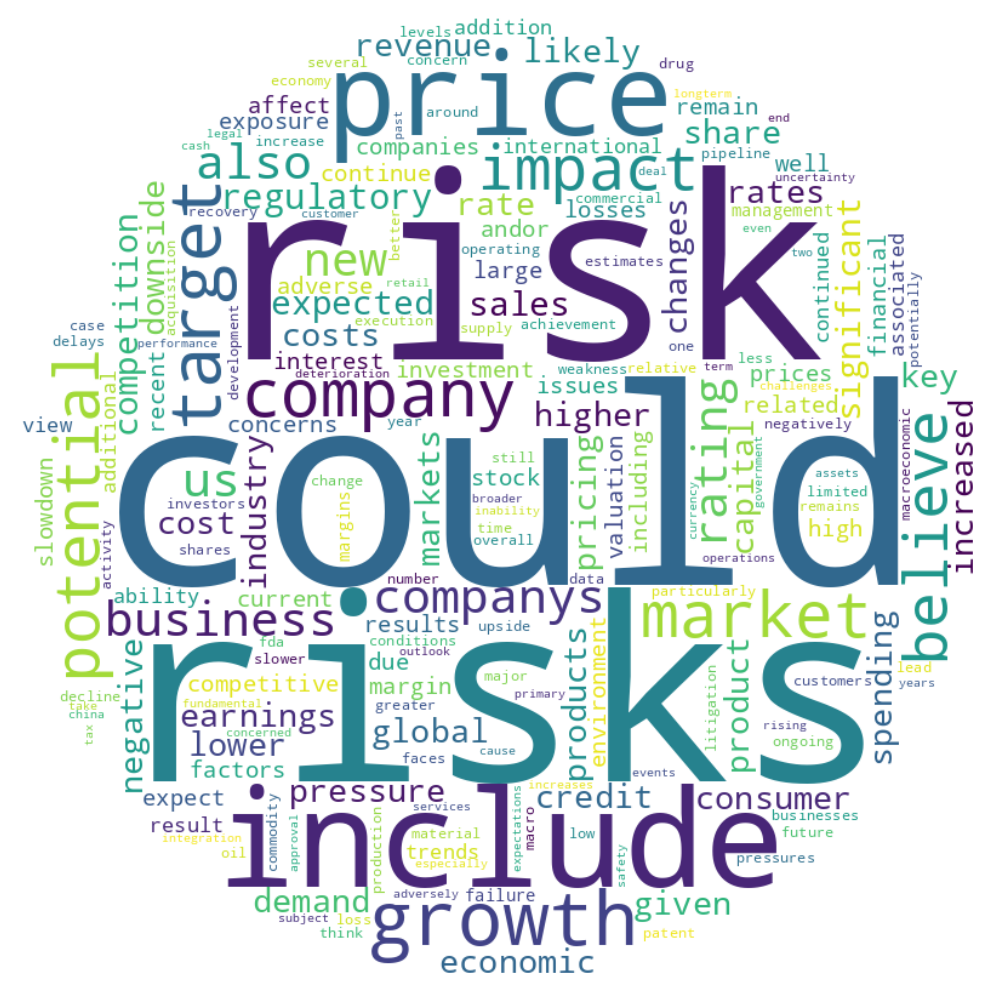}
\caption{(11) Risk Factors}
\end{subfigure}
\hfill
\begin{subfigure}[b]{0.48\textwidth}
\includegraphics[width=\textwidth]{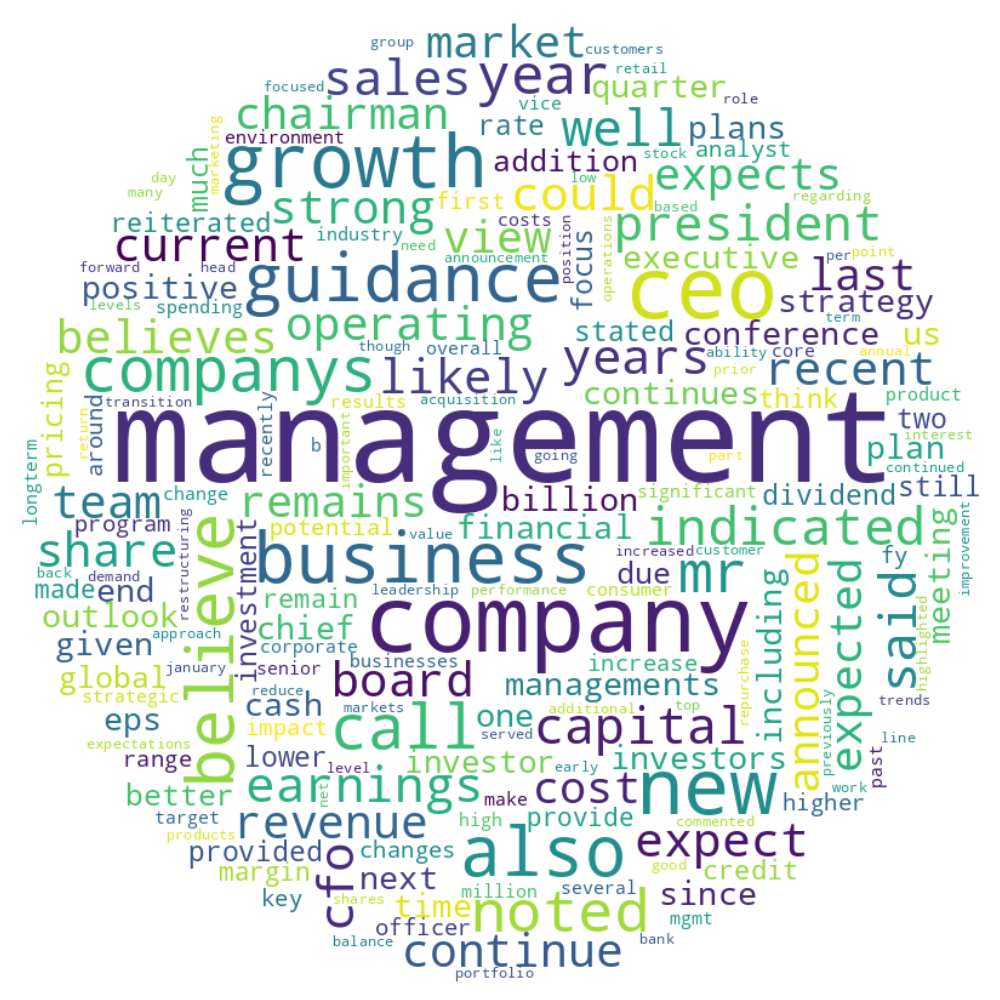}
\caption{(12) Management and Governance}
\end{subfigure}
\end{center}
\caption{Word Clouds of Analyst Report Topics}
    {\footnotesize This figure displays word clouds for 16 key topics commonly discussed in analyst reports. Each word cloud highlights the most frequently occurring terms within each topic, with word size proportional to term frequency. The topics include Executive Summary, Company Overview, Industry Analysis, Competitive Landscape, Income Statement Analysis, Balance Sheet Analysis, Cash Flow Analysis, Financial Ratios, Business Segments, Growth Strategies, Risk Factors, Management and Governance, ESG Factors, Valuation, Investment Thesis, and Appendices and Disclosures. The ``None of the Above" category is excluded from the visualization.}
\end{figure}

\newpage
\addtocounter{figure}{-1} 

\begin{figure}[H]
\begin{center}
\begin{subfigure}[b]{0.48\textwidth}
\includegraphics[width=\textwidth]{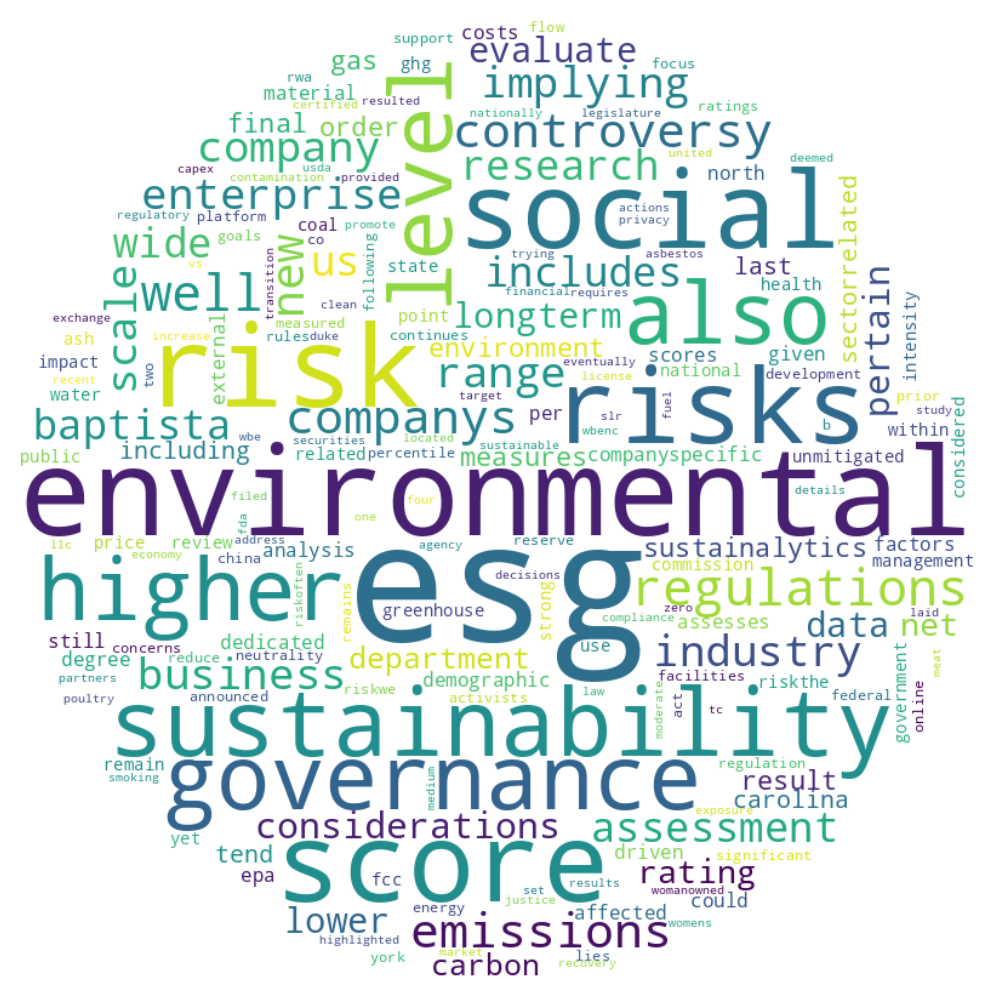}
\caption{(13) ESG Factors}
\end{subfigure}
\hfill
\begin{subfigure}[b]{0.48\textwidth}
\includegraphics[width=\textwidth]{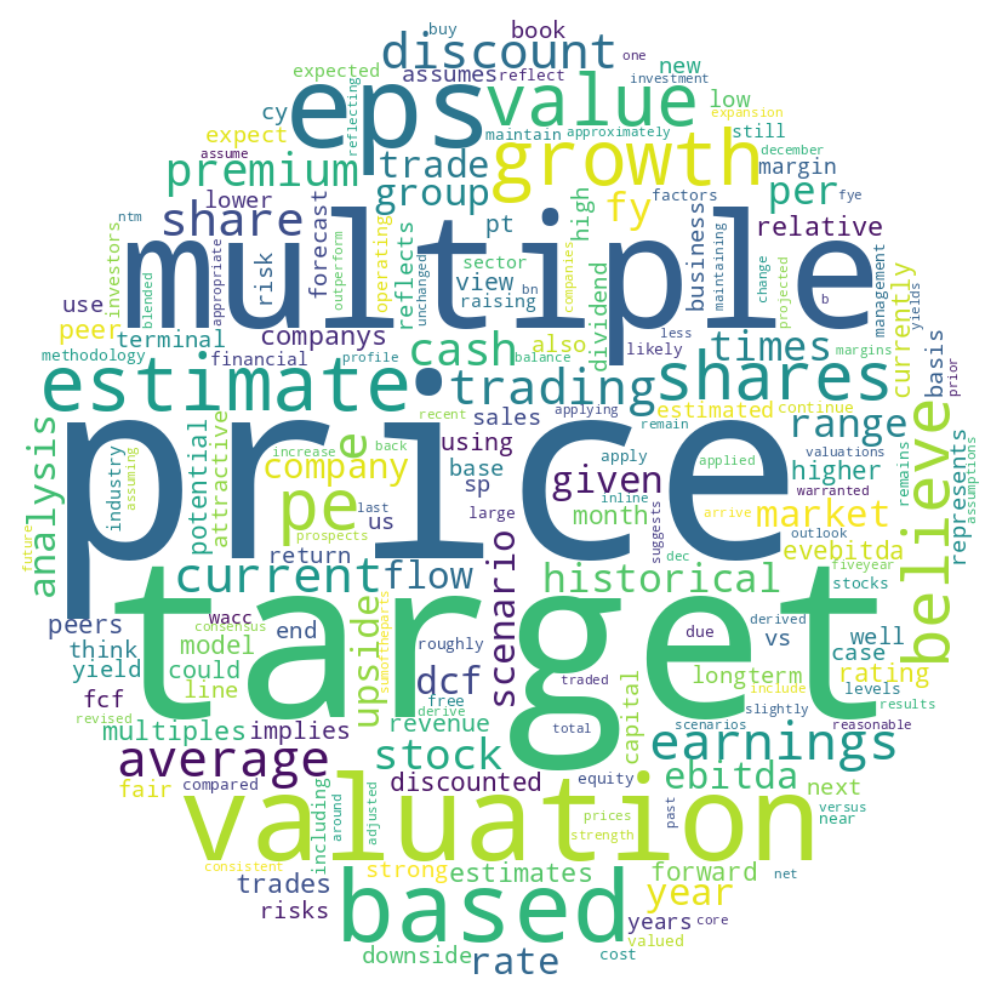}
\caption{(14) Valuation}
\end{subfigure}
\vfill
\begin{subfigure}[b]{0.48\textwidth}
\includegraphics[width=\textwidth]{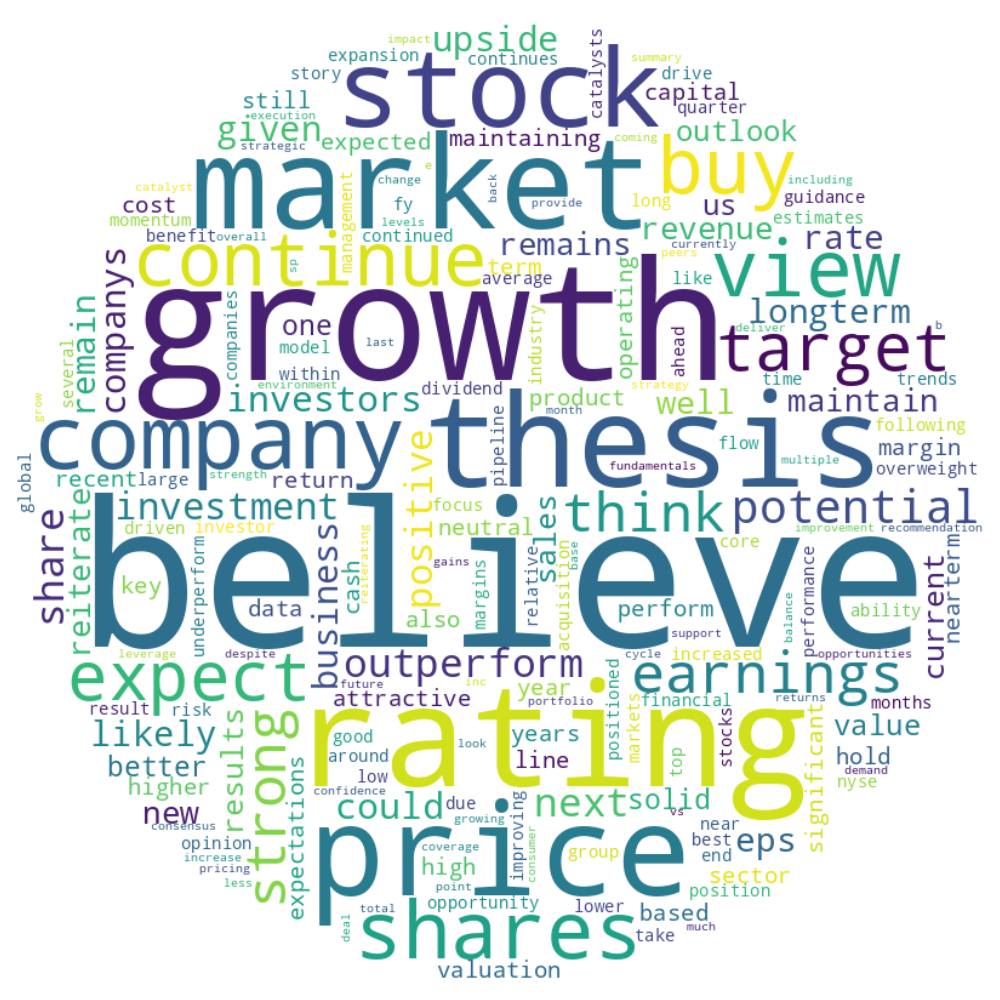}
\caption{(15) Investment Thesis}
\end{subfigure}
\hfill
\begin{subfigure}[b]{0.48\textwidth}
\includegraphics[width=\textwidth]{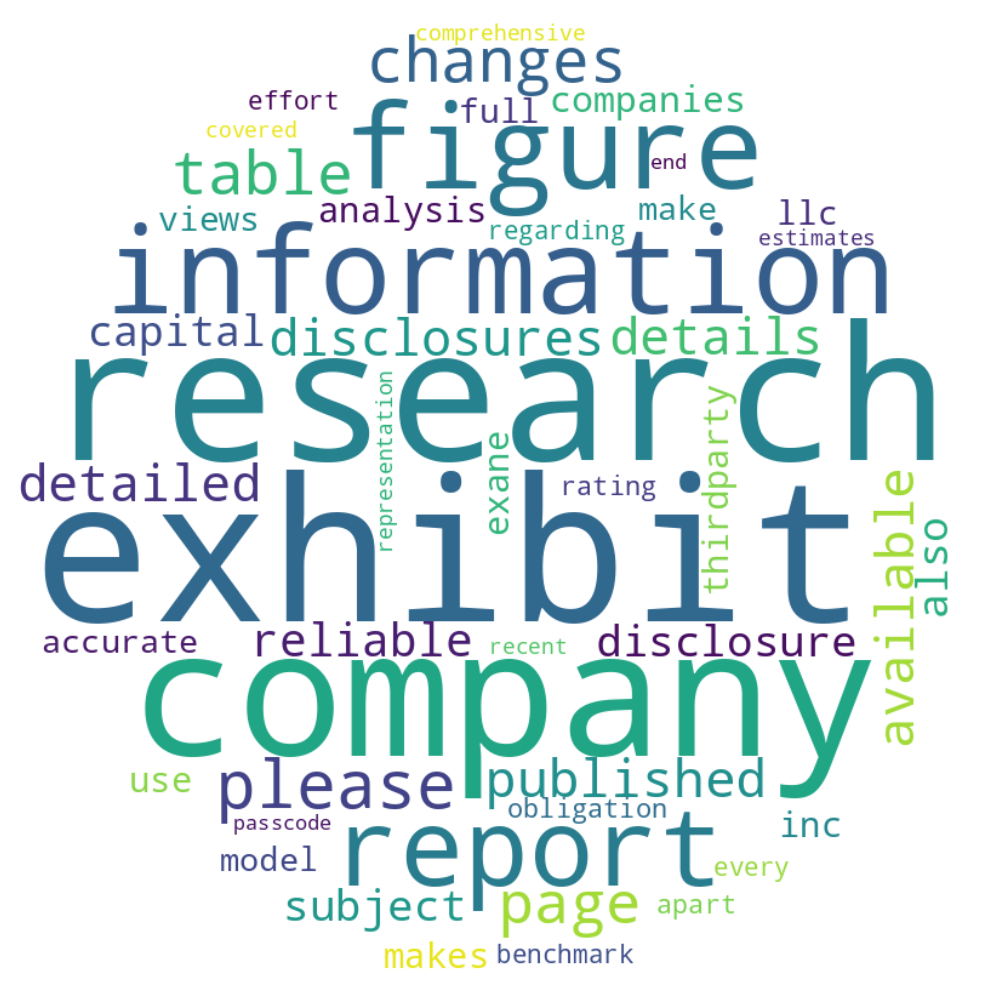}
\caption{(16) Appendices and Disclosures}
\end{subfigure}
\end{center}
\caption{Word Clouds of Analyst Report Topics}
    {\footnotesize This figure displays word clouds for 16 key topics commonly discussed in analyst reports. Each word cloud highlights the most frequently occurring terms within each topic, with word size proportional to term frequency. The topics include Executive Summary, Company Overview, Industry Analysis, Competitive Landscape, Income Statement Analysis, Balance Sheet Analysis, Cash Flow Analysis, Financial Ratios, Business Segments, Growth Strategies, Risk Factors, Management and Governance, ESG Factors, Valuation, Investment Thesis, and Appendices and Disclosures. The ``None of the Above" category is excluded from the visualization.}
\end{figure}

\newpage
\begin{figure}[H]
    \begin{center}
        \includegraphics[width=0.9\textwidth]{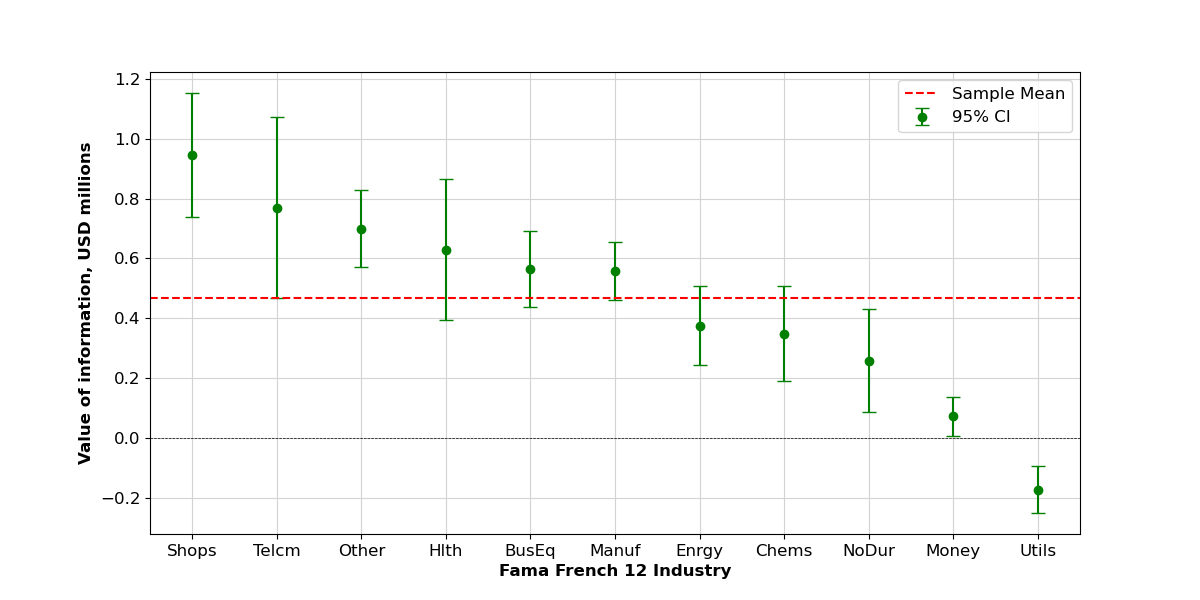}
    \end{center}
    \caption{Analyst Information Value by Industry}\label{fig:iv_ind}
    {\footnotesize 
    This figure shows the estimated information value of analyst reports across Fama-French 12 industries from 2015Q1 to 2023Q4, with values reported in millions of 2020 dollars and estimated via the delta method. The solid green line indicates the mean information value for each industry, while the green bars represent 95\% confidence intervals. The horizontal dashed red line denotes the cross-industry average.}
\end{figure}

\newpage
\begin{figure}[H]
    \begin{center}
    \includegraphics[width = \textwidth]{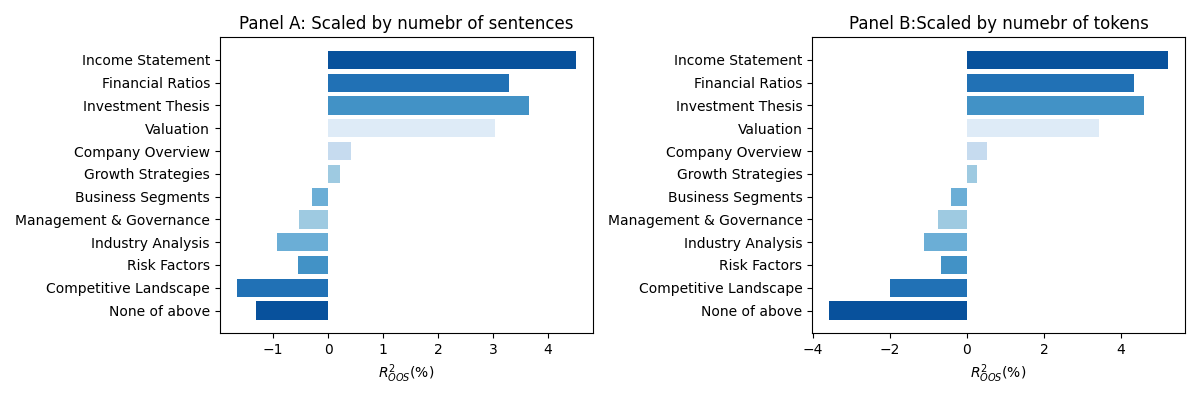}
    \end{center}
    \caption{Shapley Values Scaled by Topic Length}\label{fig:shap_scale}
    {\footnotesize 
    This figure adjusts the topic Shapley values to measure their per-sentence or per-token contribution rather than their aggregate importance, focusing on topics with over 100,000 sentences. Panel A presents the values scaled by the number of sentences in each topic, while Panel B presents them scaled by the number of tokens. This two-step scaling process involves first normalizing each topic's total Shapley value by its length (sentences or tokens) and then rescaling the results to maintain the original sum of all values, thereby preserving the model's total explanatory power.}
\end{figure}

\newpage
\begin{figure}[H]
    \begin{center}
    \includegraphics[width = \textwidth]{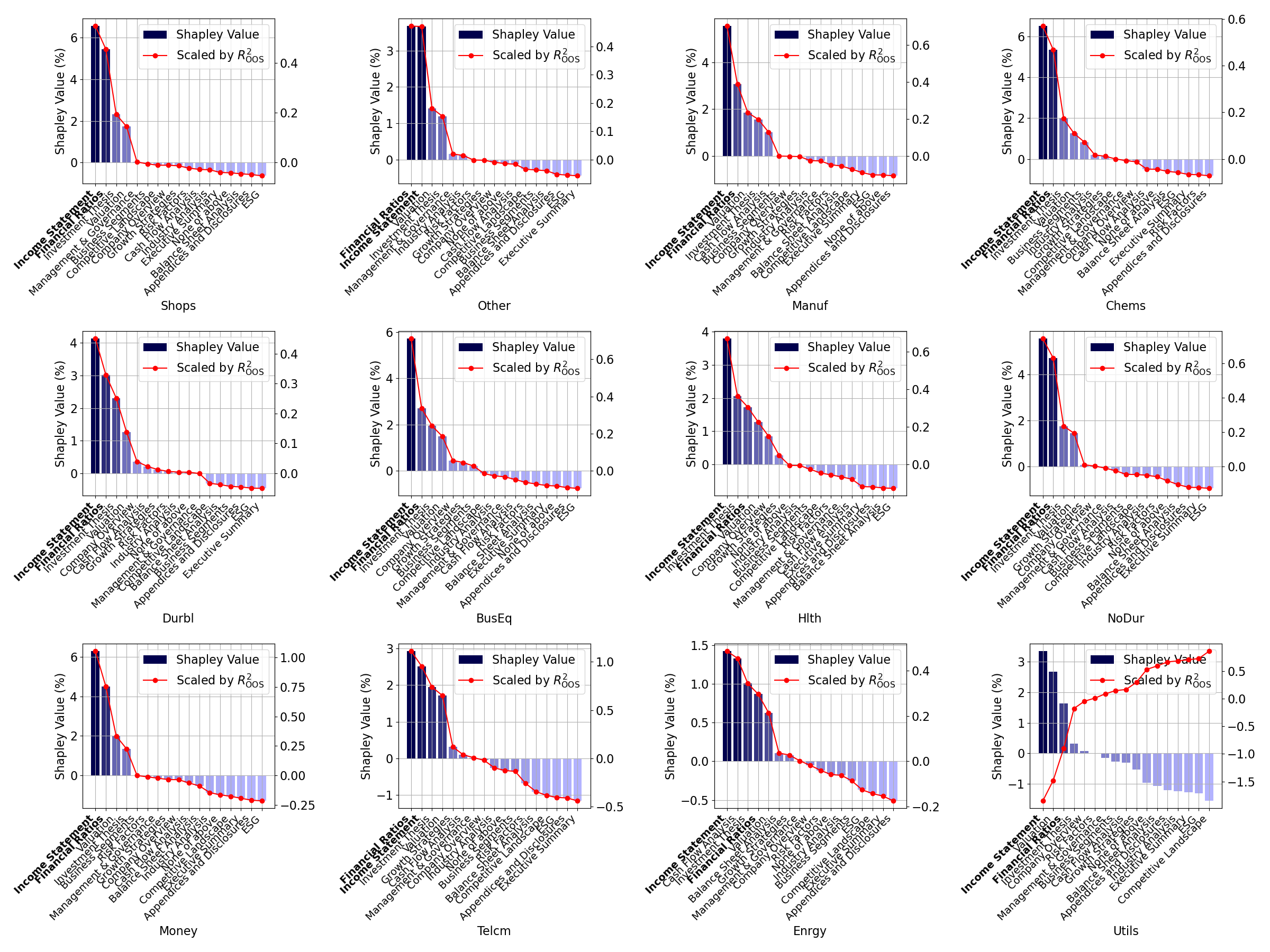}
    \end{center}
    \caption{Topic Importance via Shapley Value across Industries}\label{fig:shap_ind}
    {\footnotesize 
    This figure decomposes the information content ($R^2_{\mathrm{OOS}}$) of analyst reports into contributions from 17 topics for each of the Fama-French 12 industries. Within each industry's panel, the height of a bar represents the Shapley value for a specific topic, indicating its contribution to the total out-of-sample $R^2$ for reports in that sector. The red line shows each topic's SHAP value as a proportion of the aggregated $R^2_{\mathrm{OOS}}$.}
\end{figure}

\newpage
\begin{figure}[H]
    \begin{center}
    \includegraphics[width = 0.9\textwidth]{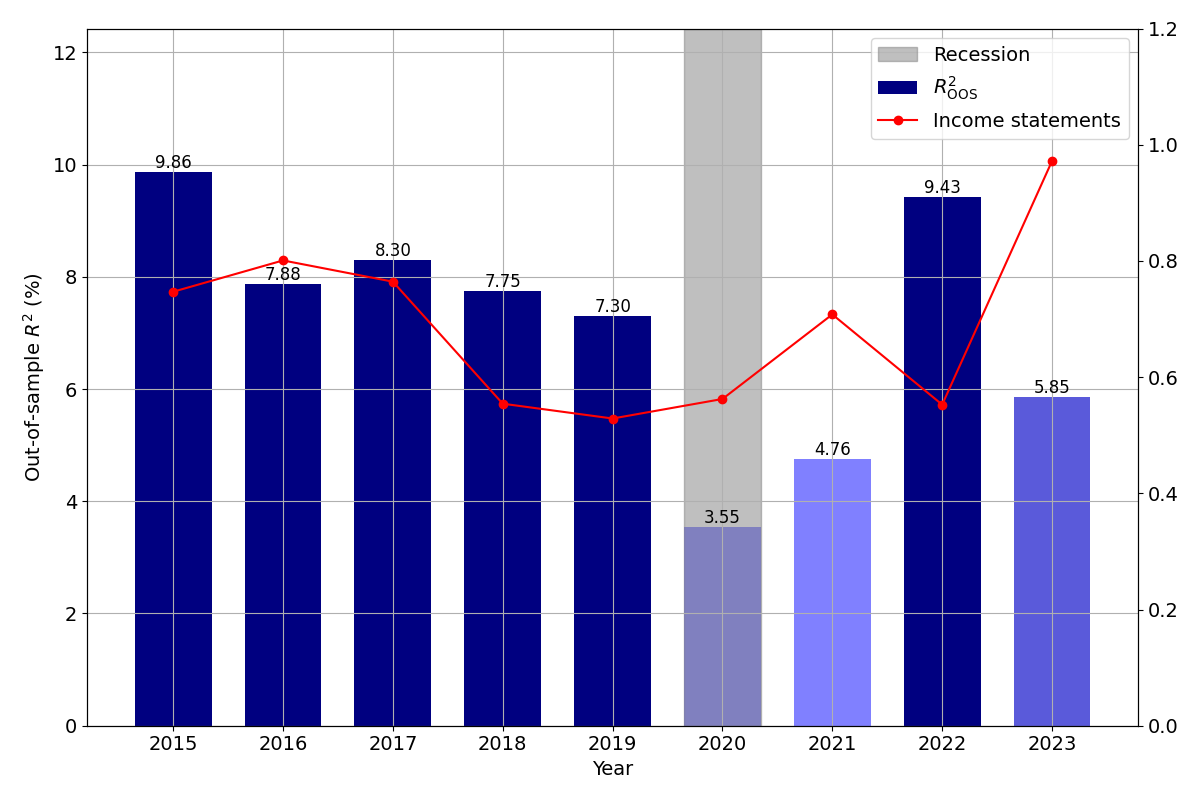}
    \end{center}
    \caption{Information Content of Sentence-Segmented Embeddings by Year}\label{fig:shap_yearly}
    {\footnotesize 
    This figure shows the evolution of the total out-of-sample explanatory power ($R^2_{\mathrm{OOS}}$) of sentence-segmented analyst reports from 2015 to 2023. Specifically, the red line isolates the contribution of the `Income Statement Analyses' topic, as measured by its annual Shapley value. The shaded area denotes the 2020 pandemic recession.}
\end{figure}

\newpage
\begin{figure}[H]
    \begin{center}
    \includegraphics[width = 0.9\textwidth]{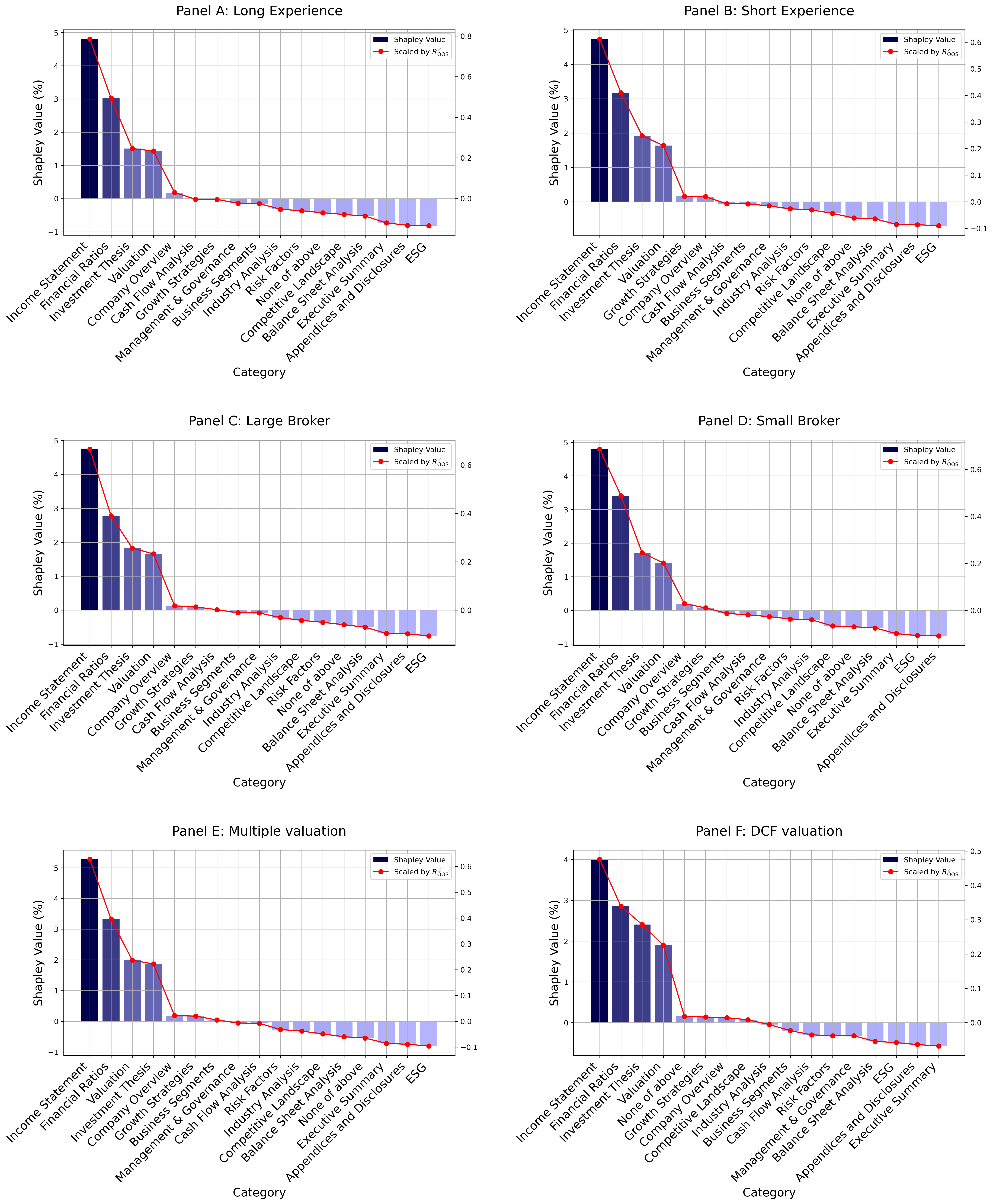}
    \end{center}
    \caption{Topic Importance via Shapley Value by Analyst Characteristics}\label{fig: shap_ana}
    {\footnotesize 
    This figure illustrates how 17 different topics contribute to the analyst report information content across six subsamples. The subsamples compare analysts with above-median (Panel A) versus below-median work experience (Panel B), large (Panel C) versus small brokerage firms (Panel D), and the primary valuation methodology used (Panel E: Multiple-Based vs. Panel F: DCF-Based). The splits for experience and brokerage size are based on the yearly median. For each group, bars represent the SHAP values of individual topics, with the red line indicating each topic's relative importance scaled by total $R^2_{\mathrm{OOS}}$.}
\end{figure}

\newpage
\begin{figure}[H]
    \begin{center}
        \includegraphics[width=0.9\textwidth]{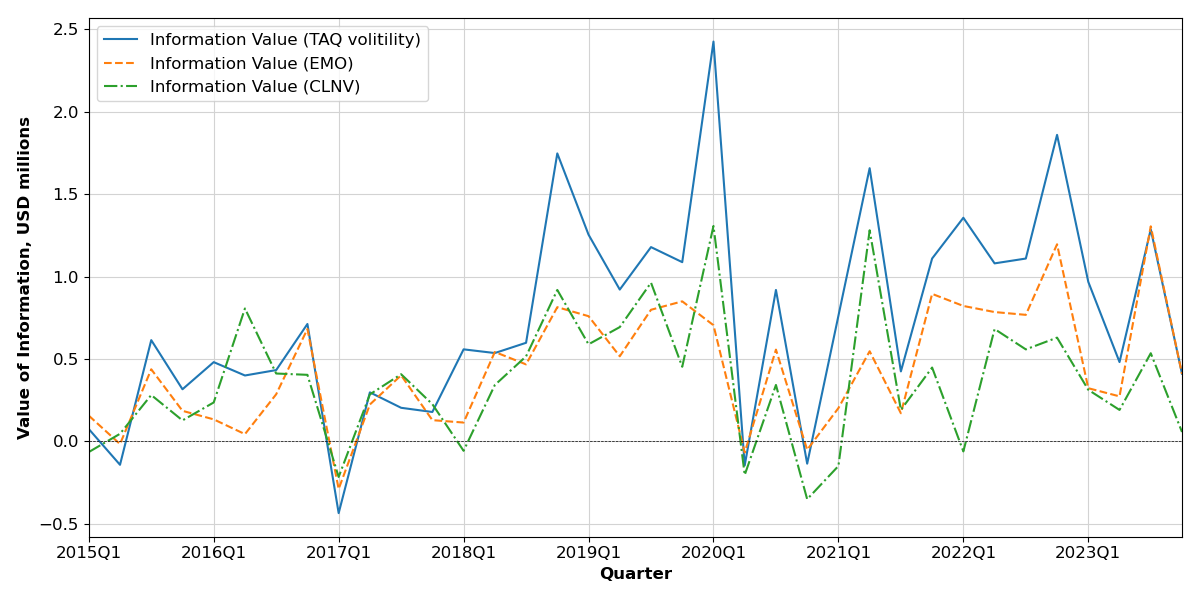}
    \end{center}
    \caption{Alternative Estimations of Analyst Information Value Over Time}\label{fig:iv_ts_robust}
    {\footnotesize
    This figure presents several alternative estimations of the quarterly analyst information value from 2015Q1 to 2024Q4, serving as a robustness check to the primary measure. These alternatives modify key components of the main calculation. The `TAQ volatility' series uses realized volatility from one-minute log returns ($\sigma_v^2$) to calculate explained return volatility as $\frac{r^2 - (r - \widehat{r})^2}{r^2} \cdot \sigma_v^2$. The `EMO' and `CLNV' series use alternative trade-signing algorithms for the price impact calculation developed by \citet{ellis2000underwriter} and \citet{lee1991inferring}, respectively. All estimates are reported in millions of 2020 dollars, with quarterly means approximated via the delta method.}
\end{figure}

\newpage
\begin{figure}[H]
    \begin{center}
        \includegraphics[width=0.9\textwidth]{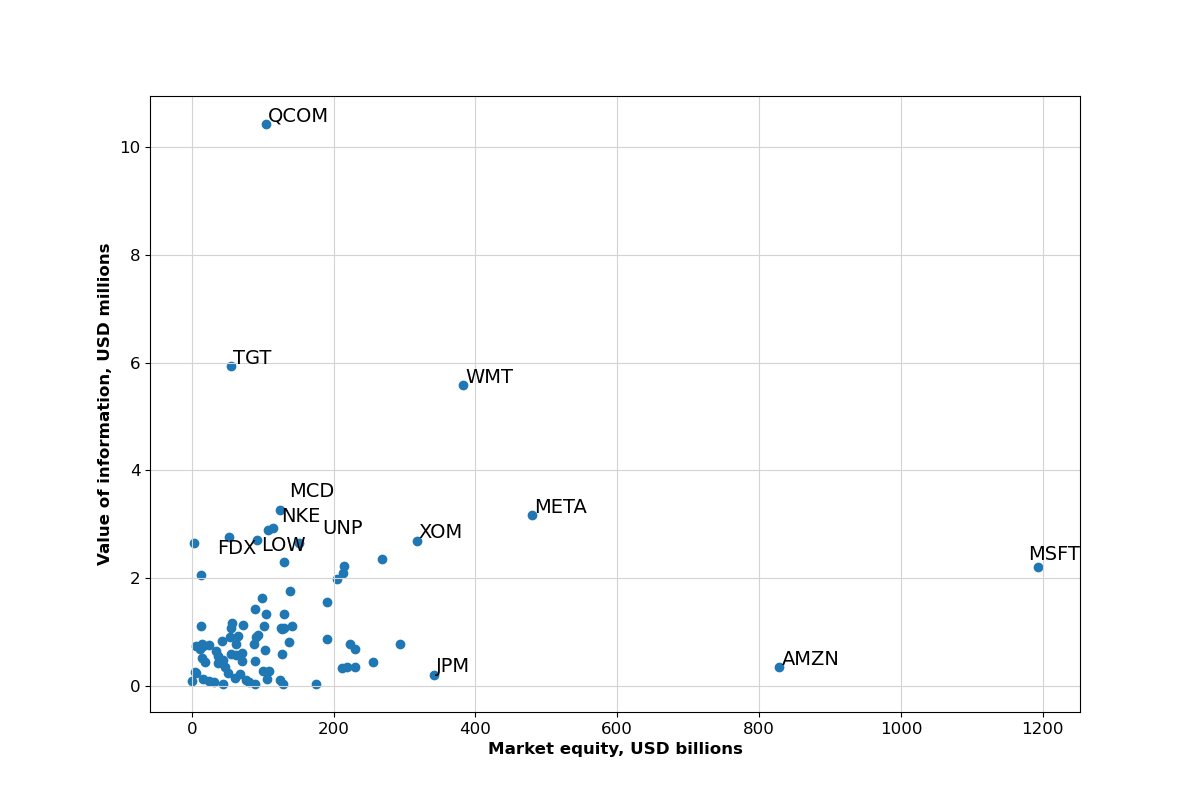}
    \end{center}
    \caption{Analyst Information Value and Stock Market Equity}\label{fig:iv_stock}
    {\footnotesize 
    This figure plots the estimated information value of analyst reports against the market equity for each individual stock, covering the period from 2015Q1 to 2023Q4. The information value (vertical axis) for each stock is estimated using the delta method and reported in millions of 2020 dollars, adjusted using the CPI. The market equity (horizontal axis) is the firm's average daily market capitalization at the time of each report release, reported in billions of dollars.}
\end{figure}

\newpage
\section*{B. Additional Tables}

\scriptsize
\begin{longtable}{>{\raggedright\arraybackslash}p{4cm}>{\raggedright\arraybackslash}p{11cm}}
\caption{Topic-Specific Sentence Examples from Analyst Reports}\label{tab: topic_example} \\
\toprule
\textbf{Topics} & \hspace{3em}\textbf{Examples} \\
\midrule
\endfirsthead
\multicolumn{2}{c}{{\tablename\ \thetable{} -- continued from previous page}} \\
\toprule
\textbf{Topics} & \hspace{3em}\textbf{Examples} \\
\midrule
\endhead
\endfoot
\bottomrule
\endlastfoot

\multirow{3}{4cm}{\textbf{Executive Summary}} & \\[-2ex]
& \begin{compactitem}
    \item Our key takeaways from CEO Jeff Immelt's presentation at EPG were: Outlook for substantial EPS growth over 2010/12 driven by abatement of credit losses (\$8-9bn in 2010E tapering to \$4bn run-rate) and CRE impair.
    \item Looking ahead, guidance was tightened, essentially framing the Street, and commentary suggests a strong outlook for the balance of 2007.
    \item Key topics: 1) Update on ABTÕs portfolio of COVID-19 tests; 2) Libre trends, Libre 2 launch and expectations for Libre 3; 3) Elective surgery trends exiting Q1, expectations for 2021 and an update on recent and upcoming new product approvals; and 4) Global trends and impact on EPD and Nutrition.
\end{compactitem} \\[0ex]
\midrule

\multirow{3}{4cm}{\textbf{Company Overview}} & \\[-2ex]
& \begin{compactitem}
    \item Oracle Corporation, founded in 1977 and headquartered in Redwood Shores, California, is one of the largest and most prominent companies in the software space -- and a technology bellwether.
    \item As it has grown, Microsoft has expanded into enterprise software with Windows Server, SQL Server, Dynamics CRM, SharePoint, Azure and Lync; hardware with the Xbox gaming/media platform and the Surface tablet; and online services through MSN and Bing.
    \item Altria Group, Inc., is the world's largest producer and marketer of consumer products, and had revenues of \$80 billion in 2002.
\end{compactitem} \\[0ex]
\midrule

\multirow{4}{4cm}{\textbf{Industry Analysis}} & \\[-2ex]
& \begin{compactitem}
    \item According to our global Immunology market model, US Psoriasis (PsO) represented a \$7.7B market in 2016 and is expected to grow at a low-teens CAGR to \$11.8B in 2019E and ~\$13B by 2021E driven by more highly effective therapies.
    \item Despite record prices, oil demand continues to grow, while supply growth lags and spare production and refining capacity is almost nonexistent.
    \item Add to that that some new advertising expectations from PricewaterhouseCoopers of a decline in ad spending of 12\% worldwide and 15\% in the U.S. for 2009 and continue to decline in 2010.
\end{compactitem} \\[0ex]
\midrule

\multirow{4}{4cm}{\textbf{Competitive Landscape}} & \\[-2ex]
& \begin{compactitem}
    \item According to Mercury Research, NVIDIA is now the 3rd largest chipset supplier (consisting of desktop and mobile chipsets, and integrated and non-integrated chipsets), shipping 5.4 million units in calendar Q3 for an 8.2\% market share, versus Intel's shipments of 51 million units (62\% share), VIA's shipments of 9.6 million units (14.4\% share), SiS's shipments of 5.3 million units (8\% share), and ATI's shipments of 4.4 million units (6.6\% share).
    \item Further, competition in the CDK-4/6 space is rising with Verzenio (abemaciclib) \& Kisqali launches placing downward pressure on Ibrance trajectory.
    \item We believe that Accenture has recognized that web services could compete directly with client/server as the new systems architecture.
\end{compactitem} \\[-3.2ex]
\midrule
\newpage
\multirow{4}{4cm}{\textbf{Financial Statement Analysis: Income Statement Analysis}} & \\[-2ex]
& \begin{compactitem}
    \item We also note that Alico cannot dividend any of its 2010 earnings to AIG, implying that the income recognized this quarter and throughout 2010 will benefit MetLife results.
    \item At the respective midpoints, sales of \$8.5 billion would be down 22\% annually and 8\% sequentially; and non-GAAP EPS would be down 39\%.
    \item The growth in operating income from Q1 2009 is largely due to recoveries in partnership income.
\end{compactitem} \\[0ex]
\midrule

\multirow{4}{4cm}{\textbf{Financial Statement Analysis: Balance Sheet Analysis}} & \\[-2ex]
& \begin{compactitem}
    \item Long-term borrowings of \$6.59 billion at September 30, 2022 were modest compared to shareholders' equity of \$37.3 billion.
    \item The company finished 2Q16 with \$21.4 billion in cash and short-term investments, up from \$15.6 Growth \& Valuation Analysis GROWTH ANALYSIS RISK ANALYSIS billion at the end of 4Q15.
    \item Inventory turns improved to 4.5x in 4Q from 4.4x in the same period last year.
\end{compactitem} \\[0ex]
\midrule

\multirow{4}{4cm}{\textbf{Financial Statement Analysis: Cash Flow Analysis}} & \\[-2ex]
& \begin{compactitem}
    \item UPS generated \$2.3 billion in operating cash for the quarter.
    \item Assuming that the company completes a large portion of its current \$1 billion stock buyback plan in 2008, we estimate that cash per share will be about \$6.20 by the end of the year.
    \item We view eBays FCF generation as relatively defensible even in the case of a revenue shortfall.
\end{compactitem} \\[0ex]
\midrule

\multirow{4}{4cm}{\textbf{Financial Statement Analysis: Financial Ratios}} & \\[-2ex]
& \begin{compactitem}
    \item The company achieves average scores on our three main measures of financial strength: leverage based on debt-to-cap, profitability and interest coverage.
    \item P/S is at 0.6x and EV/EBITDA is 7.7x.
    \item The index members currently trade at an average of 16.3-times trailing earnings, which is below the five-year average of 19-times.
\end{compactitem} \\[0ex]
\midrule

\multirow{4}{4cm}{\textbf{Business Segments}} & \\[-2ex]
& \begin{compactitem}
    \item The company is organized into three businesses, software, representing the majority of the company's total revenues, hardware systems and services.
    \item Results in 2008 also benefited from the absence of the significant level of mark-to-market losses in the company's Gas Marketing segment in 2007.
    \item The company operates five distinct segments: Americas (71\% of FY15 profit); Europe, Middle East, and Africa (4\% of FY15 profit); China and Asia Pacific (12\% of FY15 profit); Channel Development (13\% of FY15 profit).
\end{compactitem} \\[0ex]
\midrule

\multirow{3}{4cm}{\textbf{Growth Strategies}} & 
\begin{compactitem}
    \item Visa has been especially active on the acquisition front over the last several months.
    \item Combined with the Horizon assets and an emerging pipeline, there is enough in AMGN's portfolio to offset potential headwinds and allow the company to grow through its patent cycle.
    \item Specific areas where more investment may be needed: Over the past decade, ROK management often claim that Process automation represents the growth opportunity for the company, but its sales in this market have barely grown in recent years, despite its much smaller sales base compared with established incumbents.
\end{compactitem} \\[0ex] 
\midrule

\multirow{4}{4cm}{\textbf{Risk Factors}} & \\[-2ex]
& \begin{compactitem}
    \item Risks to our BUY thesis have to do with global competition, changing user behavior, global macro uncertainty, and anything else that can affect FB's relationship with members, its advertisers or its publishing partners.
    \item Risks to achieving our price target include: 1) Apple crushing PayPal; 2) increasing competition in the payments space; 3) heavy investment spending on marketing, point of sale, or technology; and 4) legislative action.
    \item In addition to the expenses incurred by patent challenges, product liability and other legal suits could occur and lead to additional liabilities and revenue loss, which could substantially change our financial assumptions.
\end{compactitem} \\[0ex]
\midrule

\multirow{4}{4cm}{\textbf{Management and Governance}} & \\[-2ex]
& \begin{compactitem}
    \item Top management changes can be unsettling, and the resulting uncertainty has caused 3M shares to decline.
    \item Chairman, President, and CEO Charles Ergen beneficially owns about 53.6\% of DISH's equity securities and has 90.5\% voting power.
    \item Bill Johnson, the present CEO of Progress Energy, will become president and chief executive officer of the new company.
\end{compactitem} \\[0ex]
\midrule

\multirow{4}{4cm}{\textbf{Environmental, Social, and Governance (ESG) Factors}} & \\[-2ex]
& \begin{compactitem}
    \item The assessment of ESG (Environmental, Social \& Governance) risks by Baptista Research includes a wide range of considerations that pertain to the long-term sustainability of a company.
    \item Sustainalytics assesses the degree to which a company's enterprise business is affected by ESG issues.
    \item Failure to adequately address social risks like labor disputes and community relations could jeopardize the company's social license to operate in certain regions.
\end{compactitem} \\[0ex]
\midrule

\multirow{4}{4cm}{\textbf{Valuation}} & \\[-2ex]
& \begin{compactitem}
    \item Our DCF derives an intrinsic value of \$100 for ABBV by discounting cash flows through 2024E, and assuming a -5\% terminal growth \& 7.6\% WACC.
    \item We value MET based on a Sum-Of-The-Parts (SOTP) analysis based on our 2021 EPS estimate and using peer comps across each business segment.
    \item Our target price is based on a five-year discounted cash flow (DCF) valuation that employs a 5\% discount rate and 20x terminal-year FCF multiple.
\end{compactitem} \\[-0.5ex]
\midrule

\multirow{3}{4cm}{\textbf{Investment Thesis}}
& \begin{compactitem}
    \item The results for the back half of the year will still be complex and confusing given the purchase accounting impacts and the full quarter of HSBC, but we do believe that there is a pay-off at the end of the road.
    \item As good as it gets: With its record multi-year backlog, Boeing's revenue profile over the rest of decade is generally considered secure, and expectations for execution and cash already appear high.
    \item Clearly, our Ford Investment Thesis, which was based in large part on our belief that Ford would be able to offset headwinds (slowing cyclical tailwinds in North America, weakness in South America, weak growth in Europe, slowing growth in China, and regulatory cost headwinds), has been thrown into question.
\end{compactitem} \\[0ex]
\midrule

\multirow{4}{4cm}{\textbf{Appendices and Disclosures}} & \\[-2ex]
& \begin{compactitem}
    \item Although the information contained in the subject report has been obtained from sources, we believe to be reliable, its accuracy and completeness cannot be guaranteed.
    \item For a complete discussion of the risk factors that could affect the market price of a company's shares, refer to the most recent Form 10-Q or 10-K that a company has filed with the Securities and Exchange Commission.
    \item The Benchmark Company, LLC makes every effort to use reliable, comprehensive information, but we make no representation that it is accurate or complete.
\end{compactitem} \\[0ex]
\end{longtable}

\newpage
\begin{table}[!htb]
\caption{Numerical Variables Description}\label{tab:numdef}
{\footnotesize This table shows the definitions of numerical measures.}
\begin{center}
\scriptsize
\tabcolsep = 0.45cm
\renewcommand{\arraystretch}{1.2}
\begin{tabularx}{\textwidth}{@{}p{4.5cm}p{11cm}@{}}
\toprule
Numerical Measures & Definition and/or sources \\
\midrule
\textbf{Panel A: Firm-Level Measures} & \\
Size &  The market value equity of the firm ($CSHOQ * PRCCQ$) at the end of the quarter prior to report release. \\
BtoM & The book value of equity ($SEQ + TXDB + ITCB -PREF$) scaled by the market value of equity ($CSHOQ * PRCCQ$) t the end of the quarter prior to the report release. \\
PriorCAR &   The cumulated 10-day abnormal return ending 2 days before release. The abnormal return is calculated as the raw return minus the buy-and-hold market value-weighted return. \\
SUE & Earnings surprise, calculated as the actual EPS minus the last consensus EPS forecast before the earnings announcement. Consensus EPS is the median value of 1-year EPS forecast within a 90-day window of all analysts following the firm. The unexpected earnings is scaled by price per share at the fiscal quarter end.\\
AbsSUE &  Absolue value of SUE, representing the distance between realized EPS and EPS expectation. \\
Miss &   Dummy variable that equals one if the actual EPS is less than the last consensus forecast, and 0 otherwise. \\
TradingVolume &  Trading volume at earnings announcement day (or the first trading day post earnings announcement), calculated as $VOL/SHROUT$. \\
DD &  The distance to default calculated following \citet{merton1974pricing}. The proxy is compiled from the National University of Singapore's Credit Research Initiative (NUS CRI). \\
Fluidity &   A measure of firms' product market competition introduced by \citet{hoberg2014product}. The data is compiled from the Hoberg and Phillips database. \\
\midrule
\textbf{Panel B: Industry-Level Measures} & \\
IndustryRecession & An indicator variable that equals one if the FF-48 industry return is negative and in the bottom quintile of FF-48 industry returns and zero otherwise. \\
\midrule
\textbf{Panel C: Macroeconomic Measures} & \\
TimeTrend  &  The number of years elapsed from the beginning of the sample. \\
\midrule
\textbf{Panel D: Report-Level Measures} & \\
$REC_{REV}$ & Recommendation revision, calculated as the current report's recommendation minus the last recommendation in I/B/E/S issued by the same analyst for the same firm. \\
$EF_{REV}$ & Earnings forecast revision, calculated as the current report's EPS forecast minus the last EPS forecast in I/B/E/S issued by the same analyst for the same firm, scaled by the stock price 50 days before the report release. \\
$TP_{REV}$ & Target price revision, calculated as the current report's target price minus the last target price in I/B/E/S issued by the same analyst for the same firm, scaled by the stock price 50 days before the report release. \\
Boldness &  An indicator variable that takes the value of 1 if the EPS forecast revision is above both the analyst's own prior forecast and the consensus forecast, or else below both, and zero otherwise. \\
SR &  Stock recommendation from I/B/E/S rating, with 1 being the most bullish (Strong Buy) and 5 being the most bearish (Sell), based on the ratings provided by the Institutional Brokers' Estimate System (I/B/E/S).\\
ERet &   12-month return forecast by scaling the 12-month price target by the stock price 1-day before release. \\
\bottomrule
\end{tabularx}
\end{center}
\end{table}

\newpage
\begin{table}[!htb]
\caption{Sumamry Statistics for Numerical Measures}\label{tab:numss}
    {\footnotesize This table reports summary statistics of numerical measures. See Table \ref{tab:numdef} for detailed variable definitions.}
    \begin{center}
    \scriptsize
    \tabcolsep = 0.23cm
    \renewcommand{\arraystretch}{1.2}
    \begin{tabularx}{\textwidth}{lYYYYYY}
\toprule
{} &   Mean &    Std. Dev. &    p25 &    P50 &     P75 &       N \\
\midrule
Miss                &   0.35 &   0.48 &   0.00 &   0.00 &    1.00 &  102776 \\
TradingVolume      &  18.19 &  21.98 &   7.24 &  11.46 &   20.29 &   99908 \\
PriorCAR          &   0.00 &   0.05 &  -0.02 &   0.00 &    0.02 &  122251 \\
Size                &   2.39 &   0.11 &   2.33 &   2.39 &    2.46 &  117534 \\
BtoM                &   0.46 &   0.43 &   0.18 &   0.32 &    0.61 &  117534 \\
DD                  &   0.00 &   0.01 &   0.00 &   0.00 &    0.00 &   86693 \\
Fluidity            &   7.17 &   3.66 &   4.40 &   6.57 &    9.44 &  114942 \\
TimeTrend          &  13.02 &   5.71 &   9.00 &  13.00 &   17.00 &  122252 \\
IndustryRecession  &   0.14 &   0.35 &   0.00 &   0.00 &    0.00 &  122252 \\
BrokerSize          &  71.92 &  56.71 &  26.00 &  53.00 &  113.00 &  119233 \\
Firm Experience     &   7.68 &   6.72 &   3.00 &   6.00 &   11.00 &  111192 \\
Number of firms     &  20.08 &  17.08 &  14.00 &  18.00 &   23.00 &  119233 \\
$REC_{REV}$            &   0.00 &   0.16 &   0.00 &   0.00 &    0.00 &  120673 \\
$EF_{REV}$             &   0.00 &   0.00 &   0.00 &   0.00 &    0.00 &   90625 \\
$TP_{REV}$             &   0.01 &   0.07 &   0.00 &   0.00 &    0.00 &   84108 \\
ERet                &   0.19 &   0.23 &   0.07 &   0.18 &    0.31 &   84108 \\
Boldness            &   0.75 &   0.43 &   1.00 &   1.00 &    1.00 &   81182 \\
SR                  &   2.77 &   0.84 &   2.00 &   3.00 &    3.00 &  122252 \\
\bottomrule
    \end{tabularx}
    \end{center}
\end{table}

\newpage
\begin{table}[!htb]
\caption{Information Content of Analyst Reports with Numbers Removed}\label{tab:rmnum}
{\footnotesize 
This table presents an analysis of analyst reports where all numerical values have been removed from the text, designed to isolate the contribution of pure qualitative information. I report the $R^2_{\mathrm{OOS}}$ from Ridge regressions predicting three-day cumulative abnormal returns ($CAR[-1,+1]$) using several input configurations: numerical forecast revisions only, embeddings from the number-free text, and a combination of both. The table also includes pairwise tests to assess whether the number-free text contains more information than the numerical revisions alone (Column (3)-(1)), and whether those revisions provide incremental value when added to the number-free text (Column (5)-(3)). T-statistics for the $R^2_{\mathrm{OOS}}$ and their differences are calculated following the procedure outlined by \citet{gu2020empirical}.}
    \begin{center}
    \scriptsize
    \tabcolsep = 0.23cm
    \renewcommand{\arraystretch}{1.2}
    \begin{tabularx}{\textwidth}{lYYYYYYYY}
\toprule
   Year & Rev only &  t-stat & Text only &  t-stat & Rev + text &  t-stat &  t-stat &  t-stat \\
\midrule
    {} &  (1) &  (2) &  (3) &  (4) &  (5) &  (6) &  (3)- (1) &  (5)- (3)  \\
\midrule
   2015 &       10.30\% &    3.59 &   13.19\% &    5.87 &           11.10\% &    3.24 &    3.97 &   -1.45 \\
   2016 &       15.47\% &   11.80 &   13.73\% &    5.00 &           17.91\% &    9.02 &   -0.81 &    3.13 \\
   2017 &        9.16\% &    5.28 &   11.66\% &    5.98 &           12.33\% &    6.12 &   10.58 &    1.56 \\
   2018 &        9.79\% &    2.98 &   11.46\% &    9.12 &           13.62\% &    7.97 &    1.22 &    4.09 \\
   2019 &        9.76\% &   15.69 &   12.61\% &   14.73 &           13.90\% &   15.17 &    4.26 &    2.79 \\
   2020 &        5.20\% &    3.43 &    4.42\% &    4.38 &            6.98\% &    5.81 &   -0.70 &    8.37 \\
   2021 &        5.31\% &    4.80 &    8.77\% &    5.06 &           10.66\% &   29.67 &    1.54 &    1.56 \\
   2022 &       10.24\% &    6.16 &   16.71\% &   10.47 &           17.90\% &    8.94 &    9.93 &    2.38 \\
   2023 &        5.09\% &    3.86 &    9.27\% &    3.60 &           10.39\% &    5.09 &    2.60 &    1.81 \\
Overall &        8.93\% &    8.51 &   10.95\% &    7.87 &           12.51\% &    8.83 &    2.40 &    3.00 \\
\bottomrule
    \end{tabularx}
    \end{center}
\end{table}

\newpage
\begin{landscape}
\begin{table}[!htb]
\caption{Information Content of Analyst Reports using Alternative Machine Learning Models}\label{tab:ml}
    {\footnotesize 
    This table compares the performance ($R^2_{\mathrm{OOS}}$) of several alternative machine learning models in estimating the information content of analyst report text. The models tested include Partial Least Squares (PLS), Extreme Gradient Boosting (XGBoost), and a series of Neural Networks (NN1-NN5) with an increasing number of hidden layers. The out-of-sample $R^2$ for each model is calculated annually using an expanding training window that begins in 2000. The `Overall' row summarizes the performance across the full 2015–2023 sample period. T-statistics for the $R^2_{\mathrm{OOS}}$ are calculated against a zero benchmark following the procedure in \citet{gu2020empirical}.}
    \begin{center}
    \scriptsize
    \tabcolsep = 0.23cm
    \renewcommand{\arraystretch}{1.2}
    \begin{tabularx}{1.4\textwidth}{p{1.5cm}YYYYYYYYYYYYYY}
\toprule
    & \multicolumn{2}{c}{PLS} & \multicolumn{2}{c}{XGBoost} & \multicolumn{2}{c}{NN1} & \multicolumn{2}{c}{NN2} & \multicolumn{2}{c}{NN3} & \multicolumn{2}{c}{NN4} & \multicolumn{2}{c}{NN5} \\
\hline
      year  & $R^2_{\mathrm{OOS}}$ & t-stat & $R^2_{\mathrm{OOS}}$ & t-stat & $R^2_{\mathrm{OOS}}$ & t-stat & $R^2_{\mathrm{OOS}}$ & t-stat & $R^2_{\mathrm{OOS}}$ & t-stat & $R^2_{\mathrm{OOS}}$ & t-stat & $R^2_{\mathrm{OOS}}$ & t-stat \\
\midrule
   2015 &              12.23\% &  10.95 &               8.87\% &  11.25 &              15.81\% &   9.72 &              13.84\% &  14.53 &              15.30\% &  14.10 &              11.58\% &  12.32 &              16.61\% &  16.95 \\
   2016 &              10.27\% &  15.59 &               6.66\% &   5.63 &              12.59\% &  24.64 &              11.78\% &  13.02 &              10.99\% &   9.10 &              12.03\% &  14.75 &              12.46\% &  17.41 \\
   2017 &               6.60\% &   6.94 &               4.92\% &   5.26 &               9.08\% &   4.64 &              10.01\% &   4.64 &               9.58\% &   4.76 &               8.57\% &   4.51 &              10.03\% &   5.49 \\
   2018 &               8.08\% &  13.37 &               5.15\% &   8.07 &              10.73\% &  26.65 &              10.54\% &  16.42 &              10.41\% &  11.43 &               9.86\% &  17.44 &              11.98\% &  13.97 \\
   2019 &              11.54\% &  38.70 &               6.32\% &  16.40 &              13.43\% &  17.37 &              15.18\% &  14.73 &              15.59\% &  18.06 &              15.89\% &  10.48 &              15.37\% &  16.94 \\
   2020 &               3.21\% &   2.16 &               3.02\% &   3.64 &               4.45\% &   2.92 &               4.73\% &   2.78 &               3.85\% &   2.27 &               4.25\% &   3.08 &               5.51\% &   4.06 \\
   2021 &               2.14\% &   1.76 &               4.30\% &   3.34 &               4.07\% &   6.44 &               5.88\% &   7.13 &               9.28\% &  13.59 &               8.47\% &   9.44 &               8.83\% &   7.67 \\
   2022 &              11.42\% &   7.40 &              10.04\% &   7.80 &              13.78\% &   8.31 &              15.67\% &   6.38 &              16.44\% &   6.64 &              17.63\% &   8.55 &              17.56\% &   7.12 \\
   2023 &               8.14\% &   9.26 &               4.23\% &   3.82 &               9.54\% &   8.50 &              12.16\% &  10.76 &              11.06\% &   8.59 &              12.18\% &  11.59 &              11.70\% &   6.78 \\
Overall &               8.26\% &   6.66 &               5.97\% &   6.76 &              10.42\% &   8.11 &              11.04\% &   7.54 &              11.18\% &   7.29 &              11.05\% &   6.62 &              12.12\% &   7.93 \\
\bottomrule
    \end{tabularx}
    \end{center}
\end{table}
\end{landscape}

\newpage
\begin{table}[!htb]
\caption{Cumulative Abnormal Returns with Alternative Windows}\label{tab:car_alt}
    {\footnotesize This table reports the $R^2_{\mathrm{OOS}}$ from Ridge regression models that use analyst report text embeddings to estimate CAR across alternative event windows. The target variables are abnormal returns centered on the report release date ($T_0$): same-day abnormal return (AR[0]), two-day cumulative return (CAR[0,+1]), and five-day cumulative return (CAR[–2,+2]). All $R^2_{\mathrm{OOS}}$ values are computed annually using an expanding training sample starting in 2000. Panel A shows results for the full sample, while Panels B and C restrict the sample to reports released within one day (excluding same day) of an earnings announcement and those outside this window, respectively. The Overall row reports the $R^2_{\mathrm{OOS}}$ and $t$-statistics for 2015-2023. The t-statistics for $R^2_{\mathrm{OOS}}$ are calculated using zero benchmarks estimation following \citet{gu2020empirical}.}
    \begin{center}
    \scriptsize
    \tabcolsep = 0.23cm
    \renewcommand{\arraystretch}{1.2}
    \begin{tabularx}{\textwidth}{p{1.5cm}YYYYYY}
\toprule
    & \multicolumn{2}{c}{AR[0]} & \multicolumn{2}{c}{CAR[0,+1]} & \multicolumn{2}{c}{CAR[-2,+2]} \\
\hline
      year  & $R^2_{\mathrm{OOS}}$ & t-stat & $R^2_{\mathrm{OOS}}$ & t-stat & $R^2_{\mathrm{OOS}}$ & t-stat \\
\midrule
\multicolumn{7}{l}{Panel A: Full sample} \\
\midrule
   2015 &               8.03\% &   8.53 &               8.24\% &   8.36 &              10.87\% &   9.40 \\
   2016 &               7.88\% &   8.27 &               6.09\% &   4.68 &               6.60\% &   3.11 \\
   2017 &               8.75\% &  14.35 &               8.56\% &   9.29 &               9.00\% &   8.78 \\
   2018 &               9.09\% &  14.95 &               9.39\% &  11.26 &               9.75\% &  10.54 \\
   2019 &               9.89\% &  99.70 &               7.68\% &   9.45 &               8.66\% &  18.93 \\
   2020 &               2.33\% &   2.33 &               1.76\% &   3.15 &               4.13\% &   2.14 \\
   2021 &               5.75\% &   7.32 &               5.80\% &   6.62 &               4.05\% &   3.57 \\
   2022 &              10.54\% &  12.32 &              11.23\% &  18.20 &              10.91\% &  13.65 \\
   2023 &               5.38\% &  15.44 &               3.39\% &   7.00 &               0.53\% &   1.48 \\
Overall &               7.55\% &   8.07 &               6.92\% &   6.77 &               7.57\% &   6.79 \\
\midrule
\multicolumn{7}{l}{Panel B: Reports issued within earnings announcement window} \\
\midrule
Overall &               9.78\% &   5.58 &               7.87\% &   5.10 &              13.14\% &    8.02 \\
\midrule
\multicolumn{7}{l}{Panel C: Reports issued beyond earnings announcement window} \\
\midrule
Overall &               5.77\% &   5.78 &               5.62\% &   5.64 &               4.96\% &   4.91 \\
\bottomrule
    \end{tabularx}
    \end{center}
\end{table}

\newpage
\begin{landscape}

\begin{table}[!htb]
\caption{Information Content of Analyst Reports across Industries}\label{tab: ind}
    {\footnotesize 
    This table reports information content of analyst report text for each of the Fama-French 12 (FF12) industries for the period of 2015–2023. Industry definitions are based on Kenneth French’s data library. T-statistics for all $R^2_{\mathrm{OOS}}$ values are calculated against a zero benchmark following the procedure in \citet{gu2020empirical}.}\vspace{-0.1in}
    \begin{center}
    \scriptsize
    \tabcolsep = 0.23cm
    \renewcommand{\arraystretch}{1.2}
    \begin{tabularx}{1.4\textwidth}{lYYYYYYYYYYYY}
\toprule
    & \multicolumn{3}{c}{Shops} & \multicolumn{3}{c}{Other} & \multicolumn{3}{c}{Manuf} & \multicolumn{3}{c}{Chems} \\
\midrule
      year  & $R^2_{\mathrm{OOS}}$ & t-stat &    N & $R^2_{\mathrm{OOS}}$ & t-stat &    N & $R^2_{\mathrm{OOS}}$ & t-stat &    N & $R^2_{\mathrm{OOS}}$ & t-stat &   N \\
\midrule
   2015 &              13.27\% &   6.03 &  657 &              17.69\% &   5.03 &  585 &              10.84\% &   3.19 &  808 &              15.05\% &   1.74 & 147 \\
   2016 &              10.43\% &   2.46 &  556 &               3.34\% &   0.48 &  593 &              10.37\% &   5.38 &  811 &              23.91\% &   3.25 & 111 \\
   2017 &              16.65\% &   3.49 &  506 &              20.45\% &   2.67 &  544 &              25.03\% &  13.64 &  722 &              11.04\% &   0.59 &  82 \\
   2018 &              10.86\% &   4.32 &  406 &              21.65\% &   3.40 &  405 &               3.32\% &   1.55 &  553 &              25.80\% &   1.87 &  77 \\
   2019 &              21.04\% &   9.24 &  394 &               7.49\% &   2.37 &  610 &              16.25\% &  12.07 &  568 &               5.00\% &   3.22 & 106 \\
   2020 &               4.39\% &   2.55 &  499 &              -0.30\% &  -0.53 &  499 &               4.06\% &   1.41 &  542 &               2.65\% &   1.60 & 115 \\
   2021 &              10.27\% &   2.13 &  312 &               9.70\% &   3.66 &  362 &               8.66\% &   4.32 &  367 &               1.46\% &   0.30 & 102 \\
   2022 &              22.75\% &   3.83 &  320 &              15.98\% &   2.66 &  437 &              14.26\% &  11.46 &  420 &               7.00\% &   2.10 & 110 \\
   2023 &              17.78\% &   1.76 &  353 &               3.49\% &   2.00 &  339 &               7.15\% &   2.71 &  360 &              19.31\% &   6.54 &  76 \\
Overall &              13.88\% &   6.03 & 4003 &              11.27\% &   3.69 & 4374 &              11.39\% &   5.57 & 5151 &              11.35\% &   3.55 & 926 \\
\midrule
    & \multicolumn{3}{c}{Durbl} & \multicolumn{3}{c}{BusEq} & \multicolumn{3}{c}{Hlth} & \multicolumn{3}{c}{NoDur} \\
\midrule
     year   & $R^2_{\mathrm{OOS}}$ & t-stat &    N & $R^2_{\mathrm{OOS}}$ & t-stat &    N & $R^2_{\mathrm{OOS}}$ & t-stat &    N & $R^2_{\mathrm{OOS}}$ & t-stat &   N \\
\midrule
   2015 &              16.83\% &   5.77 &  216 &              15.66\% &   6.41 & 1378 &               5.75\% &   2.72 & 1242 &               0.68\% &   0.31 &  241 \\
   2016 &               4.95\% &   0.79 &  209 &              10.24\% &   6.09 & 1303 &              11.07\% &   2.90 & 1185 &              17.14\% &   5.18 &  242 \\
   2017 &              -2.17\% &  -0.30 &  282 &              10.36\% &   3.12 & 1154 &               3.37\% &   2.39 & 1060 &              -2.14\% &  -0.68 &  278 \\
   2018 &              18.29\% &   5.57 &  208 &               0.32\% &   0.31 & 1002 &              12.47\% &   6.58 &  918 &               1.29\% &   0.13 &  324 \\
   2019 &              15.31\% &   2.26 &  222 &              10.17\% &   6.13 & 1036 &              10.02\% &   7.68 &  884 &              18.51\% &   2.30 &  277 \\
   2020 &               5.73\% &   0.79 &  168 &               8.11\% &   4.14 & 1100 &               2.86\% &   1.84 &  950 &              -0.37\% &  -0.28 &  265 \\
   2021 &              10.01\% &   5.31 &  150 &              11.43\% &   1.49 &  731 &               7.66\% &   4.58 &  592 &              -6.38\% &  -0.60 &  195 \\
   2022 &               7.99\% &   2.58 &  162 &              12.08\% &   4.77 &  822 &              12.05\% &   2.71 &  850 &               5.24\% &   1.11 &  187 \\
   2023 &              12.05\% &   2.37 &  140 &              10.45\% &  11.80 &  962 &               3.52\% &   1.32 &  844 &              -5.69\% &  -0.71 &  158 \\
Overall &              11.02\% &   3.24 & 1757 &               9.68\% &   6.77 & 9488 &               7.53\% &   5.71 & 8525 &               7.50\% &   1.38 & 2167 \\
\midrule
    & \multicolumn{3}{c}{Money} & \multicolumn{3}{c}{Telcm} & \multicolumn{3}{c}{Enrgy} & \multicolumn{3}{c}{Utils} \\
\midrule
     year   & $R^2_{\mathrm{OOS}}$ & t-stat &    N & $R^2_{\mathrm{OOS}}$ & t-stat &    N & $R^2_{\mathrm{OOS}}$ & t-stat &    N & $R^2_{\mathrm{OOS}}$ & t-stat &   N \\
\midrule
   2015 &              18.13\% &   4.41 & 1153 &               1.68\% &   0.00 &  403 &              -4.21\% &  -1.45 &  303 &              18.47\% &   3.69 &  401 \\
   2016 &               7.64\% &   3.53 & 1108 &             -10.18\% &  -1.22 &  329 &              15.80\% &   2.83 &  252 &              -1.84\% &  -0.23 &  310 \\
   2017 &               1.77\% &   0.46 & 1071 &               8.60\% &   5.16 &  307 &             -18.18\% &  -8.75 &  307 &             -39.46\% &  -6.99 &  315 \\
   2018 &              13.55\% &   8.48 &  879 &              10.65\% &   2.46 &  219 &               8.38\% &   1.29 &  275 &             -23.76\% &  -5.53 &  215 \\
   2019 &               2.10\% &   2.25 &  977 &               1.53\% &   0.52 &  351 &             -10.28\% &  -9.07 &  340 &              -2.36\% &  -0.27 &  193 \\
   2020 &               1.42\% &   0.76 &  952 &              -8.84\% &  -2.83 &  360 &               5.72\% &   2.62 &  434 &              -7.29\% &  -2.36 &  165 \\
   2021 &               2.62\% &   0.47 &  614 &               6.15\% &   3.27 &  188 &             -15.04\% &  -6.83 &  251 &              -4.01\% &  -1.16 &  103 \\
   2022 &               4.34\% &   2.13 &  667 &              12.84\% &   3.50 &  143 &              16.14\% &   4.44 &  286 &               8.31\% &   2.71 &  132 \\
   2023 &              -0.63\% &  -0.18 &  569 &               8.99\% &   6.89 &  182 &             -27.33\% &  -5.03 &  290 &              -6.43\% &  -1.12 &  120 \\
Overall &               5.18\% &   3.31 & 7990 &               3.32\% &   1.39 & 2482 &               1.90\% &   0.31 & 2738 &              -1.97\% &  -0.88 & 1954 \\
\bottomrule
    \end{tabularx}
    \end{center}
\end{table}
\end{landscape}

\newpage
\begin{table}[!htb]
\caption{Summary Statistics of Alternative Analyst Information Value Estimations}\label{tab:iv_ss_robust}
    {\footnotesize
    This table reports summary statistics for several alternative estimations of analyst information value for S\&P 100 stocks from 2015Q1 to 2023Q4. These alternatives modify different components of the main information value calculation in Table \ref{tab:iv_ss} (defined as explained return variance divided by price impact). The `TAQ volatility' specification calculates explained variance as $\frac{r^2 - (r - \widehat{r})^2}{r^2} \cdot \sigma_v^2$ using one-minute log returns, while the `Intraday' version excludes overnight returns from this calculation. The `EMO' and `CLNV' specifications use alternative trade-signing algorithms from \citet{ellis2000underwriter} and \citet{lee1991inferring} to calculate price impact. For each specification, the table reports the mean, standard error (SE), and 95\% and 99\% confidence intervals, with all estimates computed with the delta method and expressed in millions of 2020 dollars.}
    \begin{center}
    \scriptsize
    \tabcolsep = 0.23cm
    \renewcommand{\arraystretch}{1.2}
    \begin{tabularx}{\textwidth}{lYYYYY}
    \toprule
    {} &  Mean &    SE &        95\%CI &        99\%CI &      N \\
    \midrule
    Information value (TAQ volitility), \$M            &  0.47 &  0.05 &  [0.38, 0.56] &  [0.35, 0.58] &  17669 \\
    Information value of text (TAQ volitility), \$M    &  0.47 &  0.05 &  [0.38, 0.56] &  [0.35, 0.58] &  17672 \\
    Information value of revisions (TAQ volitility), \$M &  0.42 &  0.04 &   [0.34, 0.50] &  [0.31, 0.53] &  17672 \\
    Information value (Intraday), \$M                  &  0.42 &  0.04 &  [0.34, 0.51] &  [0.32, 0.53] &  17672 \\
    Information value of text (Intraday), \$M          &  0.38 &  0.04 &   [0.30, 0.46] &  [0.28, 0.48] &  17669 \\
    Information value of revisions (Intraday), \$M     &  0.38 &  0.04 &   [0.30, 0.46] &  [0.28, 0.48] &  17672 \\
    Information value (EMO), \$M                       &  0.34 &  0.04 &  [0.27, 0.41] &  [0.25, 0.43] &  17672 \\
    Information value of text (EMO), \$M               &  0.35 &  0.04 &  [0.27, 0.42] &  [0.25, 0.44] &  17672 \\
    Information value of revisions (EMO), \$M          &  0.34 &  0.04 &  [0.26, 0.43] &  [0.23, 0.46] &  17669 \\
    Information value (CLNV), \$M                      &  0.34 &  0.04 &  [0.26, 0.43] &  [0.23, 0.46] &  17672 \\
    Information value of text (CLNV), \$M              &  0.31 &  0.04 &  [0.23, 0.39] &  [0.21, 0.41] &  17672 \\
    Information value of revisions (CLNV), \$M         &  0.31 &  0.04 &  [0.23, 0.39] &  [0.21, 0.42] &  17672 \\
    \bottomrule
    \end{tabularx}
    \end{center}
\end{table}

\newpage
\section*{C. Machine Learning Models} \label{ml}
\normalsize

\subsection*{Ridge Regression}

Ridge regression addresses multicollinearity by adding a regularization term to the least squares objective function. The ridge regression estimator is given by:
\begin{equation}
\widehat{\beta} = \underset{\beta}{\operatorname{argmin}} \left\{ \|y_i - X_i^T \beta\|_2^2 + \alpha \|\beta\|_2^2 \right\},
\end{equation}
where \( \alpha \) is the regularization parameter that controls the trade-off between fitting the data and shrinking the coefficients. 

To find the optimal value of \( \alpha \), cross-validation is used over a grid of values ranging from \( 10^{-10} \) to \( 10^{10} \). The cross-validation process ensures that the chosen model generalizes well to unseen data, preventing overfitting while capturing the predictive power of the text embeddings.

\subsection*{Partial Least Square Regression}

To mitigate the risk of overfitting inherent in high-dimensional text embeddings, I employ Partial Least Squares (PLS) for dimensionality reduction.

The optimization problem can be expressed as follows:
\begin{equation}
    \hat{\beta} = \underset{\beta}{\operatorname{argmin}} \left( \| (\Omega^T X_i) \beta - y_i \|_2^2 \right),
\end{equation}
where $\Omega$ is a $K \times P$ transformation matrix that reduces the $K$ predictors in \( X_i \) to $P$ lower-dimensional components.

The extraction of the \( j \)-th PLS component is guided by the following objective function:
\begin{equation}
    \omega_j = \underset{\omega}{\text{argmax}} \, \text{Cov}(Y, X\omega), \quad \text{s.t.} \,\, \omega'\omega = 1, \,\, \text{Cov}(X\omega, X\omega_i) = 0 \,\, \forall i < j.
\end{equation}

In essence, this approach sequentially extracts components that maximize the covariance with the outcome variable, while ensuring orthogonality to previously extracted components.

\subsection*{Extreme Gradient Boosting}
Tree-based approaches are commonly applied in stock return forecasting literature (see, e.g., \citeay{gu2020empirical}; \citeay{cao2024man}; \citeay{bonini2023value}). XGBoost is an advanced implementation of tree-based machine learning models that builds an ensemble of decision trees. In XGBoost, each tree is added sequentially to correct the errors of previous trees. The main idea is to combine the outputs of multiple weak learners (decision trees) to create a strong learner. XGBoost incorporates regularization techniques, such as L1 (Lasso) and L2 (Ridge), to prevent overfitting and enhance model generalization.

In comparison to random forests, which build a multitude of independent trees and aggregate their predictions, XGBoost constructs trees sequentially, with each tree designed to correct the errors of the preceding ones. While random forests rely on bagging, a method that combines the predictions of various trees to reduce variance, XGBoost uses boosting, an approach that aims to reduce both bias and variance by focusing on difficult-to-predict instances in subsequent iterations. This boosting approach allows XGBoost to effectively capture and leverage the nuanced information embedded in textual data, leading to a more accurate estimation of stock returns.

\subsection*{Neural Networks}
To extend beyond the linear modeling approach, I explore the use of Neural Networks to estimate $CAR$ using text embeddings derived from analyst reports. A Neural Network can capture complex non-linear relationships between the text embeddings and the $CAR$, potentially improving the model's fitting capabilities.
Consider a three-layer Neural Network as the prediction model. The prediction problem can be formulated as follows:
\begin{equation}
f(X_i; \theta) = W_3 \sigma\left(W_2 \sigma\left(W_1 X_i + b_1\right) + b_2\right) + b_3, \label{eqnn1}
\end{equation}
where $\sigma(\cdot)$ is the ReLU activation function, $W_i$ and $b_i$ represent the weights and biases for layer $i$, respectively.
The Neural Network architecture consists of an input layer representing the text embeddings, followed by multiple hidden layers. The specific architecture employed includes 32 neurons in the first hidden layer, followed by 16, 8, 4, and 2 neurons in subsequent layers. This structure is flexible and can be adjusted by adding or removing layers as necessary to optimize performance.

The training process involves optimizing the weights and biases to minimize the loss function, which, in this case, is the mean-square loss. To regularize the model and prevent overfitting, early stopping is implemented, halting training once the validation loss ceases to decrease. Additionally, following \citet{gu2020empirical}, the models are retrained five times, and the final estimation is obtained by averaging the outputs of these five models, forming an ensemble estimation.

\section*{D. Theory}

I provide an intuition for the measure of strategic value by discussing a simple extension of \citet{kyle1985continuous} model. In the Kyle model, there is one risky asset with a payoff $\tilde{v} \sim N\left(p_0, \Sigma_0\right)$. Three types of traders exist: a strategic trader with insider information, a market maker who sets prices in a perfectly competitive market, and an uninformed trader who trades $\tilde{u} \sim N\left(0, \sigma_u^2\right)$, where $\tilde{u}$ is independent of $\tilde{v}$. Illiquidity is measured by Kyle's lambda ($\lambda$). Kyle's lambda depends on private information and liquidity trading. I extend the model by considering a case where $\phi$ percentage of variance in $\tilde{v}$ is explained by the informed trader's information.

\subsection*{Single Auction Model}

There are two periods, $t_0$ and $t_1$. The asset is traded with asymmetric information at period $t_0$, and the value $\tilde{v}$ is realized at period $t_1$. I assume without loss of generality that
$$
\tilde{v} = P_0 + \tilde{s} + \tilde{\epsilon},
$$
where $\tilde{s}$ is a mean-zero signal observed by the informed trader at $t_0$, and $\tilde{\epsilon}$ is the combination of residual information and noise. I assume that the signal $\tilde{s}$ and residual $\tilde{\epsilon}$ are uncorrelated, that is, $\sigma_v^2 = \sigma_s^2 + \sigma_\epsilon^2$.

Let
$$
\phi=\frac{\operatorname{var}(\tilde{s})}{\operatorname{var}\left(\tilde{v}\right)} = \frac{\sigma_s^2}{\sigma_v^2} .
$$

This measure $\phi$ is the "R-square" of projecting $\tilde{v}$ on $\tilde{s}$, or the percentage of explainable variance in $\tilde{v}$ using signal $\tilde{s}$.

After observing $\tilde{s}$, the informed trader submits a market order $\tilde{x}=X(\tilde{s})$, and the uninformed trader trades a zero-mean random variable $\tilde{z}$ that is normally distributed and independent of $\tilde{v}$. The profit of the informed trader is given by: $\tilde{\pi}=(\tilde{v}-\tilde{p})\tilde{v}$. The insider has a rational guess of $P(\tilde{x}+\tilde{u})$ and understands that his order $\tilde{x}$ will move the price against him.

The market maker observes the order flow $\tilde{y} \stackrel{\text{def}}{=} \tilde{x} + \tilde{z}$ and determines the equilibrium price $\tilde{p}=P(\tilde{x}+\tilde{u})$ to break even. The assumptions for the market maker are risk neutrality and perfect competition, which drives the profits for market makers to zero.

\subsection*{Equilibrium}

An equilibrium is a set of $X$ and $P$ satisfying
$$
E[\tilde{\pi}(X, P) \mid \tilde{s}=s],
$$
$$
\tilde{p}(X, P)=E[\tilde{v} \mid \tilde{x}+\tilde{u}].
$$

The first condition is profit maximization, stating that given the market maker's pricing rule, the insider chooses a strategy $X$ maximizing her conditional expected profit, taking into account the pricing rule. The second condition is market efficiency. Given the insiders' trading strategy, the market maker sets the price to be the expected value of the security.

Conjecture:
$$
\begin{aligned}
& P(\tilde{y})=\mu+\lambda \tilde{y}, \\
& X(\tilde{s})=\alpha+\beta \tilde{s}.
\end{aligned}
$$

The profit of the insider can be written as
$$
\begin{aligned}
E[\tilde{\pi}(X, P) \mid \tilde{s}=s] & = E[(\tilde{v}-\mu-\lambda(\tilde{u}+x)) x \mid \tilde{s}=s] \\
& =(P_0 + s + -\mu-\lambda x) x.
\end{aligned}
$$

Traders take into account that her order flow will move the price against her, which serves to restrain her position size.

Solving for optimal profit, I get
$$
x^*=\frac{P_0 + s-\mu}{2 \lambda}=\alpha+\beta v.
$$

Hence, I can express $\alpha$ and $\beta$ as
$$
\begin{gathered}
\beta=\frac{1}{2 \lambda}, \\
\alpha=\frac{P_0-\mu}{2 \lambda}= (P_0-\mu) \beta.
\end{gathered}
$$

When the market maker puts a higher weight on the order flow in setting the price, the trader puts a lower weight on her information.

I now look at the price-setting rule as
$$
\mu+\lambda y=E\{\tilde{v} \mid \alpha+\beta \tilde{s}+\tilde{u}=y\}.
$$

Essentially, the market maker observes a normally distributed signal about
$$
\begin{aligned}
\mu + \lambda y 
&= E[\tilde{v} \mid y] \\
&=\bar{v}+\frac{\operatorname{cov}(\tilde{v}, \alpha+\beta \tilde{s}+\tilde{u})}{\operatorname{var}(\alpha+\beta \tilde{s}+\tilde{u})}(\beta \tilde{s}-\beta \bar{s}+\tilde{u}) \\
&= p_0 + \frac{\beta \sigma_s^2}{\beta^2 \sigma_s^2 + \sigma_u^2} (y - \alpha).
\end{aligned}
$$

Hence, I can express $\lambda$ and $\mu$ as
$$
\begin{aligned}
& \lambda=\frac{\beta \sigma_s^2}{\beta^2 \sigma_s^2 + \sigma_u^2},\\
& \mu= p_0-\alpha \lambda.
\end{aligned}
$$

There is a unique linear equilibrium given by

$$
\begin{aligned} 
& \mu= P_0, \\ 
& \lambda=\frac{\sigma_s}{2 \sigma_u}, \\ 
& \alpha= 0, \\ 
& \beta=\frac{\sigma_u}{\sigma_s}.
\end{aligned}
$$

\subsection*{Discussion}

\paragraph{Kyle's Lambda}
The parameter $\lambda$ is universally known as Kyle's lambda. Formally, it is the impact on the equilibrium price of a unit order. Its reciprocal ($1 / \lambda$) measures the liquidity (or depth) of the market. If $1 / \lambda = \frac{2 \sigma_u}{\sigma_s}$ is larger, then the market is more liquid, either because there is less private information in $\sigma_s$ or there is more liquid trade in the sense of $\sigma_u$.

\paragraph{Value of Private Information} 

Notice that the equilibrium strategy of the informed trader is $\tilde{x}=\beta s$. The unconditional expected gain of the informed trader is
$$
\begin{aligned}
\mathrm{E}[\tilde{x}[\tilde{v}-p(\tilde{x}+\tilde{z})]] & =\beta \mathrm{E}[s(\tilde{v}-\mu-\lambda \beta s -\lambda \tilde{u})] \\
& =\beta(1-\lambda \beta) \sigma_s^2 \\
& =\frac{\sigma_s \sigma_u}{2} \\
& =\frac{\phi \sigma_v^2}{4\lambda}.
\end{aligned}
$$

The expected gain for the informed trader is maximized when she has more private information or when there is more liquidity trading. On the other hand, liquidity traders incur losses equivalent to the informed trader's gains, but they accept these losses due to other motives for trading. The equilibrium price ensures that market makers do not profit or lose in expectation.

\section*{E. Delta Method Approximation}

I use the delta method to approximate the expected value and variance of a ratio of two random variables. This method relies on a first-order Taylor expansion. Specifically, consider two random variables $X$ and $Y$ with means $\mu_X$ and $\mu_Y$, and variances $\sigma_X^2$ and $\sigma_Y^2$, respectively. We are interested in the ratio $Z = \frac{X}{Y}$ and want to approximate the mean $E[Z]$ and the variance $\text{Var}(Z)$.

\subsection*{Delta Method Approximation}

The key idea is to approximate the function $g(X, Y) = \frac{X}{Y}$ using a Taylor series expansion around the means $\mu_X$ and $\mu_Y$.
For the function $g(X, Y)$, I use a first-order Taylor expansion around $(\mu_X, \mu_Y)$:
\[
g(X, Y) \approx g(\mu_X, \mu_Y) + \left. \frac{\partial g}{\partial X} \right|_{(\mu_X, \mu_Y)} (X - \mu_X) + \left. \frac{\partial g}{\partial Y} \right|_{(\mu_X, \mu_Y)} (Y - \mu_Y).
\]
The partial derivatives of $g(X, Y) = \frac{X}{Y}$ are
\[
\frac{\partial g}{\partial X} = \frac{1}{Y}, \quad \frac{\partial g}{\partial Y} = -\frac{X}{Y^2}.
\]
Evaluating these at $(\mu_X, \mu_Y)$, I get
\[
\left. \frac{\partial g}{\partial X} \right|_{(\mu_X, \mu_Y)} = \frac{1}{\mu_Y}, \quad \left. \frac{\partial g}{\partial Y} \right|_{(\mu_X, \mu_Y)} = -\frac{\mu_X}{\mu_Y^2}.
\]
Substituting the partial derivatives into the Taylor expansion, I obtain
\[
\frac{X}{Y} \approx \frac{\mu_X}{\mu_Y} + \frac{1}{\mu_Y} (X - \mu_X) - \frac{\mu_X}{\mu_Y^2} (Y - \mu_Y).
\]
Taking expectations on both sides gives
\[
E\left[ \frac{X}{Y} \right] \approx \frac{\mu_X}{\mu_Y} + \frac{1}{\mu_Y} E[X - \mu_X] - \frac{\mu_X}{\mu_Y^2} E[Y - \mu_Y].
\]
Since $E[X - \mu_X] = 0$ and $E[Y - \mu_Y] = 0$, the approximation can be simplified to
\[
E\left[ \frac{X}{Y} \right] \approx \frac{\mu_X}{\mu_Y}.
\]
Using the delta method, the variance of $Z=\frac{X}{Y}$ is approximated by
$$
\operatorname{Var}(Z) \approx\left(\frac{\partial g}{\partial X}\right)^2 \operatorname{Var}(X)+\left(\frac{\partial g}{\partial Y}\right)^2 \operatorname{Var}(Y)+2\left(\frac{\partial g}{\partial X}\right)\left(\frac{\partial g}{\partial Y}\right) \operatorname{Cov}(X, Y) .
$$
Substituting the partial derivatives yields
$$
\operatorname{Var}\left(\frac{X}{Y}\right) \approx \frac{\operatorname{Var}(X)}{\mu_Y^2}+\frac{\mu_X^2 \operatorname{Var}(Y)}{\mu_Y^4}-2 \frac{\mu_X \operatorname{Cov}(X, Y)}{\mu_Y^3} .
$$

\subsection*{Application to Value of Information}
In the context of the provided study, I estimate the mean value of information for a subsample $s$ using the delta method.

Let $\widehat{\Omega}_{it} = \frac{\widehat{\sigma}_{it}^2}{\widehat{\lambda}_{i t} / P_{i t_-}} = \frac{r_{i f}^2-\left(r_{i t}-\frac{\sum_{j=1}^N \widehat{r}_{i j t}}{N}\right)^2}{\widehat{\lambda}_{i t} / P_{i t_-}}$ be the value of information for stock $i$ on date $t$. Define the mean and variance over subsample $s$ as
\[
\mu_{\nu s} = \frac{1}{|s|} \sum_{it \in s} \left(r_{i f}^2-\left(r_{i t}-\frac{\sum_{j=1}^N \widehat{r}_{i j t}}{N}\right)^2\right),
\]
and the mean price impact per dollar over subsample $s$ as
\[
\mu_{\lambda s} = \frac{1}{|s|} \sum_{it \in s} \frac{\widehat{\lambda}_{it}}{P_{i t_-}}.
\]
Using the delta method approximation for the mean of the ratio, the mean value of information over subsample $s$ is given by
\[
E\widehat{\Omega}_s \approx \frac{\mu_{\nu s}}{\mu_{\lambda s}}.
\]
The variance of the ratio $\widehat{\Omega}_{it} = \frac{\widehat{\sigma}_{it}^2}{\widehat{\lambda}_{i t} / P_{i t_-}}$ over the subsample $s$ is estimated as
\[
\text{Var}\left( \widehat{\Omega}_s \right) \approx \frac{1}{\mu_{\lambda s}^2} \left( \Sigma_{\nu s} + \frac{\mu_{\nu s}^2}{\mu_{\lambda s}^2} \Sigma_{\lambda s} - 2 \frac{\mu_{\nu s}}{\mu_{\lambda s}} \Sigma_{\nu \lambda s} \right),
\]
where $\Sigma_{\nu s}$, $\Sigma_{\lambda s}$, and $\Sigma_{\nu \lambda s}$ are defined as
\[
\Sigma_{\nu s} = \frac{1}{|s|} \sum_{it \in s} \left( \widehat{\sigma}_{it}^2 - \mu_{\nu s} \right)^2,
\]
\[
\Sigma_{\lambda s} = \frac{1}{|s|} \sum_{it \in s} \left( \frac{\widehat{\lambda}_{it}}{P_{i t_-}} - \mu_{\lambda s} \right)^2,
\]
\[
\Sigma_{\nu \lambda s} = \text{Cov} \left( \widehat{\sigma}_{it}^2, \frac{\widehat{\lambda}_{it}}{P_{i t_-}} \right).
\]

This provides a first-order approximation for the mean and variance of the value of information over the subsample $s$ using the delta method.

\end{document}